\documentclass[aps,twocolumn,floatfix]{revtex4}
\usepackage{graphicx}
\usepackage{amsmath}
\usepackage{color}
\newcommand{\be}{\begin{equation}}
\newcommand{\ee}{\end{equation}}
\newcommand{\bea}{\begin{eqnarray}}
\newcommand{\eea}{\end{eqnarray}}
\newcommand{\bs}{\begin{split}}
\newcommand{\bse}{\begin{subequations}}
\newcommand{\ese}{\end{subequations}}
\newcommand{\ecp}{${\rm EuCo_2P_2}$}
\newcommand{\tcs}{${\rm ThCr_2Si_2}$}

\begin{document}
 
\title{Magnetic Structure and Magnetization of Helical Antiferromagnets in High Magnetic Fields Perpendicular to the Helix Axis at Zero Temperature}
\author {David C.\ Johnston} 
\affiliation {Ames Laboratory and Department of Physics and Astronomy, Iowa State University, Ames, Iowa 50011}

\date{\today}

\begin{abstract}

The zero-temperature angles of magnetic moments in a helix or sinusoidal fan confined to the $xy$~plane, with respect to an in-plane magnetic field $H_x$ applied perpendicular to the $z$~axis of a helix or fan, are calculated for commensurate helices and fans with field-independent turn angles $kd$ between moments in adjacent layers of the helix or fan using the classical $J_0$-$J_{1}$-$J_{2}$ Heisenberg model. For $0<kd<4\pi/9$, first-order transitions from helix to a fan structure occur at fields~$H_{\rm t}$ as previously inferred, where the fan is found to be approximately sinusoidal.  However, for \mbox{$4\pi/9 \leq kd \leq \pi$,} different behaviors are found depending on the value of~$kd$ and these properties vary nonmonotonically with~$kd$. In this $kd$~range, the change from helix to fanlike structure is usually a crossover with no phase transition between them, although first-order transitions are found for $kd = 3\pi/5$ and $8\pi/11$ and a second-order transition for $kd = 3\pi/4$.  At a critical field $H_{\rm c}$, the fan or fanlike structures exhibit a second-order transition to the paramagnetic state.  The $H_{\rm c}$ for a helix undergoing a field-induced change to a fan or fanlike structure is found to be the same as for a sinusoidal fan with the same $kd$ and interlayer interactions.  Analytical expressions for $H_{\rm c}$ versus~$kd$ are presented.  We also calculated the average $x$-axis moment per spin~$\mu_{x{\rm ave}}$ versus $H_x$ for helices and fans with crossovers and phase transitions between them.  When smooth helix to fanlike crossovers occur in the range $4\pi/9 \leq kd \leq \pi$, $\mu_{x{\rm ave}}$ exhibits an S-shape behavior with increasing $H_x$.  This predicted behavior is consistent with $\mu_{x{\rm ave}}(H_x)$ data previously reported by Sangeetha, {\it et al.}\ [Phys.\ Rev.\ B {\bf 94}, 014422 (2016)] for single-crystal \ecp\ possessing a helix ground state with $kd\approx 0.85\pi$.  The low-field magnetic susceptibility and the ratio $H_{\rm t}/H_{\rm c}$ are calculated analytically or numerically versus $kd$ for helices, and are shown to approach the respective known limits for $kd\to0$.

\end{abstract}

\maketitle

\section{Introduction}

\begin{figure}
\includegraphics [width=2.in]{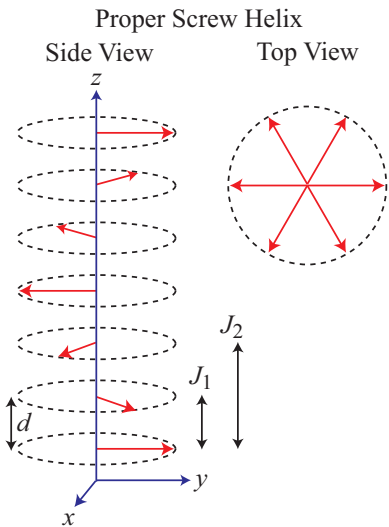}
\caption {(Color online) Generic helical AFM structure \cite{Johnston2012}.  Each arrow represents a layer of moments perpendicular to the $z$~axis that are ferromagnetically aligned within the $xy$ plane and with interlayer separation $d$.  The wave vector {\bf k} of the helix is directed along the $z$~axis.  The magnetic moment turn angle between adjacent magnetic layers is $kd$.  The nearest layer and next-nearest layer exchange interactions $J_{1}$ and $J_{2}$, respectively, within the $J_0$-$J_{1}$-$J_{2}$ Heisenberg MFT model are indicated.}
\label{Fig:J0_Jz1_Jz2_model_helix}
\end{figure}

The unified molecular field theory (MFT) for a spin lattice containing identical crystallographically-equivalent spins treats the magnetic and thermal properties of collinear and coplanar noncollinear Heisenberg antiferromagnets (AFMs) on the same footing \cite{Johnston2012, Johnston2015, Johnston2015b}.  This formulation has the added advantage that the theory is expressed in terms of readily measured quantities such as the spin~$S$, the ordering temperature $T_{\rm N}$, the magnetic susceptibility $\chi$ at $T=0$ and at $T_{\rm N}$, the Weiss temperature $\theta_{\rm p}$ in the Curie-Weiss law describing $\chi(T>T_{\rm N})$, and for a planar helix as shown in Fig.~\ref{Fig:J0_Jz1_Jz2_model_helix}, the turn angle $kd$ along the helix $z$~axis between adjacent layers of ferromangetically (FM) aligned spins in zero magnetic field~$H$\@. Here $k$ is the wavevector of the helix along the $z$~axis and $d$ is the distance between adjacent layers of moments aligned FM in the $xy$ plane.  The theory quantitatively describes the thermal and helical magnetic properties of \ecp\ single crystals, which is therefore considered to be a prototype for a helical AFM obeying the unified MFT \cite{Sangeetha2016}.   The MFT was recently extended to include the influences of magnetic-dipole and uniaxial magnetocrystalline anisotropies on the thermal and magnetic properties of Heisenberg AFMs \cite{Johnston2016, Johnston2017}.

In zero field the angle~$\phi$ between the $x$~axis and the FM-aligned moments within the $xy$~plane in layer~$n$ is given by the linear relation
\be
\phi_{n0} = n\,kd.
\label{Eq:phin0Helix}
\ee
The above MFT was used to derive the anisotropic $\chi(T)$ and thermal properties of collinear and coplanar noncollinear AFMs in zero or low field at $T\leq T_{\rm N}$.  In addition the average magnetic moment per spin $\mu_{z{\rm ave}}$ with high fields applied parallel to the helix $z$~axis and the associated critical field were derived.  However, in those studies the average magnetic moment $\mu_{x{\rm ave}}$ of a helical AFM structure versus high in-plane magnetic field $H_x$ was not calculated.

Previous work on the influence of a large $H_x$ on the classical magnetic structure of a helix at $T=0$ indicated that with increasing $H_x$, the circular hodograph on the right side of Fig.~\ref{Fig:J0_Jz1_Jz2_model_helix} described by Eq.~(\ref{Eq:phin0Helix}) first becomes distorted, and then a transition to a fan structure may occur at a field $H_{\rm t}$ in which the twofold rotational symmetry axis of the fan is aligned with the $x$~axis \cite{Nagamiya1962, Kitano1964, Nagamiya1967}.  It was also established that the wavevector of a helix changes when a large in-plane field is applied \cite{Nagamiya1962}.  A numerical study of the phase diagram in the plane of the nearest- and next-nearest layer interactions $J_{1}$ and~$J_{2}$, respectively (see Fig.~\ref{Fig:J0_Jz1_Jz2_model_helix}), was carried out including the field-dependent helix or fan wavevector \cite{Robinson1970}.

From analysis of the short-range order at finite temperature calculated by the transfer matrix method and zero-temperature calculations of the minimum energy of commensurate configurations, it was concluded that when the turn angle is in the range $0 < kd < \pi/2$, the helix to fan transition is first order, whereas for $\pi/2 < kd < \pi$ the change is continuous \cite{Carazza1991}.  We find differences from this conclusion in both $kd$ ranges.  In particular, in the range $0 < kd < \pi/2$, for $kd = 4\pi/9$, we find a smooth crossover between the helix and fan phases with no phase transition.  In the range $\pi/2 < kd < \pi$, in addition to smooth crossovers with no phase transitions, we find first- and second-order transitions between the helix and fan phases for $kd = 8\pi/11$ and $kd = 3\pi/4$, respectively.

When a helix undergoes a transition to a sinusoidal fan structure in the $xy$~plane in a high field~$H_x$, perpendicular to the helix $z$~axis, the angle $\phi_{n}$ of ordered moment $\vec{\mu}_n$ with respect to the positive $x$~axis is
\be
\phi_{n} = \phi_{\rm max}\sin(n\,kd),
\label{Eq:phiFan}
\ee
where $\phi_{\rm max}>0$ is the amplitude of the fan.  We will show that under the assumption that the helix and fan wavevectors are the same and do not depend on the field, when the helix undergoes a first-order transition to a fan phase with increasing field, in general the fan phase is not sinusoidal, although the actual fan structure can be rather close to this structure depending on the value of the helix and fan wavevector.  In addition we find many instances where the distorted helix instead undergoes a continuous evolution instead of a phase transition into a fan or fanlike structure.

\begin{figure}
\includegraphics [width=3.3in]{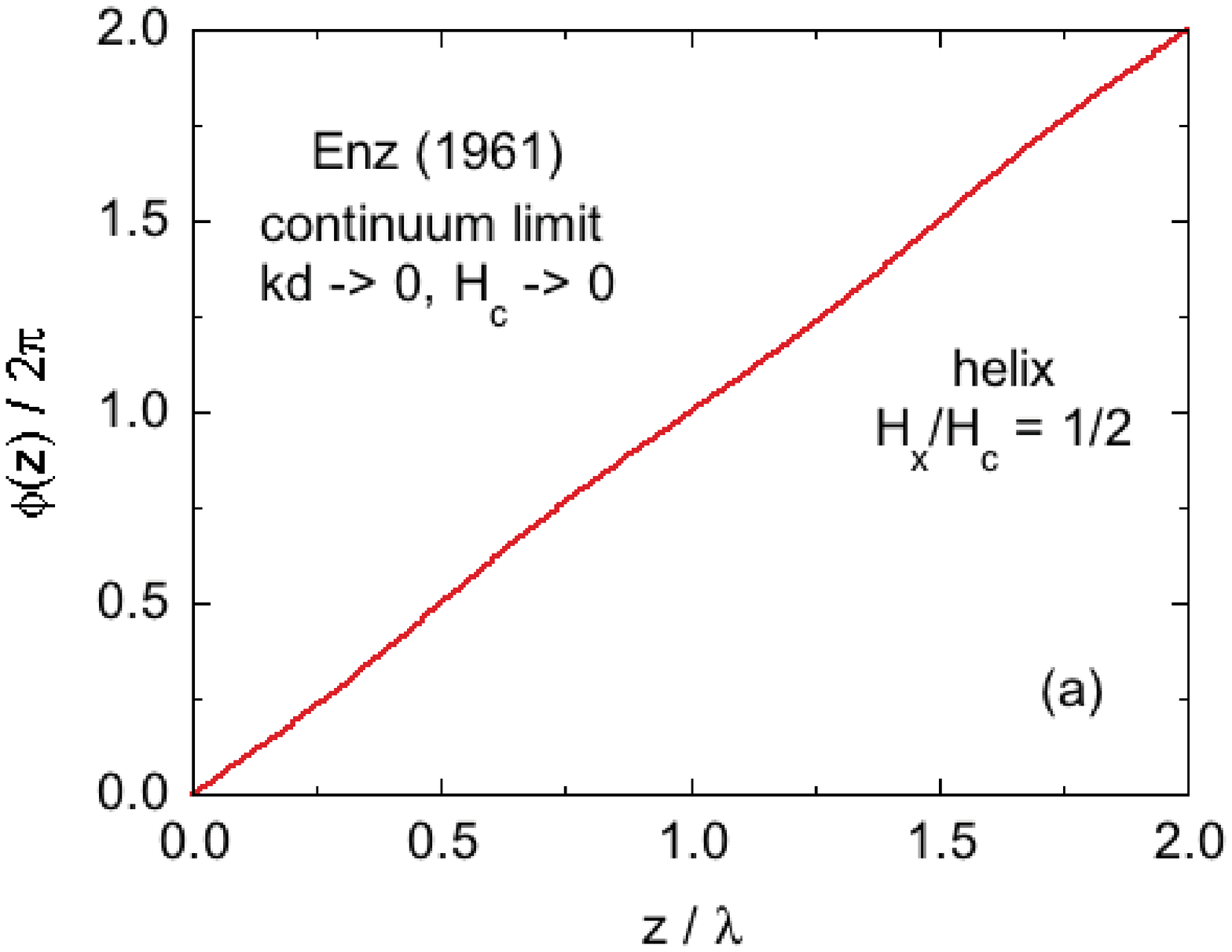}
\includegraphics [width=3.3in]{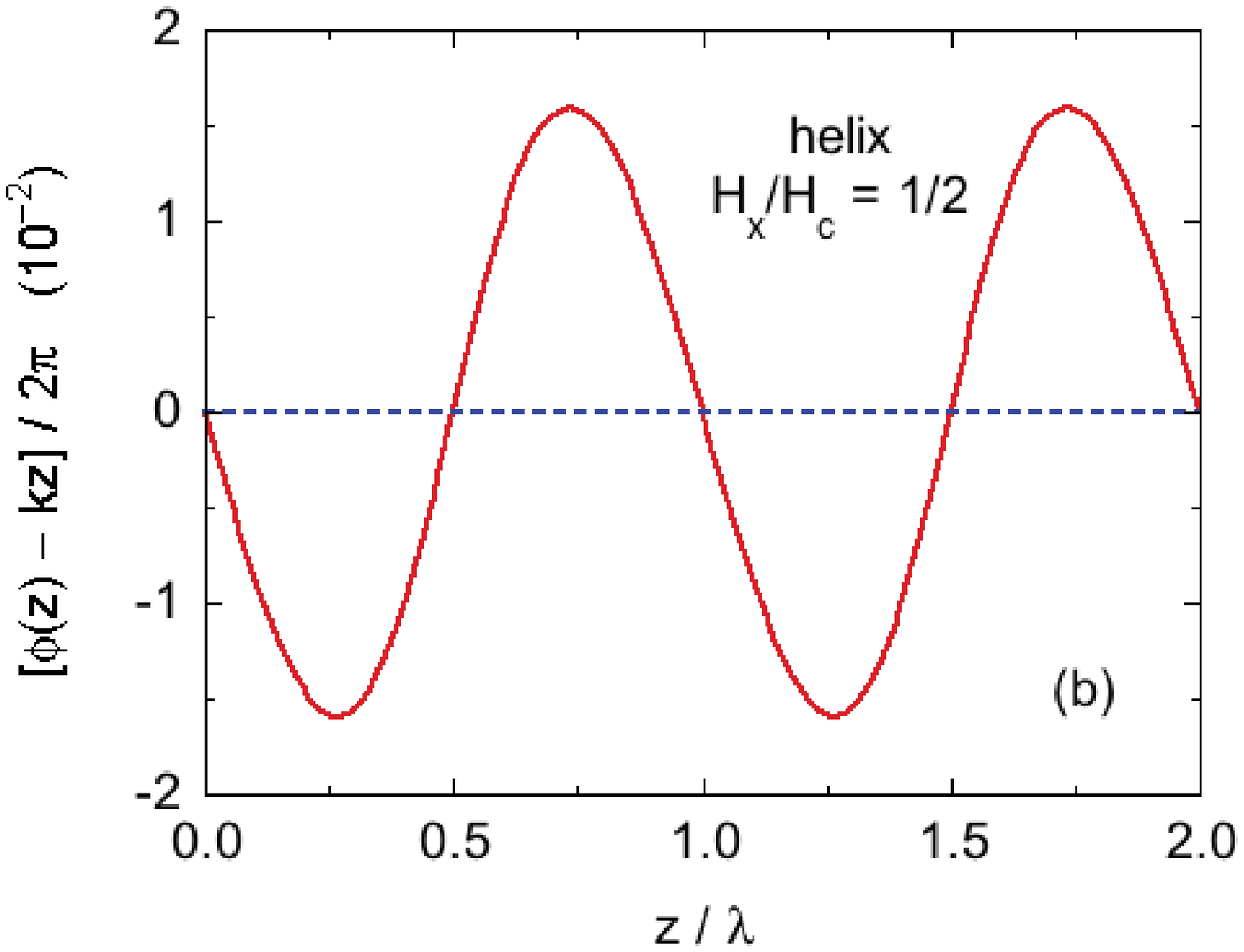}
\includegraphics [width=3.3in]{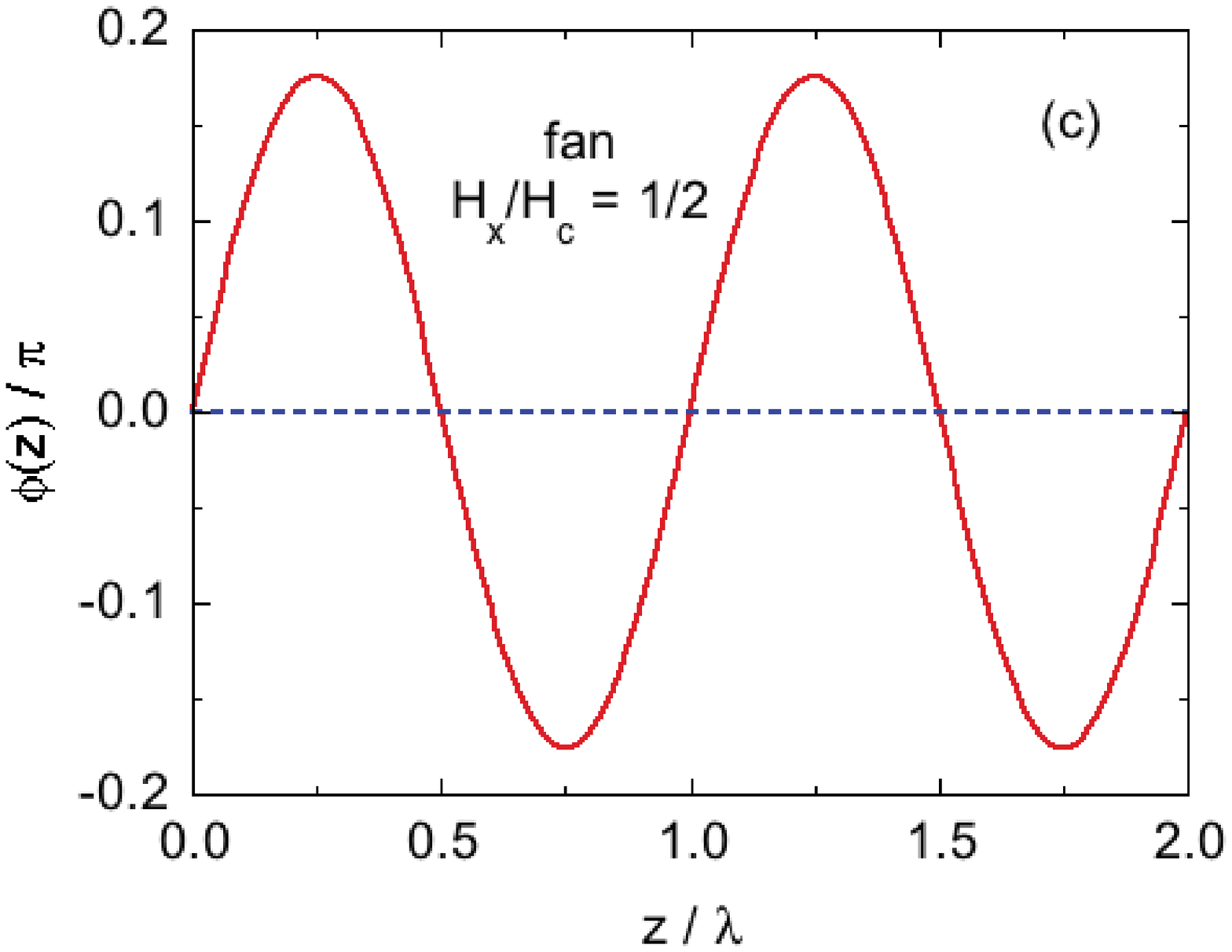}
\caption {(Color online) (a)~Phase angle $\phi(z)$ with respect to the $+x$~axis of the moments in planes at height~$z$ along a helix with an infinitesimal angle $kd$ between adjacent moment layers versus $z/\lambda$ in an applied field $h_x/h_{\rm c} = 1/2$, which is the boundary between the first-order transition with field between the helix and fan phases.  (b)~The near-sinusoidal modulation of the linear background $kz$ phase of the helix in the field $h_x/h_{\rm c} = 1/2$.  (c)~The phase~$\phi$ of the moments in planes along the $z$~axis in the fan structure versus $z/\lambda$ with $h_x/h_{\rm c} = 1/2$.  These data were generated using the information in Ref.~\cite{Enz1961}}
\label{Fig:muxVShx_Enz_hxhc12}
\end{figure}

Information about the helix and fan structures versus applied field at zero temperature has been provided for the continuum case in which $kd\to0$ (almost ferromagnetic, see Fig.~\ref{Fig:J0_Jz1_Jz2_model_helix}) \cite{Enz1961}.  In the helix structure in a field, the angle $\phi(z)$ of the $xy$-plane-oriented moments with respect to the positive $x$~axis versus position along the helix $z$~axis is given by the linear term~(\ref{Eq:phin0Helix}) plus an approximately sinusoidal modulation described by \cite{Enz1961}
\be
\phi(z) = 2\pi\frac{z}{\lambda} - \frac{1}{5}\left(\frac{H_x}{H_{\rm c}}\right)\sin\phi(z)\qquad\Big(0\leq \frac{H_x}{H_{\rm c}} \leq 1/2\Big),
\label{Eq:phzHelix}
\ee
where $H_{\rm c}$ is the critical field at which the $x$~component of the moment reaches the saturation value.  The value $H_{\rm t} = H_{\rm c}/2$ is the field at which the structure changes from helix to fan in a first-order transition. The result for $H_x/H_c = 1/2$ for maximum modulation of the linear term versus $z/\lambda$ is shown in Fig.~\ref{Fig:muxVShx_Enz_hxhc12}(a).  The slightly distorted sine-wave modulation is shown in Fig.~\ref{Fig:muxVShx_Enz_hxhc12}(b), which has an amplitude of only 1.6\% of $2\pi$.  For the fan phase, the $\phi(z)$ is given by
\bse
\label{Eqs:phizFan}
\be
\phi(z) = \phi_{\rm max}\sin(2\pi z/\lambda) \quad (1/2 \leq H_x/H_{\rm c} \leq 1),
\ee
where the amplitude $\phi_{\rm max}$ of the sinusoidal fan is
\be
\phi_{\rm max} = \frac{4}{3}\left[\left(\frac{H_{\rm c}}{H_x}\right)^{1/2} - 1\right]\quad \left(\frac{1}{2}\leq \frac{H_x}{H_{\rm c}} \leq 1\right).
\ee
\ese
A plot of $\phi/\pi$ versus $z/\lambda$ for $H_x/H_{\rm c} = 1/2$ for the fan phase is shown in Fig.~\ref{Fig:muxVShx_Enz_hxhc12}(c).

The transition from the distorted helix phase with the moment angle in Eq.~(\ref{Eq:phzHelix}) and Figs.~\ref{Fig:muxVShx_Enz_hxhc12}(a) and~\ref{Fig:muxVShx_Enz_hxhc12}(b) to that of the fan in Eqs.~(\ref{Eqs:phizFan}) and Fig.~\ref{Fig:muxVShx_Enz_hxhc12}(c) is qualitatively similar to the transition from a periodically-modulated linear behavior to oscillating behavior for the simple pendulum with decreasing kinetic energy of the pendulum bob \cite{Belendez2007, Lima2010}.

In the above studies, the magnetic phase diagram at temperature $T=0$ showing the regions of stability of the helix and fan phases within the exchange-interaction parameter space was of primary interest, with very few presentations of magnetization versus field data.  Here we significantly extend these studies to provide a detailed comprehensive study of the $\phi_n(H_x)$ and $\mu_{x{\rm ave}}(H_x)$ from $H_x=0$ to $H_x=H_{\rm c}$ for a wide variety of discrete rational $kd$~values where we assume that $kd$ is not affected by the applied field.  A planar anisotropy is assumed to be present that is strong enough that the ordered moments remain in the $xy$~plane for the applied field ranges considered here.

\begin{figure}
\includegraphics [width=3.3in]{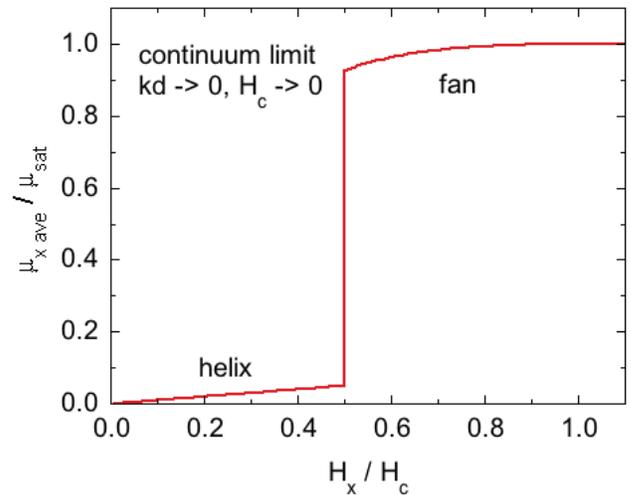}
\caption {(Color online) Our results for the reduced average moment $\mu_{x{\rm ave}}/\mu_{\rm sat}$ versus reduced field $H_x/H_{\rm c}$ for a helix and fan with an infinitesimal difference~$kd$ between the moment directions in adjacent moment layers.  A first-order transition between the helix and fan structures occurs at $H_x/H_{\rm c} = 1/2$ \cite{Enz1961}.}
\label{Fig:muxVShx_Enz}
\end{figure}

For example, for the helix and fan structures in the continuum limit $kd\to0$ discussed above that was studied in 1961 \cite{Enz1961}, the $\mu_{x{\rm ave}}(H_x)$ was not presented.  However, this can be obtained by averaging $\cos[\phi(z/\lambda)]$ for that structure over one wavelength~$\lambda$ for each value of $H_x/H_{\rm c}$.  Our result is shown for both the helix region $0\leq H_x/H_{\rm c} \leq 1/2$ and the fan region $1/2\leq H_x/H_c \leq 1$ in Fig.~\ref{Fig:muxVShx_Enz}.  For the helix phase, the magnetization is proportional to field with a susceptibility $(\mu_{x{\rm ave}}/\mu_{\rm sat})/(H_x/H_{\rm c}) = 1/10$.  In the fan phase, $\mu_{x{\rm ave}}$ for $H_x/H_{\rm c} = 1/2$ is already near saturation and varies nonlinearly upon increasing $H_x/H_{\rm c}$ to the value of unity.  A first-order transition in $\mu_{x{\rm ave}}(H_x)$ between the helix and fan phases occurs.  The data in this figure are very similar to those in Fig.~\ref{Fig:PhiMuxEnN12kd2PiOn12Helix} below for our smallest discrete value $kd=\pi/6$.


In addition to the intrinsic interest in the properties of helices and fans at high transverse fields, a primary motivation for the present work was to enable the high-field $\mu_{x{\rm ave}}(H_x)$ of real helix compounds at temperatures low compared with their~$T_{\rm N}$ to be fitted by theory.  \ecp\ has the \tcs\ structure with space group $I4/mmm$, with the Eu$^{+2}$ cations with spin $S=7/2$ and spectroscopic splitting factor $g=2$ occupying a body-centered tetragonal sublattice.  Neutron diffraction measurements on a single crystal demonstrated AFM ordering of the Eu spins at $T_{\rm N} = 66.5(5)$~K with no contribution from the Co atoms \cite{Reehuis1992}. The magnetic structure is a planar helix with the Eu ordered moments aligned in the $ab$~plane of the tetragonal structure ($xy$~plane here), with the helix axis along the perpendicular $c$~axis ($z$~axis here).  The value of the AFM propagation vector corresponds to a turn angle $kd$ between the ordered moments in adjacent layers of the helix at low temperatures given by
\be
kd\,(15~{\rm K}) = 0.852(4)\pi.
\label{Eq:kdNeuts}
\ee
This value of~$kd$ with $\pi/2 < kd < \pi$ indicates that the dominant {\it interlayer} interactions are AFM  \cite{Johnston2012, Johnston2015}, and dominant ferromagnetic (FM) {\it intralayer} interactions are then inferred from $\chi(T)$ measurements \cite{Sangeetha2016}.  

\begin{figure}
\includegraphics [width=3.4in]{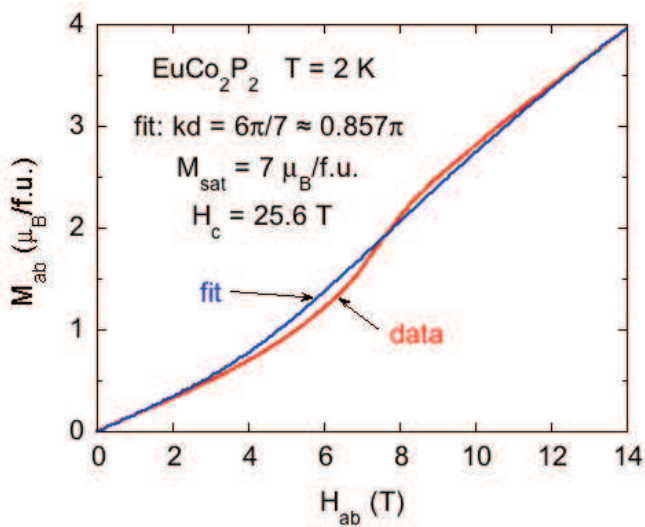}
\caption {(Color online) Magnetization $M_{ab}$ versus transverse applied $ab$-plane magnetic field $H_{ab}$ at a temperature of 2~K for tetragonal ${\rm EuCo_2P_2}$ with a zero-field $c$-axis helical magnetic structure (red data)  \cite{Sangeetha2016}.  The fit to the data by the $M_{xy}(H_{xy})$ prediction for a helix with turn angle $6\pi/7$ with saturation magnetization $M_{\rm sat} = 7~\mu_{\rm B}$/f.u.\ and critical field $H_{\rm c} = 25.6$~T is shown as the blue curve.}
\label{Fig:EuCo2P2_Mab_data_fit}
\end{figure}

A detailed study of the magnetic and thermal properties of \ecp\ single crystals was carried out recently \cite{Sangeetha2016}.  The low-field $\chi_{ab}$ below $T_{\rm N}$ was analyzed in terms of the above unified MFT and this theory was found \cite{Sangeetha2016} to accurately fit the data with a $kd$ value similar to the neutron result in Eq.~(\ref{Eq:kdNeuts}).  However, the high-field $M_{ab}(H_{ab})$ data, shown in Fig.~\ref{Fig:EuCo2P2_Mab_data_fit} \cite{Sangeetha2016}, did not agree with the conventional wisdom that instead of an S-shaped metamagnetic feature as seen in the data, a first-order transition should occur with a distinct discontinuity in $M_{ab}(H_{ab})$ at a transition field $H_{\rm t}<H_{\rm c}$, as in Fig.~\ref{Fig:muxVShx_Enz}.

This conundrum called for new calculations of the high-field $\mu_{x{\rm ave}}(H_x)$ for the helix.  Our calculations provide clear predictions for comparison with experiment. They are also a benchmark for comparing the magnetic structures and magnetization versus field data within this model with the properties of more complicated systems such as helices and fans with a field-dependent~$kd$ and/or including quantum and/or additional anisotropy effects to see what changes these additional features create in the results.  As a preview of our results, shown in Fig.~\ref{Fig:EuCo2P2_Mab_data_fit} is a fit to the $M_{ab}(H_{ab})$ data for \ecp\ by our theoretical prediction with $kd = 0.86\pi$, close to the value in Eq.~(\ref{Eq:kdNeuts}), which semiquantatively reproduces the S~shape of the data at the observed field.

Section~\ref{Sec:Theory} gives an outline of the general theory we use to calculate the field dependences of the moment angles and in-plane magnetization versus applied in-plane field for the helix and sinusoidal fan.  We assume for simplicity that the turn angle~$kd$ is independent of in-plane field and that the helix and fan are commensurate with the spin lattice as noted above.  In order to calculate the magnetization versus field for a helix or fan, it is necessary that the number of moment layers per wavelength be an integer, which in turn requires that the wave vector be commensurate with the spin lattice.  However, from calculations on commensurate helices/fans, we obtain results such as in Eq.~(\ref{Eq:EnHelix}) and Fig.~\ref{Fig:Helix_Zero_Field_Energies} and in Eqs.~(\ref{Eqs:hcVals}) below that we infer also apply to incommensurate wave vectors. The results for the dependences of the moment angles and magnetization of a sinusoidal fan structure are presented in Sec.~\ref{Sec:GenFanResults}.  Here we first discuss their dependences on $kd$ and the ratio $J_{12} \equiv J_{1}/J_{2}$.  Then we specialize to the cases where $J_{12}$ takes the value associated with a helix with the same~$kd$.  This allows direct comparison of the sinusoidal fan properties with the high-field fan phase of the helix.  This is of special interest because the moment angles in the field-induced fan originating from the helix are solved independently rather than enforcing a sinusoidal fan relationship on those angles.  Therefore the field-induced fan may not be sinusoidal and hence a comparison of the moment angles in that fan and the magnetization with the corresponding properties of a strictly sinusoidal fan is of significant interest.  The moment angles in the helix and the transverse magnetization versus transverse field are derived for specific values of $kd$ in Sec.~\ref{Sec:HelixFieldDep}.  We find that for $4\pi/9 \leq kd \leq \pi$, the evolution of the moment angles with increasing field from a distored helix to a fan or fanlike structure with increasing $kd$ can be either a crossover, a second-order transition, or a first-order transition with no obvious dependence of their order versus~$kd$.  For this reason, plots of the phase angles and magnetization versus applied field are given for many $kd$ values.  A summary of our results on helices with field-independent wavevector (that may transition to fans with the same wavevector) is given in Sec.~\ref{HelixDatSumm}, where a phase diagram is  constructed versus~$kd$.

\section{\label{Sec:Theory} Theory}
In this paper, we use the one-dimensional $J_0$-$J_{1}$-$J_{2}$ model \cite{Johnston2012,Johnston2015} for both the helix and fan phases, where $J_0$ is the sum of the Heisenberg exchange interactions between a representative spin and the other spins in the same $xy$ layer, $J_{1}$ is the sum of the interactions between the spin in one layer and the spins in either of the two nearest-neighbor layers along the $z$~axis, and $J_{2}$ is the sum of the interactions of the spin with the spins in either of the two next-nearest neighbor layers, as shown in  Fig.~\ref{Fig:J0_Jz1_Jz2_model_helix}.  In this model, the energy $E_n$ of a spin~$S$ with ordered moment~$\vec{\mu}_n$ consists of the sum of the exchange and Zeeman terms, given in general by \cite{Johnston2015}
\bea
E_n &=& \frac{S}{2}\bigg\{SJ_0 + 2SJ_{1} \big[\cos(\phi_{n+1} - \phi_n) \label{Eq:EGen}\\
&& +\ \cos(\phi_{n-1} - \phi_n)\big] + 2SJ_{2}\big[\cos(\phi_{n+2} - \phi_n) \nonumber\\
&& +\ \cos(\phi_{n-2} - \phi_n)\big]\bigg\} - \mu_n H_x \cos\phi_n,\nonumber
\eea
where $S$ is the spin angular momentum of a moment in units of $\hbar$, the prefactor of 1/2 is present because the exchange interaction energy between a pair of spins is equally shared between them, and $\phi_i$ is the angle between moment~$i$ and the positive $x$~axis.  A positive (negative)~$J$ is AFM (FM). All energies are normalized by (positive) $SJ_{2}$.  Also, we define the variables
\be
J_{02} \equiv \frac{SJ_0}{SJ_{2}} = \frac{J_0}{J_{2}}, \quad J_{12} \equiv \frac{J_{1}}{J_{2}}, \quad h_x =\frac{g\mu_{\rm B}H_x}{J_{2}},
\ee
where $g$ is the spectroscopic splitting factor and $\mu_{\rm B}$ is the Bohr magneton.  The magnitude $\mu_n$ of each spin~$n$ at zero temperature is
\be
\mu_{n} = \mu_{\rm sat} = gS\mu_{\rm B},
\ee
where $\mu_{\rm sat}$ is the saturation moment of a spin.  Then Eq.~(\ref{Eq:EGen}) becomes
\bse
\label{Eqs:EnEAve}
\bea
\frac{E_n}{SJ_{2}} &=& \frac{1}{2}\bigg\{J_{02} + 2J_{12} \big[\cos(\phi_{n+1} - \phi_n) \label{Eq:EGen2}\\
&& +\ \cos(\phi_{n-1} - \phi_n)\big] + 2\big[\cos(\phi_{n+2} - \phi_n) \nonumber\\
&& +\ \cos(\phi_{n-2} - \phi_n)\big]\bigg\} - h_x \cos\phi_n.\nonumber
\eea
The average energy per spin is
\be
\frac{E_{\rm ave}}{SJ_{2}} = \frac{1}{n_\lambda}\sum_{n=0}^{n_\lambda-1} \frac{E_n}{SJ_{2}},
\label{Eq:EAveGen}
\ee
\ese
where $n_\lambda$ is the number of FM-aligned layers per helix or fan wavelength~$\lambda$ along the $z$~axis.

For a sinusoidal fan structure, the only variable to solve for is the amplitude $\phi_{\rm max}(h_x, kd)$ which is obtained by minimizing the energy with respect to $\phi_{\rm max}$ for given values of $h_x$ and $kd$.  For a helix, the ground-state moment configuration is determined by minimizing $E_{\rm ave}$ with respect to the $n_\lambda$ values of $\phi_i$ in a helix for given values of $h_x$ and $kd$.  For both the fan and helix $\lambda$ is assumed to be independent of $h_x$ since $kd$ is assumed to be.  We will see that if a helical structure is assumed for low fields, a fan or fanlike structure, if they occur, is automatically generated with increasing $h_x$ when solving for the $\phi_i(h_x)$ that minimize $E_{\rm ave}$.  Thus we do not assume a sinusoidal fan structure {\it a priori} for a field-induced fan.  Indeed, we find that the fan or fanlike structures obtained are never perfectly sinusoidal except near the reduced critical field $h_{\rm c}$.  This observation is inferred from the fact that $h_{\rm c}$ for the field-induced fan or fanlike structure is identical to that for the sinusoidal fan calculated separately.

In the following two sections we apply the above general expressions first to the helix and then separately to the perfectly sinusoidal fan.

\subsection{\label{Sec:HelixThy} Helix Phases}

In this paper we assume that a helix is commensurate with the spin lattice and that its wavelength~$\lambda = n_\lambda d$ is independent of the applied field and contains $n_\lambda$ layers, where $d$ is the distance between FM-aligned moment layers along the $z$~axis (see Fig.~\ref{Fig:J0_Jz1_Jz2_model_helix}).  The commensurate wave vector $k$ of the helix is given in general by
\be
k = \frac{2\pi m}{\lambda} = \frac{2\pi m}{n_\lambda d},
\ee
where $m$ is a positive integer and $m/n_\lambda$ is an irreducible fraction.  The reason that the variable integer $m$ is included is because the helix or fan may be incommensurate for $m = 1$ but commensurate for $m>1$ (see Table~\ref{Tab:FanGndStateE} below).  Thus the magnitude of the turn angle between adjacent moment layers along the helix is
\be
kd =  \frac{2\pi m}{n_\lambda},
\label{Eq:kdDef}
\ee
where $2m/n_\lambda \leq 1$ so that $kd \leq \pi$. A $kd$ satisfying $\pi < kd < 2\pi$ would correspond to a helix with $kd^* = 2\pi - kd$ that would have the opposite helicity but the same magnetization versus $x$-axis field response.

\begin{figure}
\includegraphics [width=3.4in]{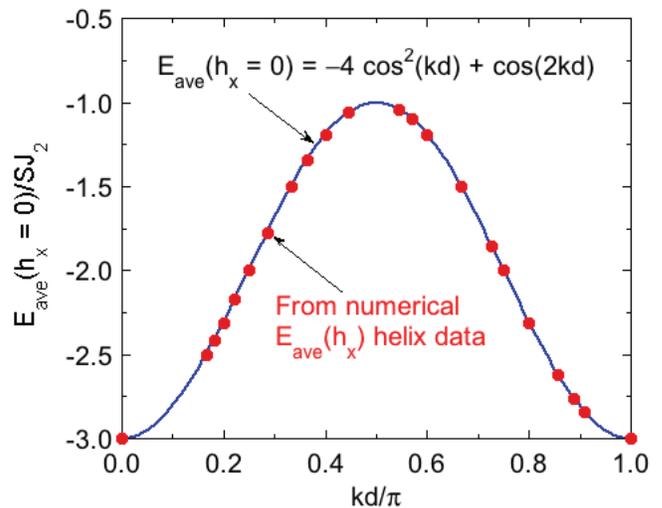}
\caption {(Color online) Average energy per spin $E_{\rm ave}$ normalized by $SJ_{2}$ in field $h_x = 0$ versus helix turn angle~$kd$.  The solid blue curve is the exact variation predicted in Eq.~(\ref{Eq:EnHelix}) with $J_{02}=0$, whereas the red filled circles are from numerical calculations of $E_{\rm ave}(h_x)/SJ_{2}$ carried out later in Sec.~\ref{Sec:HelixFieldDep}.  }
\label{Fig:Helix_Zero_Field_Energies}
\end{figure}

The phase differences in Eq.~(\ref{Eq:EGen2}) for a helix in zero field are
\be
\phi_{n\pm1}-\phi_n = \pm kd, \qquad \phi_{n\pm2}-\phi_n = \pm 2kd.
\label{Eq:PhaseDiffs}
\ee
Using the phase differences in Eq.~(\ref{Eq:PhaseDiffs}), the energy per spin in Eq.~(\ref{Eq:EGen2}) becomes
\bse
\label{Eqs:HelixEnergy}
\be
\frac{E_n}{SJ_{2}} = \frac{1}{2}\left[J_{02} +2J_{12}\cos(kd) +2\cos(2kd) \right] - h_x \cos\phi_n.
\label{Eq:EnHelix0}
\ee
Minimizing the energy with respect to $kd$ for $h_x=0$ gives \cite{Nagamiya1962, Johnston2015}
\be
J_{12} = -4 \cos(kd) \qquad (h_x=0),
\label{Eq:Coskd2}
\ee
so Eq.~(\ref{Eq:EnHelix0}) becomes
\bea
\frac{E_n}{SJ_{2}} = \frac{1}{2}\left[J_{02} -8\cos^2(kd) +2\cos(2kd) \right] \label{Eq:EnHelix}\\
(h_x=0)\nonumber.
\eea
\ese
This variation is plotted as a solid blue curve in Fig.~\ref{Fig:Helix_Zero_Field_Energies} after setting the reference energy $J_{02}=0$.  Also shown as red filled circles are the values abtained from numerical calculations of $E_{\rm ave}(h_x)$ discussed later in Sec.~\ref{Sec:HelixFieldDep} for which the $kd$ values are listed below  in Table~\ref{Tab:Helix/FanData}.  The variation in Fig.~\ref{Fig:Helix_Zero_Field_Energies} is symmetric about $kd=\pi/2$.  Thus 
\be
E_{\rm ave}\left(h_x=0,\frac{\pi}{2}-kd\right) = E_{\rm ave}\left(h_x=0,\frac{\pi}{2}+kd\right).
\label{Eq:ELeftRight}
\ee
This equality is seen to describe the particular pairs of discrete values $kd = 1/5,\ 4/5;\ 1/4,\ 3/4;\ 1/3,\ 2/3$; and~$2/5,\ 3/5$ for which we calculated $E_{\rm ave}(h_x)$.  Thus for every helix with $0 \leq kd < \pi/2$, there is another helix with $\pi/2 < kd \leq \pi$ with the same energy at $h_x=0$.   

\begin{figure}
\includegraphics [width=3.in]{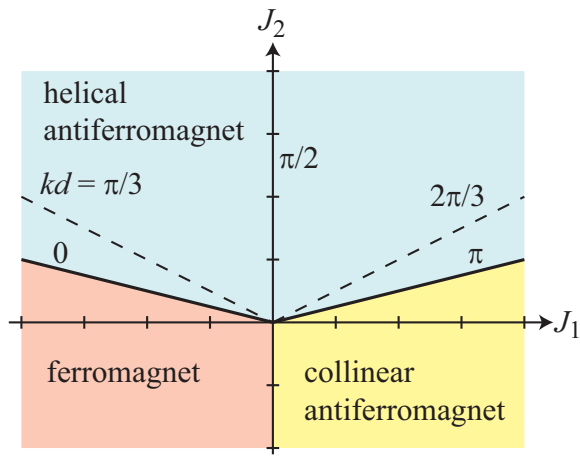}
\caption {(Color online) Zero-temperature phase diagram of the $J_0$-$J_{1}$-$J_{2}$ Heisenberg model \cite{Johnston2015}.  The phase angle $\pi/2$ does not correspond to a helix because it would require $J_{1} = 0$, resulting in two noninteracting AFM sublattices.}
\label{Fig:J0Jz1Jz2PhaseDiag}
\end{figure}

From Eq.~(\ref{Eq:Coskd2}), the allowed domain of $J_{12}$ for any helix within the present model is
\be
0 < \big|J_{12}\big| \leq 4,
\label{Eq:Jz1Jz2HelixRange}
\ee
where $J_{2}>0$ and $J_{12}$ can be of either sign but $J_{12} = 0$ is excluded.  The phase diagram in the $J_{1}$-$J_{2}$ plane at $T=H=0$ for the $J_0$-$J_{1}$-$J_{2}$ model with $m=1$ in Eq.~(\ref{Eq:kdDef}) is shown in Fig.~\ref{Fig:J0Jz1Jz2PhaseDiag} \cite{Johnston2015}.  The competing phases are the FM phase $(kd=0)$, the collinear A-type AFM phase $(kd=\pi)$, and the helical phase $(0<kd<\pi)$.  The value of the reduced exchange constant $J_{02}$ in Eqs.~(\ref{Eqs:HelixEnergy}) is irrelevant to the phase diagram.

We describe the angle~$\phi_n$ of an ordered moment in an $xy$~layer of a helix with respect to the positive $x$~axis by
\bea
\phi_n &=& n\,kd + \Delta\phi_n(h_x)\qquad ({\rm odd}~n_\lambda),\label{Eqs:phinHelixDefs}\\
\phi_n &=& \left(n+\frac{1}{2}\right)\,kd + \Delta\phi_n(h_x)\qquad ({\rm even}~n_\lambda),\nonumber
\eea
where $n = 0,\ 1,\ 2,\ \ldots,\ n_\lambda-1$.  The reason for the difference between these two equations is that for odd~$n_\lambda$, one moment is parallel to the field $h_x$ and is unaffected by it, whereas for even $n_\lambda$, one moment would also have to be antiparallel to the field and would not respond to the field because there would be no torque on it. 

\begin{figure}
\includegraphics [width=2in]{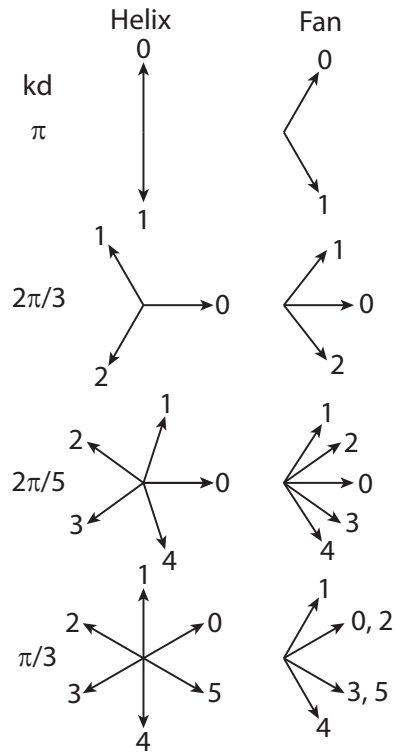}
\caption {(Color online) Illustration of the numbering of helix and fan moments in hodographs with different values of interlayer turn angle~$kd$.  The numbers are the subscripts~$n$ of the angles $\phi_n$ in Eqs.~(\ref{Eqs:phinHelixDefs}) and~(\ref{Eq:Phi}), respectively.  Here we took $\phi_{\rm max}$ in Eq.~(\ref{Eq:Phi}) for a fan to be 60$^\circ$.  The $+x$ axis along which the transverse magnetic field is applied is directed towards the right and the $z$~axis is pointed out of the page.  However, for a helix changing to a fan with increasing field, the numbering of the moments in the helix is retained for all fields.}
\label{Fig:HelixFanAngles}
\end{figure}

For either even or odd $n_\lambda$, the expressions for $\phi_n$ in Eqs.~(\ref{Eqs:phinHelixDefs}) give a symmetrical distribution of moments above and below the $x$~axis in the $xy$~plane, either with or without an applied field~$h_x$, as shown for several helix examples in zero field in Fig.~\ref{Fig:HelixFanAngles}.  That is, for every moment at angle $0 < \phi_n < \pi$ there is another moment at angle $-\phi_n$. This also holds in the presence of any $x$-axis field.  Hence the number of $\Delta\phi_n(h_x)$ values to be solved for by minimizing the helix energy in Eq.~(\ref{Eq:EnHelix}) at a given $h_x$ is either $n_\lambda/2$ (for even~$n_\lambda$) or $(n_\lambda-1)/2$ (for odd~$n_\lambda$).  The multidimensional minimization of the energy to determine the unique values of $\Delta\phi_n$ with $\pi < \Delta\phi_n < 0$ was carried out using the {\tt FindMinimum} utility of {\tt Mathematica} for $3 \leq n_\lambda \leq 11$ (odd~$n_\lambda$) or $2 \leq n_\lambda \leq 12$ (even~$n_\lambda$).  Note that when a field-induced transition or continuous evolution from helix to fan occurs, we label the angles $\phi_n$ by the notation in Eq.~(\ref{Eqs:phinHelixDefs}) and not by the notation for fans in Eq.~(\ref{Eq:Phi}) below.  

Once the angles $\phi_n(h_x)$ are determined, the reduced $x$-axis components $\mu_{nx}$ of the ordered moments $\vec{\mu}_n$ and the average $x$-axis moment per spin $\mu_{x{\rm ave}}$ are obtained for either the helix or fan as
\bea
\mu_{nx}/\mu_{\rm sat} &=& \cos\phi_n, \label{munx}\\
\mu_{x{\rm ave}}/\mu_{\rm sat} &=& \frac{1}{n_\lambda}\sum_{n=0}^{n_\lambda-1}\mu_{nx}.\nonumber
\eea

A helix is usually the stable phase at low fields with respect to a fan or fanlike phase.  The only exceptions are for $kd=\pi/2$, for which a helix phase does not exist, and for $kd=\pi$ and~$2\pi/3$ for which the helix and fan phases are identical if the values of $J_{12} = 4$ and~2 for the helix, respectively,  are used in the formulas for the fan (see following section).  The low-field susceptibility of the helix $\chi_x \equiv \lim_{h_x\to0}[\mu_{x{\rm ave}}(h_x)/\mu_{\rm sat}]/h_x$ can be obtained from the low-field $\mu_{x{\rm ave}}(h_x)/\mu_{\rm sat}$ magnetization versus field calculations.  The more accurate method used here for a helix with a given value of~$kd$ is to express the average energy for the helix in Eq.~(\ref{Eq:EGen2}) in terms of the $\phi_n$ and $h_x$ variables and then minimize the energy with respect to each of the $\phi_n$ with $0< \phi_n < \pi$ which gives $n_\lambda/2$ (even $n_\lambda$) or $(n_\lambda-1)/2$ (odd $n_\lambda$) equations.  Each of these is Taylor expanded about $\phi_n=0$ and $h_x=0$ and only the first-order terms retained.  Then the $n_\lambda/2$ or $(n_\lambda-1)/2$ linear equations are easily solved for the $\phi_n$ versus $h_x$ using {\tt Mathematica} and from those one obtains $\mu_{x{\rm ave}}$ to first order in $h_x$ using Eqs.~(\ref{munx}) from which $\chi_x$ can be obtained to arbitrary accuracy.  In Table~\ref{Tab:Helix/FanData} below, we quote $\chi_x$ to six figures and also include the exact analytic expressions for $\chi_x(kd)$ if obtained automatically by {\tt Mathematica} using the above method.

\subsection{\label{Sec:FanThy} Sinusoidal Fan Phases}

At sufficiently high fields, the solutions for the $\phi_n(h_x)$ for a (distorted) helix structure change into values one may associate with a fanlike structure.  This change can be either a continuous crossover, or a second-order transition for $4\pi/9 \lesssim kd \leq \pi$, or a first-order transition for $0 < kd \lesssim 4\pi/9$.  However, because the $\phi_n$ values of the fanlike structure are determined from exact calculations for the helix (assuming that $kd$ does not depend on field), it is not {\it a priori} clear that these solutions and the respective $\mu_{\rm ave}(h_x)$ values correspond precisely to those for a perfectly sinusoidal fan with a $J_{12}$ of the helix with the same $kd$ value as given in Eq.~(\ref{Eq:Coskd2}).  Therefore in this section we first discuss how to compute the $\phi_n(h_x)$ and $\mu_{x\rm ave}(h_x)$ of sinusoidal fans with arbitrary  values of~$J_{12}$ and~$kd$.  Then we specialize to the values $J_{12} = -4\cos(kd)$ for comparison with the properties of the corresponding high-field fanlike phases of the helix with turn angle~$kd$.

For a sinusoidal fan along the $z$~axis with the moments aligned in the $xy$~plane, the phase angle $\phi_n$ with respect to the $+x$ axis versus moment layer number~$n$ for a field-independent wavelength is defined as
\bea
\phi_n &=& \phi_{\rm max}\sin(n kd) \hspace{0.72in} ({\rm odd}~n_\lambda), \label{Eq:Phi}\\
\phi_n &=& \phi_{\rm max}\sin\left[\left(n + \frac{1}{2}\right)kd\right]\quad ({\rm even}~n_\lambda),\nonumber\\
&& (n = 0,~1,~\ldots, n_\lambda - 1),\nonumber
\eea
which differentiates between odd and even~$n_\lambda$ for the same reasons as in Eqs.~(\ref{Eqs:phinHelixDefs}), where the amplitude of the sinusoidal modulation of the phase is $0 \leq \phi_{\rm max} \leq \pi$ which is determined for general values of the parameters $J_{12}$ and~$kd$ by minimizing the average energy per spin for each value of the reduced field~$h_x$.  The variations of $\phi_n$ with~$n$ for several fan examples are shown in Fig.~\ref{Fig:HelixFanAngles}.  

The energy of moment~$n$ where $kd$ is assumed independent of field is obtained from Eq.~(\ref{Eq:EGen2}) as
\be
\frac{E_n}{SJ_{2}} = \frac{1}{2}\big[J_{02} + 2J_{12}\cos(kd) + 2\cos(2kd) \big] - h_x \cos\phi_n,
\label{Eq:EnFan}
\ee
where the turn angle $kd$ between adjacent moment layers along the $z$~axis is given in Eq.~(\ref{Eq:kdDef}), and the average energy per wavelength of the fan is given by Eq.~(\ref{Eq:EAveGen}) in terms of the $E_n$ values in Eq.~(\ref{Eq:EnFan}). Thus for given values of $J_{12},\ kd$, and $h_x$, the single remaining parameter $\phi_{\rm max}$ is found by minimizing~$E_{\rm ave}$.

However, for $J_{12} = h_x = 0$, the values of $\phi_{\rm max}$ are determined by solving $\mu_{x{\rm ave}} = 0$ for $\phi_{\rm max}$ using Eqs.~(\ref{munx}) (i.e., there is no net magnetization of the fan).  The results are listed in Table~\ref{Tab:FanDataH0T0} for $n_\lambda = 2\pi/kd$ values fom 2 to 23 for comparison with results below for nonzero $J_{12}$ and~$h_x$.  One sees that the values asymptote rapidly to an $n_\lambda\to\infty$ limit given in the table caption.  From Eqs.~(\ref{Eq:EAveGen}) and~(\ref{Eq:EnFan}), the average energies of all sinusoidal fans with $J_{12} = h_x = 0$ but different~$kd$ have the same value $J_{02}/2$.

\begin{table}
\caption{\label{Tab:FanDataH0T0} Numerical values and/or analytic forms for the angular amplitude $\phi_{\rm max}$ for $J_{12} = h_x = 0$ versus number $n_\lambda$ of moment layers per wavelength of the fan along the $z$~axis with turn angle $kd = 2\pi/n_\lambda$.  The values of $\phi_{\rm max}$ for both even and odd $n_\lambda$ asymptote for large $n_\lambda$ to the first zero of the zero-order Bessel function given by  \mbox{$j_{0,1} = 2.404\,825\,557\,695\,772\,768\,621\,\cdots$.}}
\begin{ruledtabular}
\begin{tabular}{cccc}
$n_\lambda$ 	& $\phi_{\rm max}$		& $n_\lambda$ 	& $\phi_{\rm max}$		\\
		& (rad)	& 	& (rad)					\\
\hline 
2  &  1.570\,796\,326\,794\,90  &  3  &   2.418\,399\,152\,312\,29  \\ 
4  &  2.221\,441\,469\,079\,18  &  5  &   2.404\,831\,434\,267\,50  \\
6  &  2.392\,123\,788\,172\,31  &  7  &   2.404\,825\,558\,225\,55  \\
8  &  2.404\,470\,919\,537\,39  &  9  &   2.404\,825\,557\,695\,79  \\
10  & 2.404\,819\,681\,417\,96  &  11  &  2.404\,825\,557\,695\,77  \\
12  & 2.404\,825\,491\,997\,91  &  13  &  2.404\,825\,557\,695\,77  \\
14  & 2.404\,825\,557\,165\,99  &  15  &  2.404\,825\,557\,695\,77  \\
16  & 2.404\,825\,557\,692\,54  &  17  &  2.404\,825\,557\,695\,77  \\
18  & 2.404\,825\,557\,695\,76  &  19  &  2.404\,825\,557\,695\,77  \\
20  & 2.404\,825\,557\,695\,77  &  21  &  2.404\,825\,557\,695\,77  \\
22  & 2.404\,825\,557\,695\,77  &  23  &  2.404\,825\,557\,695\,77  \\
	&  Analytic Forms		\\
2	&	$\pi/2$ 			\\
3 	&	$4\pi/3^{3/2}$		\\
4	&  	$\pi/\sqrt{2}$  	\\
6	&  	$4\arctan\left(\sqrt{2\sqrt{3} - 3}\right)$ \\
\end{tabular}
\end{ruledtabular}
\end{table}

\section{\label{Sec:GenFanResults} Field-Dependent Results: Sinusoidal Fan Phase}

\subsection{General Results}

\begin{table*}
\caption{\label{Tab:FanGndStateE} The dependence on $kd$ and~$J_{12}$ of the amplitude $\phi_{\rm max}$ of the sinusoidal  fan, the reduced critical field $h_{\rm c}$, and the physically-allowed range of the parameter~$J_{12}$ for the fan are listed for specific values of $kd$ and $n_\lambda$, all determined by minimizing the energy of the fan.  In the allowed fan $J_{12}$ range, a minimum value corresponds to the value at which $h_{\rm c} = 0$, and the maximum value to the value at which a ferromagnetic $\mu_{x{\rm ave}}$ in $h_x=0$ goes to zero.  For $kd= 6\pi/11,\ 4\pi/7$, and~$3\pi/5$, only FM solutions are found.  Shown in the last column is the value of $J_{12}$ of a helix with the same value of $kd$.  A positive (negative) value of $J_{12}\equiv J_{1}/J_{2}$ corresponds to an AFM (FM) value of the nearest-layer exchange~$J_{1}$. }
\begin{ruledtabular}
\begin{tabular}{lcccccc}
$kd$	&	$n_\lambda$ &	$\phi_{\rm max}(h_x=0)$	&	$\phi_{\rm max}(h_x,J_{12})$	&	$h_{\rm c}$	& Physical Fan $J_{12}$ Range	& $J_{12}$ for Helix\\
\hline 
$\pi$	&	2Ê 	&	$\pi/2$			&	$\arccos(h_x/4J_{12})$		&	$4 J_{12}$		&Ê 	$J_{12}\geq 0$		& Ê 							4 				\\
$10\pi/11$&	11Ê 	&	depends on $J_{12}$		&	depends on $J_{12}$		&	$0.3179 + 3.9191 J_{12}$	&Ê 	$J_{12}\geq 0.5213 $		  Ê	&	3.83797	 	\\
$8\pi/9$	&	11Ê 	&	depends on $J_{12}$		&	depends on $J_{12}$		&	$0.4685 + 3.8793 J_{12}$	&Ê 	$J_{12}\geq 0.6956 $		  Ê	&	3.75877	 	\\
$6\pi/7$	&	7Ê 	&	depends on $J_{12}$		&	depends on $J_{12}$		&	$0.7534 + 3.8020 J_{12}$	&Ê 	$J_{12}\geq 0.9160 $			&	3.60388	 	\\
$5\pi/6$	&	12Ê 	&	depends on $J_{12}$		&	depends on $J_{12}$		&	$1.0005 + 3.7321 J_{12}$	&Ê 	$J_{12}\geq 1.0050 $		  Ê	&	3.46410	 	\\
$4\pi/5$	&	5Ê 	&	depends on $J_{12}$		&	depends on $J_{12}$		&	$1.3823 + 3.6182 J_{12}$	&Ê 	$J_{12}\geq 1 $		  Ê		&	3.23607	 	\\
$3\pi/4$	&	8Ê 	&	depends on $J_{12}$		&	depends on $J_{12}$		&	$2.0005 + 3.4142 J_{12}$	&Ê 	$J_{12} \geq 0.5858$		  Ê	&	2.82843	 	\\
$8\pi/11$	&	11Ê 	&	depends on $J_{12}$		&	depends on $J_{12}$		&	$2.2851 + 3.3098 J_{12}$	&	$J_{12}\geq 0.3441$				&  	2.61944	\\
$2\pi/3$	&	3Ê 	&	$4\pi/3^{3/2}$		&	$\frac{2}{\sqrt{3}}\arccos\left[\frac{h_x-(1+J_{12})}{2(1+J_{12})}\right]$	&Ê 	$3(1+J_{12})$		&	$J_{12}\geq - 1$	&  2	\\
$3\pi/5$	&	10Ê 	&	depends on $J_{12}$		&	depends on $J_{12}$		&	$3.6186 + 2.6178 J_{12}$	&Ê 	$J_{12}\geq -1.3823$ (FM)	  Ê	&	1.23607	 	\\
$4\pi/7$	&	7Ê 	&	depends on $J_{12}$		&	depends on $J_{12}$		&	$3.8024 + 2.4451 J_{12}$	&Ê 	$J_{12}\geq -1.5551$ (FM)	  Ê	&	0.890084	 	\\
$6\pi/11$	&	11Ê 	&	depends on $J_{12}$		&	depends on $J_{12}$		&	$3.9195 + 2.2846 J_{12}$	&Ê 	$J_{12}\geq -1.7156$ (FM)	  Ê	&	0.569259	 	\\
$\pi/2$	&	4Ê 	&	$\pi/\sqrt{2}$	&	$\sqrt{2}\arccos\left[\frac{h_x}{2(2+J_{12})}\right]$	&	$2(2+J_{12})$	&Ê $J_{12}\geq -2$Ê 		&Ê		0	\\
$2\pi/5$	&	5Ê 	&	depends on $J_{12}$		&	depends on $J_{12}$		&	$3.6186 + 1.3819 J_{12}$ &Ê $-2.6185\leq J_{12}\leq 1$ Ê	&		$-1.23607$		\\
$4\pi/11$	&	11Ê 	&	depends on $J_{12}$		&	depends on $J_{12}$		&	$3.3095 + 1.1687 J_{12}$ &Ê $-2.8317\leq J_{12}\leq 0.6980$ Ê	&		$-1.66166$		\\
$\pi/3$	&	6Ê 	&	$2\arccos\left(\frac{3-J_{12}}{6}\right)$	&	$2\arccos\left(\frac{3-J_{12}+h_x}{6}\right)$&	$3+J_{12}$	&Ê $-3\leq J_{12}\leq 3(2-\sqrt{3})$	&	$-2$	\\
		&	(6)Ê 	&							&						&		&Ê $-3\leq J_{12}\leq 0.803848$			 &		\\
$2\pi/7$	&	7Ê 	&	depends on $J_{12}$		&	depends on $J_{12}$		&	$2.4455 + 0.7530 J_{12}$	&Ê 	$-3.2477\leq J_{12}\leq -0.0917$  Ê&	$-2.49396$	 	\\
$\pi/4$	&	8Ê 	&	depends on $J_{12}$		&	depends on $J_{12}$		&	$2.0005 + 0.5858 J_{12}$	&Ê 	$-3.4151\leq J_{12}\leq -0.5858$  Ê&	$-2.82843$		\\
$2\pi/9$	&	9Ê 	&	depends on $J_{12}$		&	depends on $J_{12}$		&	$1.6532 + 0.4679 J_{12}$	&Ê 	$-3.5332\leq J_{12}\leq -0.9961$ 	&Ê Ê $-3.06418$		\\
$\pi/5$	&	10Ê	&	depends on $J_{12}$		&	depends on $J_{12}$		&	$1.3826 + 0.3820 J_{12}$	&  Ê	$-3.6197\leq J_{12}\leq -1.3706$  Ê& Ê 	$-3.23607$		\\
$2\pi/11$	&	11Ê 	&	depends on $J_{12}$		&	depends on $J_{12}$		&	$1.1697 + 0.3175 J_{12}$	&Ê  	$-3.6840\leq J_{12}\leq -1.6933$  Ê&Ê Ê $-3.36501$		\\
$\pi/6$	&	12Ê 	&	depends on $J_{12}$		&	depends on $J_{12}$		&	$1.0007 + 0.2680 J_{12}$	& Ê 	$-3.7338\leq J_{12}\leq -1.9689$ Ê &	$-3.46410$		\\
$2\pi/13$	&	13Ê 	&	depends on $J_{12}$		&	depends on $J_{12}$		&	$0.8643 + 0.2291 J_{12}$	& Ê 	$-3.7732\leq J_{12}\leq -2.2036$ Ê & Ê	$-3.54182$		\\
$\pi/7$	&	14Ê 	&	depends on $J_{12}$		&	depends on $J_{12}$		&	$0.7537 + 0.1981 J_{12}$	& 	$-3.8043\leq J_{12}\leq -2.4037$	&	$-3.60388$		\\
$2\pi/15$	&	15Ê 	&	depends on $J_{12}$		&	depends on $J_{12}$		&	$0.6623 + 0.1729 J_{12}$	& 	$-3.8299\leq J_{12}\leq -2.5748$	&	$-3.65418$		\\
$\pi/8$	&	16Ê 	&	depends on $J_{12}$		&	depends on $J_{12}$		&	$0.5863 + 0.1523 J_{12}$	& 	$-3.8510\leq J_{12}\leq -2.7215$	&	$-3.69552$		\\
$2\pi/17$	&	17Ê 	&	depends on $J_{12}$		&	depends on $J_{12}$		&	$0.5225 + 0.1351 J_{12}$	& 	$-3.8685\leq J_{12}\leq -2.8480$	&	$-3.72989$		\\
$\pi/9$	&	18Ê 	&	depends on $J_{12}$		&	depends on $J_{12}$		&	$0.4684 + 0.1206 J_{12}$	& 	$-3.8835\leq J_{12}\leq -2.9576$	&	$-3.75877$		\\
$2\pi/19$	&	19Ê	&	depends on $J_{12}$		&	depends on $J_{12}$		&	$0.4221 + 0.1083 J_{12}$	& 	$-3.8963\leq J_{12}\leq -3.0528$	&	$-3.78327$		\\
$\pi/10$	&	20Ê	&	depends on $J_{12}$		&	depends on $J_{12}$		&	$0.3825 + 0.0979 J_{12}$	& 	$-3.9073\leq J_{12}\leq -3.1361$	&	$-3.80423$		\\
$2\pi/21$	&	21Ê	&	depends on $J_{12}$		&	depends on $J_{12}$		&	$0.3479 + 0.0888 J_{12}$	& 	$-3.9171\leq J_{12}\leq -3.2093$	&	$-3.82229$		\\
$\pi/11$	&	22Ê 	&	depends on $J_{12}$		&	depends on $J_{12}$		&	$0.3179 + 0.0810 J_{12}$	& 	$-3.9251\leq J_{12}\leq -3.2738$	&	$-3.83797$		\\
$2\pi/23$	&	23Ê 	&	depends on $J_{12}$		&	depends on $J_{12}$		&	$0.2915 + 0.0741 J_{12}$	& 	$-3.9327\leq J_{12}\leq -3.3309 $	&	$-3.85167$		\\
0& $\infty$		&					&						&						&								&	$-4$				\\
\end{tabular}
\end{ruledtabular}
\end{table*}

Values of $\phi_{\rm max}(h_x=0)$ versus $kd = 2\pi m/n_\lambda$ were calculated for values of $kd$ arising from the different combinations of $m$ and~$n_\lambda$ in Eq.~(\ref{Eq:kdDef}) listed in Table~\ref{Tab:FanGndStateE}.  The critical field $h_{\rm c}$ was usually determined numerically from the criterion $h_{\rm c} = \lim_{\phi_{\rm max}\to0}(h_x)$ using the {\tt FindRoot} utility of {\tt Mathematica}.  The results in general depend on both $J_{12}$ and $h_x$.  We also find that physical solutions only exist for a range of $J_{12}$ values that depend on the value of~$kd$.  For each value of $kd$, we found that $h_{\rm c}$ varies linearly with $J_{12}$ over the physical range of $J_{12}$ values listed in Table~\ref{Tab:FanGndStateE}.  Exact analytic expressions are obtained for $\phi_{\rm max}$ and $h_{\rm c}$ versus $J_{12}$ and $h_x$ for $kd = \pi,\ 2\pi/3,\ \pi/2,$ and~$\pi/3$, as listed.  Interestingly, the values of $\phi_{\rm max}(h_x=0)$ for $kd = \pi,\ 2\pi/3$, and $\pi/2$ are seen to be identical with the corresponding values in Table~\ref{Tab:FanDataH0T0} for $J_{12}=0$ and $h_x=0$.

For some values of $kd$, only FM-polarized solutions of $\mu_{x{\rm ave}}(h_x=0)$ are obtained.  This is not relevant to cases where a helix with the given~$kd$ transitions to a fan with the same~$kd$ and~$J_{12}$ at finite~$h_x$, at which a finite magnetization of the fan would naturally occur.  The lower limit of a $J_{12}$ range corresponds to the value at which the critical field $h_{\rm c} \to 0$, whereas the upper limit (if present) is the maximum value at which a FM value of $\mu_{x{\rm ave}} \to 0$ (PM state), above which the FM value of $\mu_{x{\rm ave}}$ at $h_x=0$ would become negative.  In the last column of Table~\ref{Tab:FanGndStateE} is the $J_{12}$ value of a helix with the same value of $kd$ as for the fan according to Eq.~(\ref{Eq:Coskd2}). One sees that for each value of $kd$, the helix value of $J_{12}$ lies within the physical range of $J_{12}$ for the fan.  We have verified that within the physical range of $J_{12}$, the derived values of $\phi_{\rm max}$ correspond to minima (rather than maxima) of the energy.  Once $\phi_{\rm max}(h_x,\ J_{12})$ is determined, the magnetization isotherms $\mu_{x{\rm ave}}/\mu_{\rm sat}$ versus~$h_x$ at $T=0$ can be generated for different values of $J_{12}$ within the physical range using Eqs.~(\ref{munx}) and~(\ref{Eq:Phi}).

\begin{figure}
\includegraphics [width=3.3in]{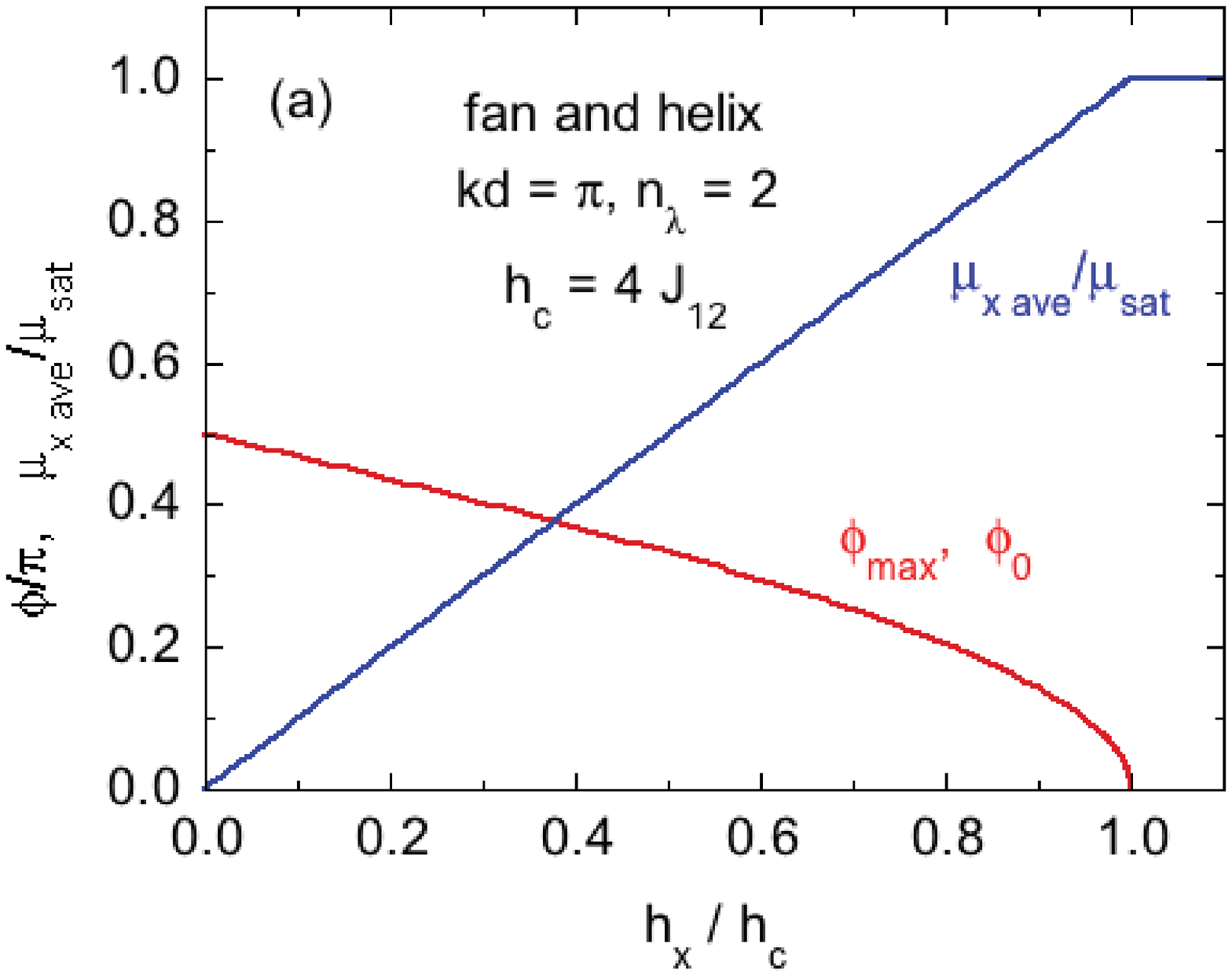}
\includegraphics [width=3.3in]{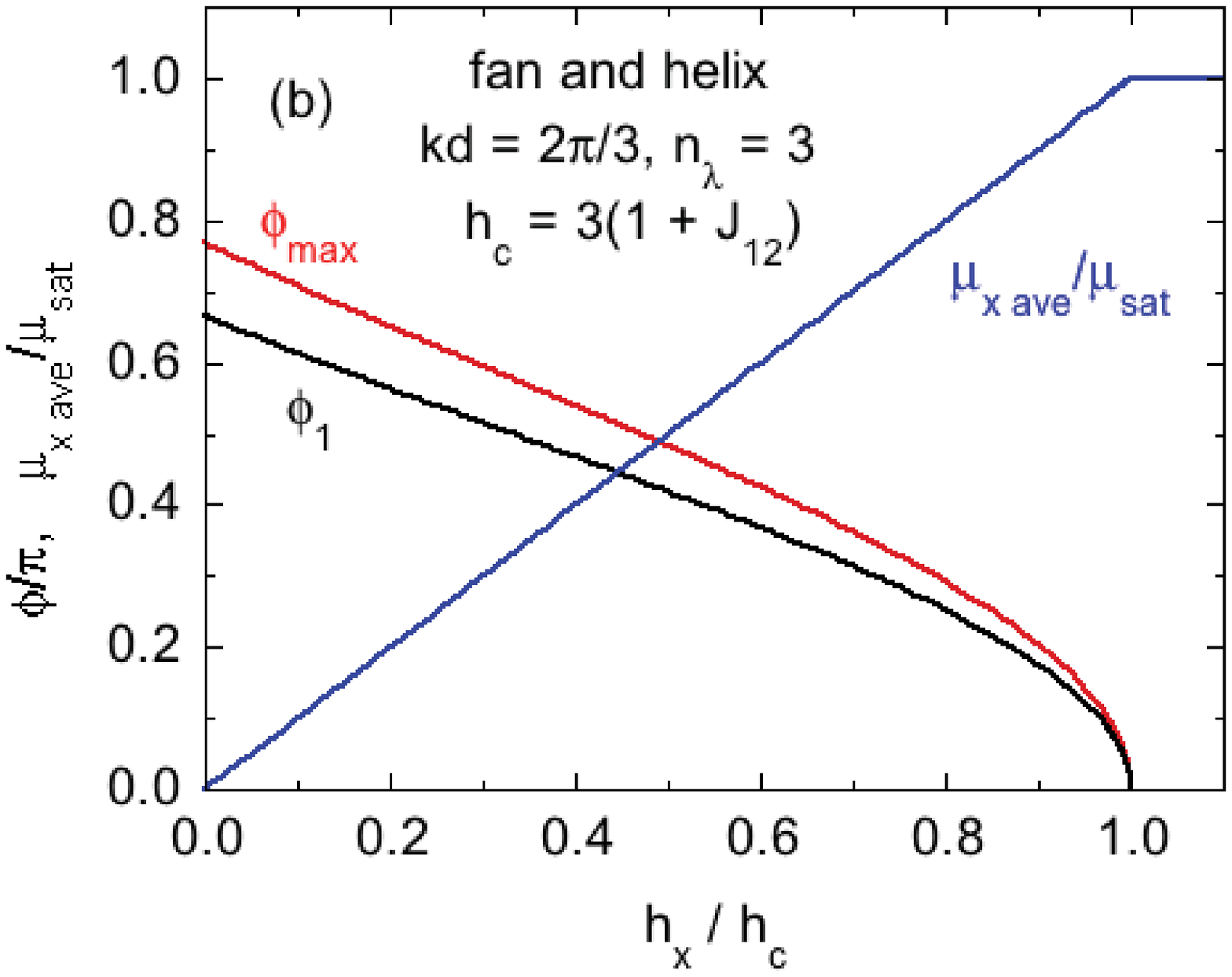}
\includegraphics [width=3.3in]{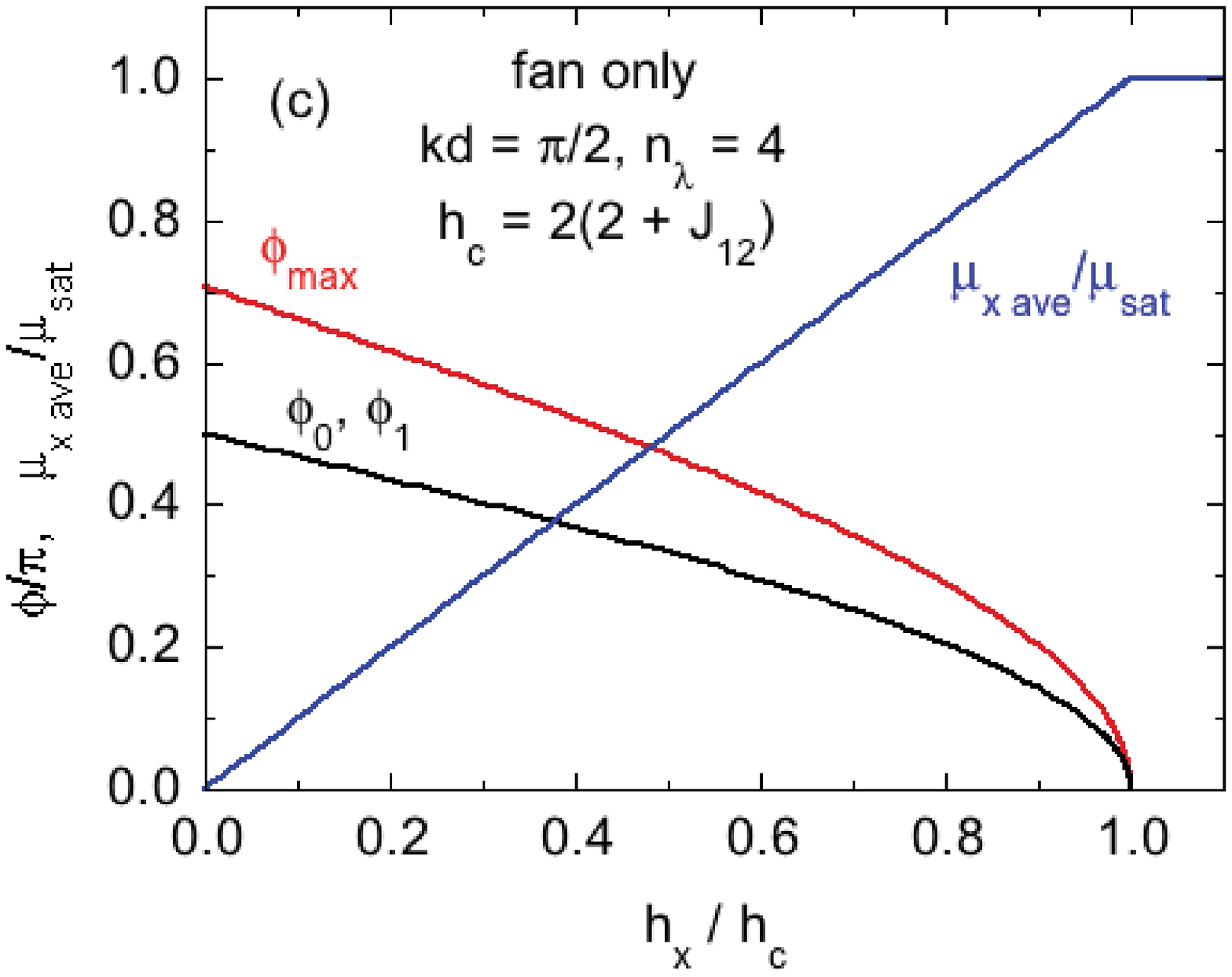}
\caption {(Color online) The amplitude $\phi_{\rm max}$ of the sinusoidal fan and the moment angles $\phi_n$ with $0 \leq \phi_n \leq \pi$ with respect  to the $+x$ axis, both normalized by $\pi$~rad, and the normalized average $x$-axis moment component per spin $\mu_{x{\rm ave}}/\mu_{\rm sat}$ versus reduced $x$-axis field $h_x/h_{\rm c}$ for fans with turn angles $kd$ given by (a)~$kd=\pi$, (b)~$kd=2\pi/3$,  and (c)~$kd=\pi/2$.  The subscripts $n$ in $\phi_n$ are defined in Eqs.~(\ref{Eq:Phi}).  The fan properties in~(a) and~(b) with the $kd$ and $J_{12}$ values of the respective helices are identical with those of the helices.  The parameters of the fan in~(c) have no helix counterpart, as explained in the text.}
\label{Fig:FanDatN2_kdPi}
\end{figure}

\begin{figure}
\includegraphics [width=3.4in]{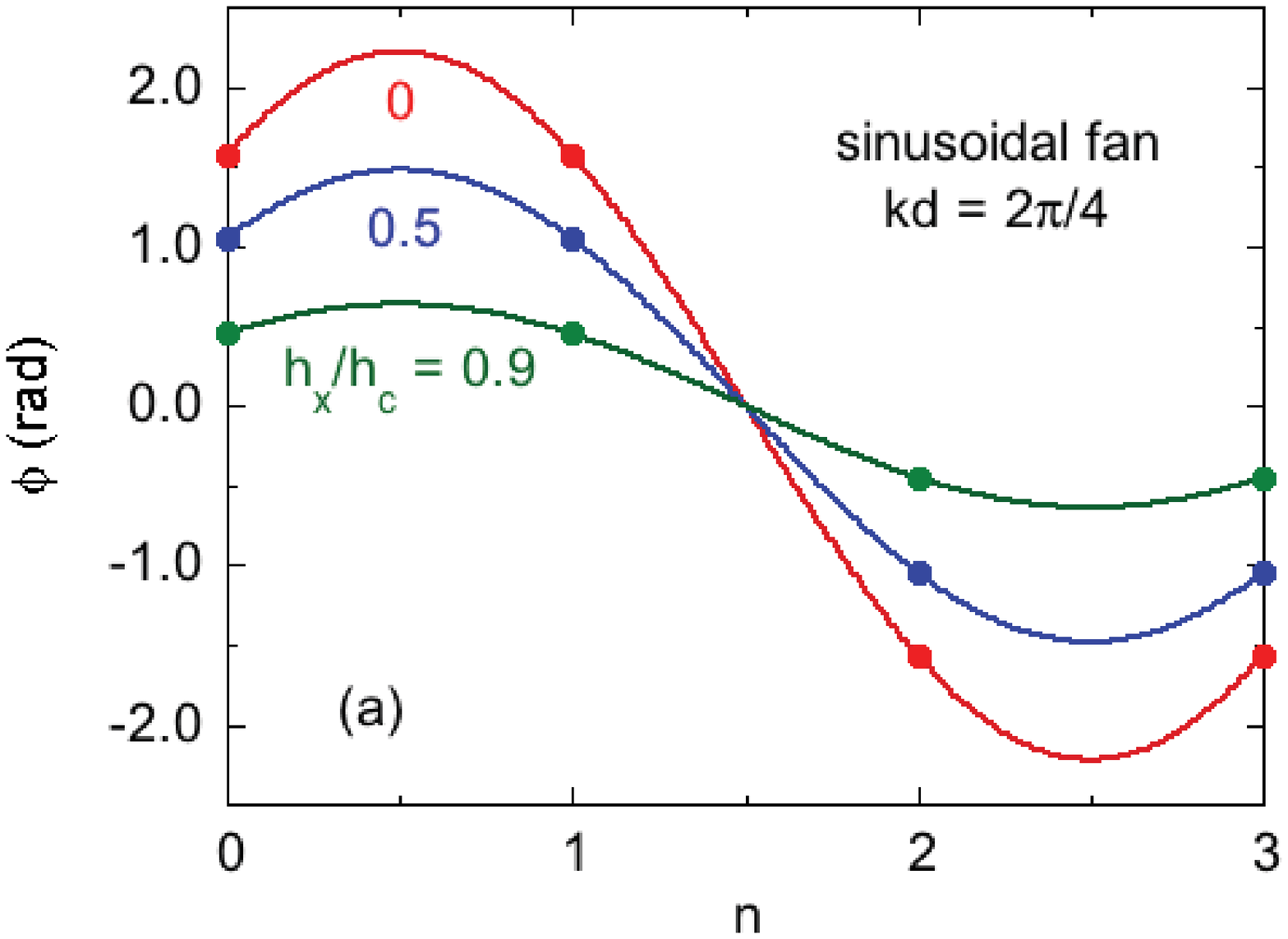}
\includegraphics [width=3.4in]{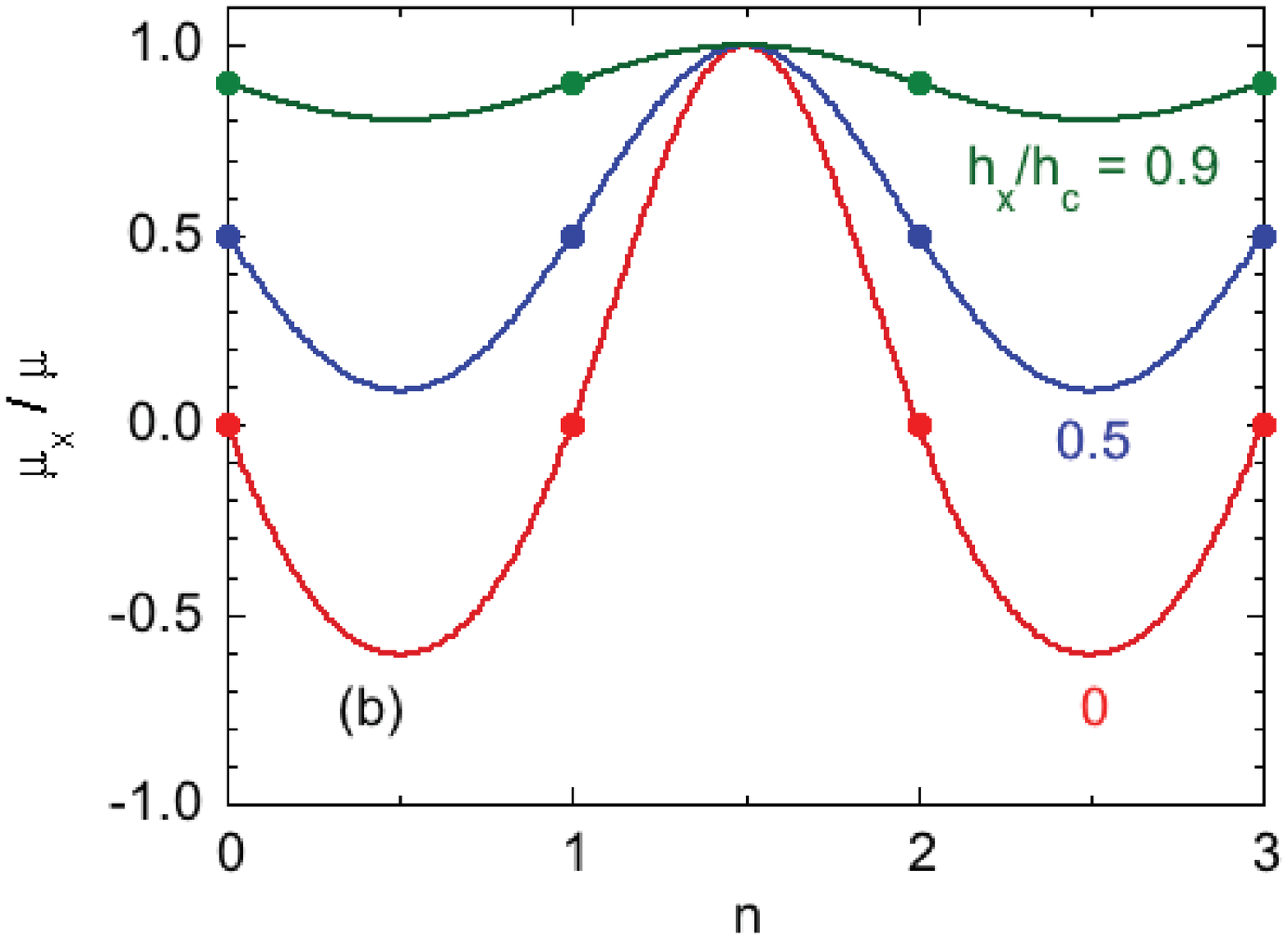}
\caption {(Color online) Plots of (a) $\phi_n$ and (b) $\mu_{x}$ versus $n$ for a fan with $kd=\pi/2$ and $n_\lambda = 4$ for three values of the reduced field $h_x/h_{\rm c}$, showing that the physical moments with integer~$n$ lie on sinusoids described by Eq.~(\ref{Eq:Phi}).}
\label{Fig:Phi_mux_N4_vs_n}
\end{figure}

To illustrate sinusoidal fan structures and magnetizations, we present some representative plots of them.  Shown in Fig.~\ref{Fig:FanDatN2_kdPi} are plots of $\phi_{\rm max}/\pi$, the angles $\phi_n/\pi$ of moments~$n$ with $0 \leq \phi_n \leq \pi$, and the average $x$-axis moment per spin $\mu_{x{\rm ave}}$ versus the reduced field $h_x/h_{\rm c}$ for (a)~$kd=\pi$, $n_\lambda = 2$, (b)~$kd=2\pi/3$, $n_\lambda = 3$, and (c)~$kd=\pi/2$, $n_\lambda = 4$.  The indices~$n$ of the $\phi_n$ are the same as in, and dictated by,  Eq.~(\ref{Eq:Phi}), examples of which are shown in Fig.~\ref{Fig:HelixFanAngles}.  In all plots such as these shown in this paper, there exist moment angles that are the negatives of the ones shown at each $h_x$ (see  Fig.~\ref{Fig:HelixFanAngles}).  The $\mu_{x{\rm ave}}$ is seen to be proportional to $h_x$ for each set of parameters shown.  This is a general characteristic of the sinusoidal fan phase.

The plots in Figs.~\ref{Fig:FanDatN2_kdPi}(a)--\ref{Fig:FanDatN2_kdPi}(c) are valid for all $J_{12}$ values within the physical ranges listed in Table~\ref{Tab:FanGndStateE} for the given sets of parameters.  The dependences on $J_{12}$ are taken into account via the normalization of the horizontal axes by the $J_{12}$-dependent critical fields $h_{\rm c}$.  We also emphasize that the behaviors in Figs.~\ref{Fig:FanDatN2_kdPi}(a) and~\ref{Fig:FanDatN2_kdPi}(b) are found to be identical with those of helices with the same special parameters and hence exhibit no phase transitions versus field other than that at the reduced critical field~$h_{\rm c}$.  The fan in Fig.~\ref{Fig:FanDatN2_kdPi}(c) with a turn angle of $kd = \pi/2$ has no helix counterpart, because according to Eq.~(\ref{Eq:Coskd2}) that would require that $J_{1z}=0$ which would result in two noninteracting sets of next-nearest-neighbor collinear AFM layers, each with a response to a field given by the behavior for $kd=\pi$ in Fig.~\ref{Fig:FanDatN2_kdPi}(a).  Shown in Fig.~\ref{Fig:Phi_mux_N4_vs_n} are plots of $\phi_n$ versus $n$ for $kd=\pi/2$ with $n_\lambda = 4$, which illustrate that even though the fan is sinusoidal the actual angles for the moments may not appear to be so.

In the fan phases appearing above the transition field $h_{\rm t}$ of a helix, the values of $\phi_n$ are not {\it a priori} prescribed as in Eq.~(\ref{Eq:Phi}) for the sinusoidal fan.  Instead, the set of $\phi_n$ for a given $kd$ is found by multidimensional minimization of the energy with the $\phi_n$ as independent variables as discussed previously and could therefore be nonsinusoidal.  However, we infer that the fan structure of the helices above~$h_{\rm t}$ obtained by energy minimization is indeed sinusoidal for $h_x\to h_{\rm c}$, since $h_{\rm c}$ for the field-induced fan is found to be identical to that of the sinusoidal fan by itself.  On the other hand, at smaller $h_x$ values in the fan phase, deviations from the predictions of the moment angles for sinusoidal fans are found as illustrated in Sec.~\ref{Sec:HelixFieldDep} below.

\subsection{Fans with Helix $J_{12}$ Values}

\begin{figure*}
\includegraphics [width=3.4in]{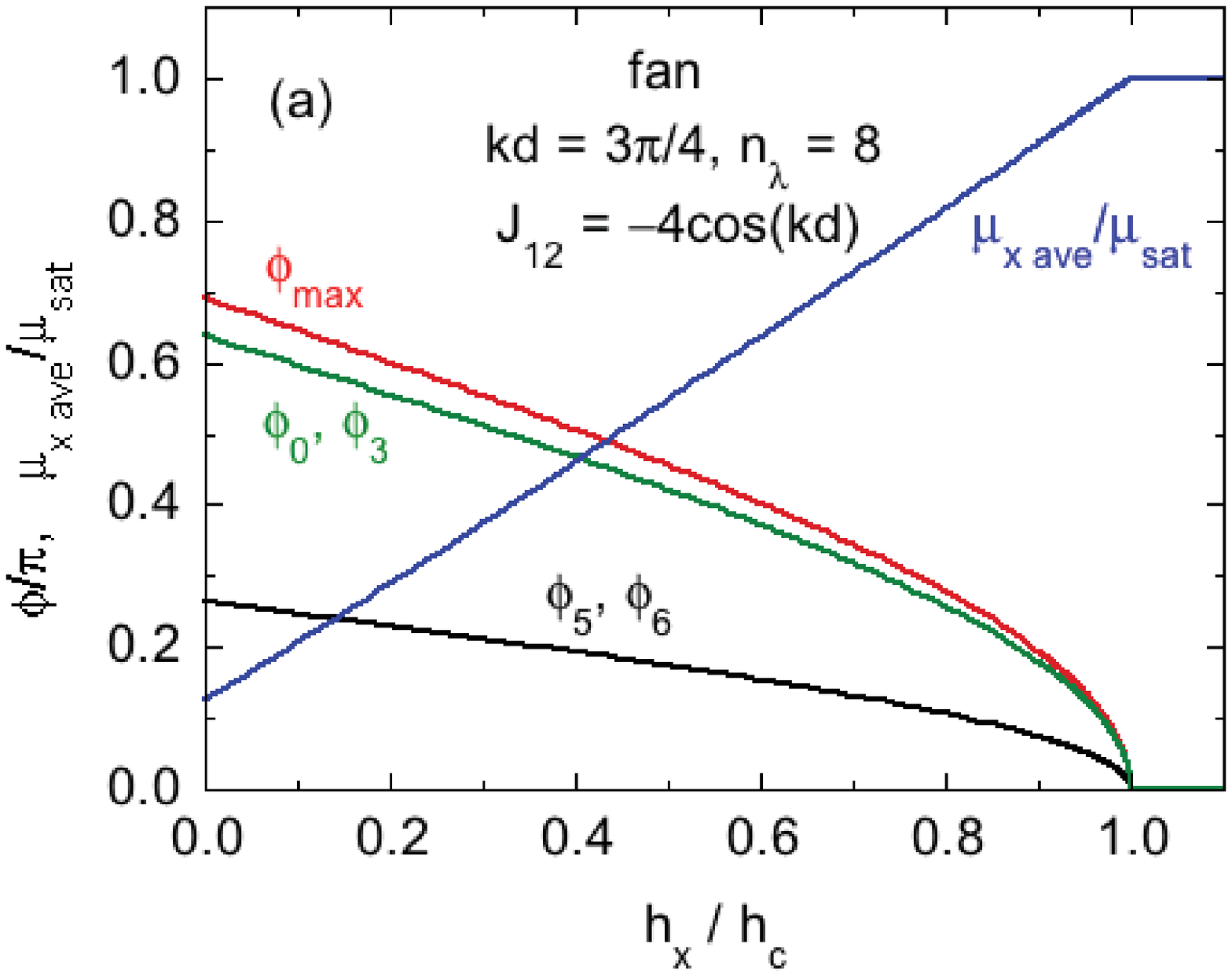}\includegraphics [width=3.4in]{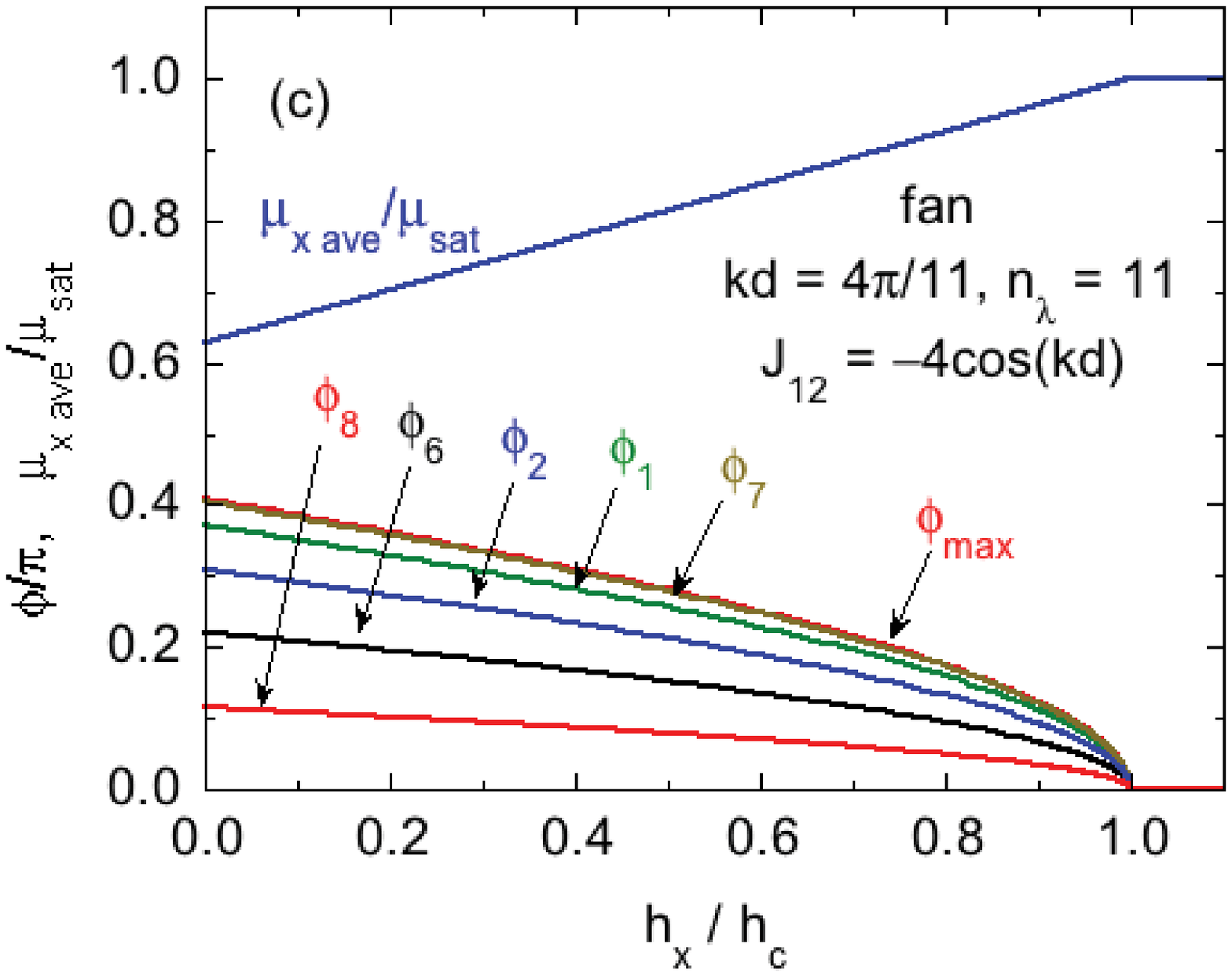}
\includegraphics [width=3.4in]{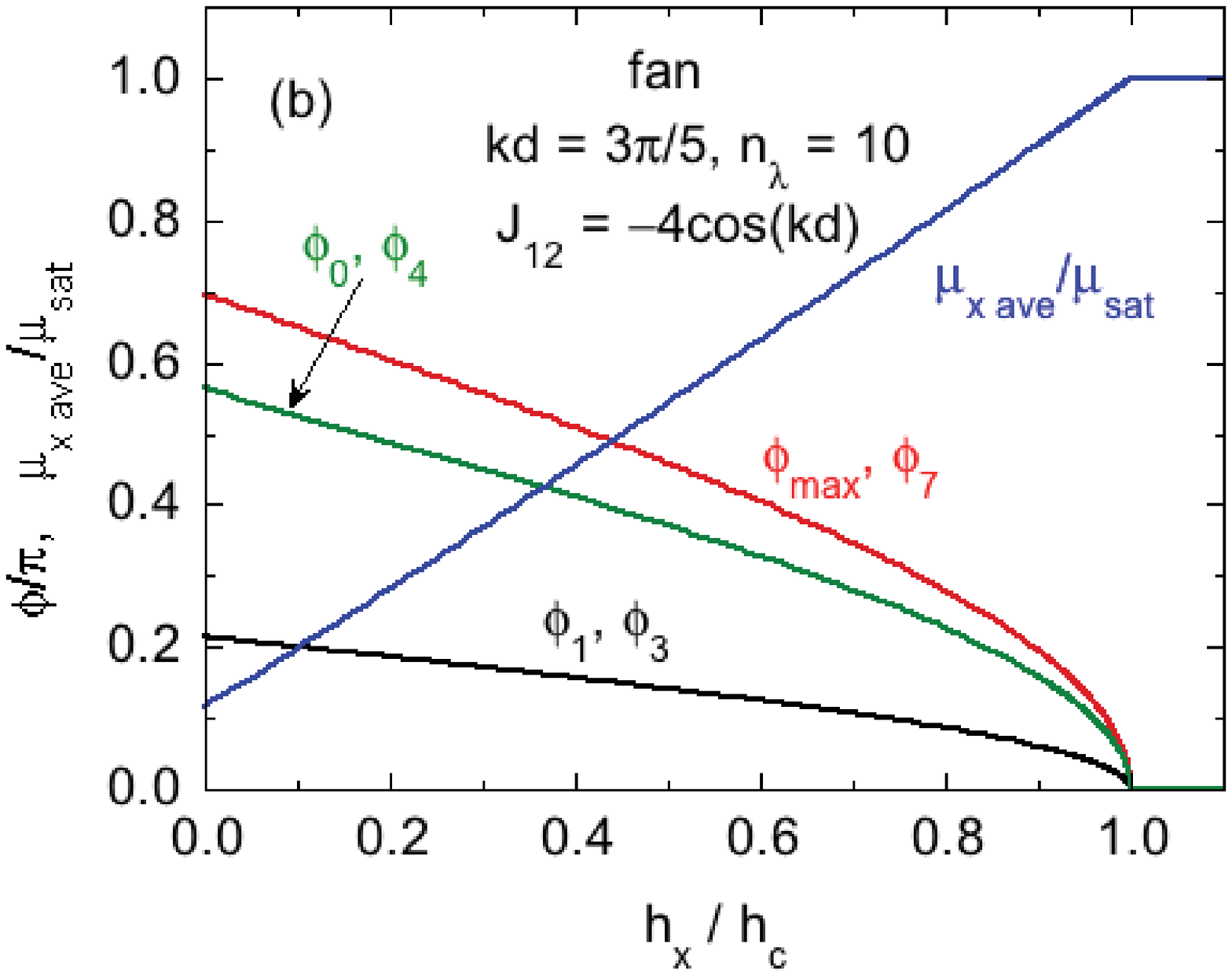}\includegraphics [width=3.4in]{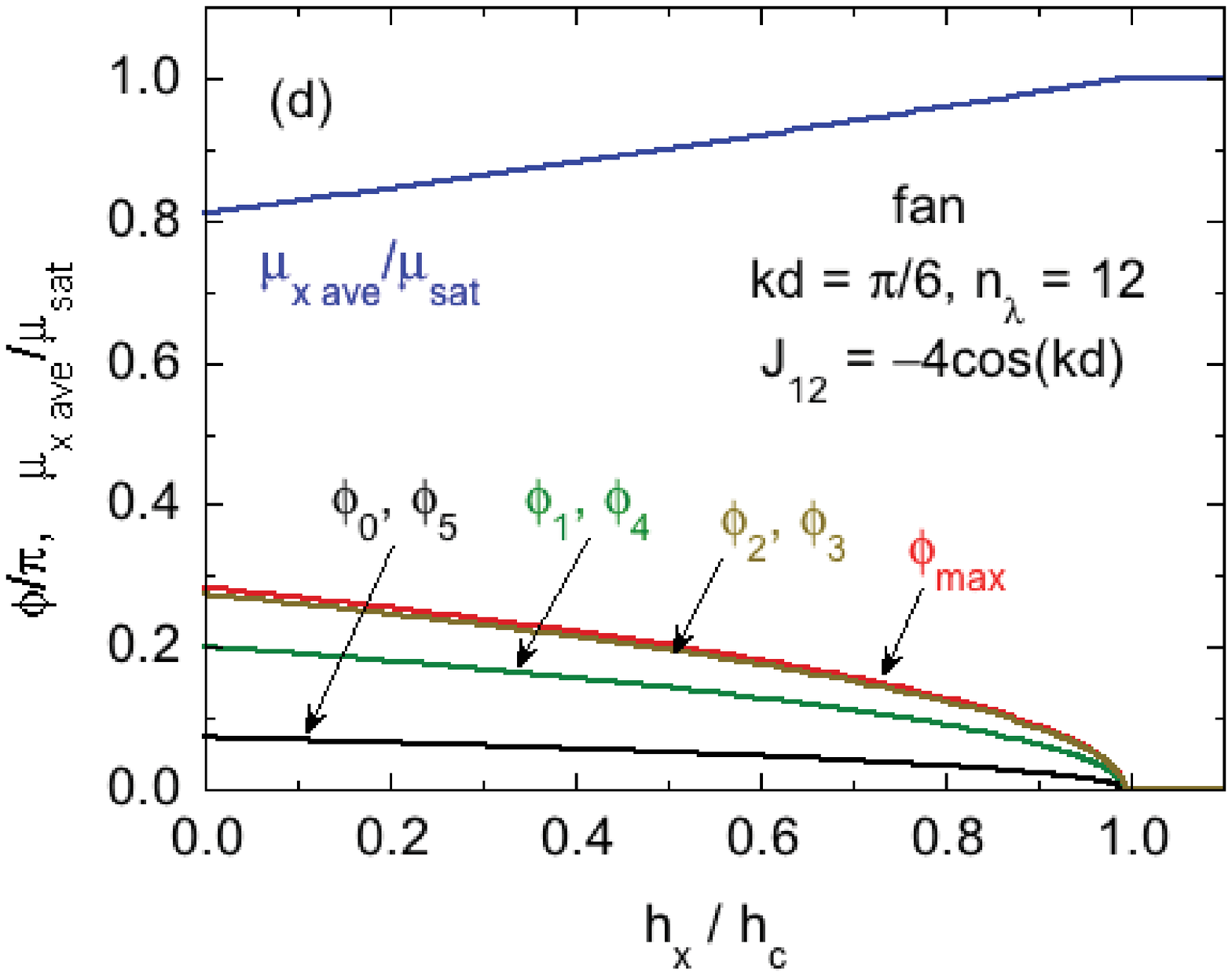}
\caption {(Color online) Plots as in Fig.~\ref{Fig:FanDatN2_kdPi} of sinusoidal fans with (a)~$kd=3\pi/4$, $n_\lambda = 8$, (b)~$kd=3\pi/5$, $n_\lambda = 10$,  (c)~$kd=4\pi/11$, $n_\lambda = 11$, and (d)~~$kd=\pi/6$, $n_\lambda = 12$.  For each $kd$ the value of $J_{12}$ must also be specified.  These are chosen to be the values for the helices with the same $kd$ values using Eq.~(\ref{Eq:Coskd2}) as given in each figure. The sinusoidal fan properties are similar to those of helices with the same $kd$ and~$J_{12}$ values above their respective helix to fan transition fields.  }
\label{Fig:FanAnglesHelix_kd3PiOn4}
\end{figure*}

Helix structures can undergo a transition from the helix phase to a fan phase at sufficiently large reduced fields~$h_x$ \cite{Nagamiya1962}.  Therefore, a particularly interesting case is when $kd$ for the fan is the same as for a commensurate helix and when $J_{12}$ is the same as given by Eq.~(\ref{Eq:Coskd2}) for a helix with turn angle~$kd$.  Special cases of these equalities are illustrated in Figs.~\ref{Fig:FanDatN2_kdPi}(a) and~\ref{Fig:FanDatN2_kdPi}(b).  Although in general the wavelength of a helix along the $z$~axis depends on $h_x$, contrary to the assumption of this paper, when $h_x$ approaches the critical field $h_{\rm c}$, the wave vector of the fan is the same ($kd$) as for the corresponding helix in zero field \cite{Nagamiya1962}.  This is consistent with our result that $h_c$ for the sinusoidal fan is the same as that found from energy minimization of the helix in the field-induced fan regime, both with the same values of $kd$ and $J_{12}$, as shown in the following Sec.~\ref{Sec:HelixFieldDep}.

Plots of the fan amplitudes~$\phi_{\rm max}$, moment angles~$\phi_n$ and reduced average $x$-axis moments $\mu_{x{\rm ave}}/\mu_{\rm sat}$ versus the reduced field ratio~$h_x/h_{\rm c}$ for $J_{12}$ values corresponding to helix turn angles~$kd = 3\pi/4,\ 3\pi/5,\ 4\pi/11$, and~$\pi/6$ are presented in Fig.~\ref{Fig:FanAnglesHelix_kd3PiOn4}.  Two of these $kd$ values are greater than $\pi/2$ (AFM $J_{1}$) and two are smaller (FM $J_{1}$) and were chosen to be representative of values for helices that show clear helix to fan transitions (either first or second order) in both ranges of $kd$ as shown later.  For $kd\gtrsim 4\pi/9$, we find that the helix-to-fanlike structure change is usually but not always a smooth crossover rather than a phase transition.

We obtain the reduced critical field $h_{\rm c}$ for the particular values $J_{12} = -4\cos(kd)$ of the sinusoidal fan as follows.  The cosine terms in the average energy expression~(\ref{Eq:EGen2}) for $E_n$ are expanded to fourth order in the arguments $\phi_n$ given in Eq.~(\ref{Eq:Phi}).  Then the average energy $E_{\rm ave}$ in Eq.~(\ref{Eq:EAveGen}) is minimized with respect to $\phi_{\rm max}$, yielding an expression for $\phi_{\rm max}$ in terms of $kd,\  n_\lambda$, and the reduced field $h_x$.  The fan to PM phase transition occurs when $\phi_{\rm max}\to 0$.  Solving for $h_x \equiv h_{\rm c}$ in the expression $\phi_{\rm max}(kd,n_\lambda,h_x)=0$ yields the following solutions for $h_{\rm c}$ in terms of $kd$ and $n_\lambda$.  The solutions for odd or even~$n_\lambda$ are
\bse
\label{Eqs:hcFan}
\be
h_{\rm c} = \frac{\rm Num(odd~or~even)}{\rm Denom(odd~or~even)},
\ee
\begin{widetext}
where
\bea
{\rm Num(odd)} &=& 8 \sin^4(kd/2) \Big\{-3 + 4 n_\lambda + 3 \cos(2 kd\, n_\lambda) + \csc(kd) \big\{-\sin(3 kd) + \sin[kd (3 - 2 n_\lambda)]\big\}\\
&& -\ 2 \big\{5 \cos(kd\, n_\lambda) +\ 4 \cos[kd(1 - n_\lambda)]\big\} \cot(kd) \sin(kd\, n_\lambda)\Big\},\nonumber\\
{\rm Denom(odd)} &=& -1 + 2 n_\lambda + \csc(kd) \sin[kd(1 - 2 n_\lambda)],\nonumber\\
\nonumber\\
{\rm Num(even)} &=& 8 \sin^4(kd/2) \Big\{4 n_\lambda \sin(kd) + \cos(2 kd\, n_\lambda) \sin(2 kd)\\
&& -\ [4 + 4 \cos(kd) + \cos(2 kd)] \sin(2 kd\, n_\lambda) - \sin[2 kd (1 + n_\lambda)]\Big\},\nonumber\\
{\rm Denom(even)} &=& 2 n_\lambda \sin(kd) - \sin(2 kd\, n_\lambda).\nonumber
\eea
\end{widetext}
\ese

\begin{table*}
\caption{\label{TabFanData} Critical field~$h_{\rm c}$ for second-order transitions of the fan to the paramagnetic phase using the $J_{12}$ for the helix with the same turn angle~$kd$ obtained using Eq.~(\ref{Eq:Coskd2}).  Both analytical and numerical values for $h_{\rm c}$ are given for each $kd$ value. }
\begin{ruledtabular}
\begin{tabular}{ccccc|ccccc}
$kd/\pi$	&	$kd/\pi$	&$n_\lambda$ &	 $h_{\rm c}$	& $h_{\rm c}$	& $kd/\pi$	&	$kd/\pi$	&	$n_\lambda$ 	&	 $h_{\rm c}$	& $h_{\rm c}$	\\
&	&	&	(helix $J_{12}$)	& (helix $J_{12}$)	& 	&	&	&	(helix $J_{12}$)	& (helix $J_{12}$)\\
\hline 
11/12  &  0.916667	&  24  & $16\cos^4(\pi/24)$   &  1.54595e+01  &  5/12	&  0.416667  	&  24	& $16\sin^4(5\pi/24)$   	&  2.19740e+00  \\
10/11  &  0.909091  &  11  & $16\cos^4(\pi/22)$   &  1.53585e+01  &  2/5 	&  0.4  		&  5  	& $5(3-\sqrt{5})/2$  	&  1.90983e+00  \\
9/10	&  0.9  		&  20  & $16\cos^4(\pi/20)$   &  1.52265e+01  &  8/21  &  0.380952  	&  21  	& $16\sin^4(4\pi/21)$	&  1.61117e+00  \\
8/9	&  0.888889  	&  9	  & $16\cos^4(\pi/18)$   &  1.50496e+01  &  3/8  	&  0.375  	&  16  	& $16\sin^4(3\pi/16)$  	&  1.52432e+00  \\
7/8	&  0.875  	&  16  & $16\cos^4(\pi/16)$   &  1.48052e+01  &  4/11  &  0.363636  	&  11  	& $16\sin^4(2\pi/11)$   	&  1.36696e+00  \\
6/7	&  0.857143  	&  7   & $16\cos^4(\pi/14)$   &  1.44547e+01  &  6/17  &  0.352941  	&  17  	& $16\sin^4(3\pi/17)$  	&  1.22882e+00  \\
5/6	&  0.833333  	&  12  & $7+4\sqrt{3}$   	&  1.39282e+01  &  8/23  &  0.347826  	&  23  	& $16\sin^4(4\pi/23)$  	&  1.16612e+00  \\
9/11	&  0.818182  	&  22  & $16\cos^4(\pi/11)$   &  1.35609e+01  &  1/3  	&  0.333333  	&  6  	&    	1			&  1  		\\
4/5	&  0.8  		&  5   & $5(3+\sqrt{5})/2$    &  1.30902e+01  &  6/19  &  0.315789  	&  19  	& $16\sin^4(3\pi/19)$	&  8.21024e$-$01  \\
7/9	&  0.777778  	&  18  & $16\cos^4(\pi/9)$    &  1.24757e+01  &  4/13  &  0.307692 	&  13  	& $16\sin^4(2\pi/13)$  	&  7.46272e$-$01  \\
10/13&  0.769231  	&  13  & $16\cos^4(3\pi/26)$  &  1.22292e+01  &  3/10  &  0.3  		&  20  	& $16\sin^4(3\pi/20)$   	&  6.79684e$-$01  \\
3/4	&  0.75  		&  8   & $2(3+2\sqrt{2})$    	&  1.16569e+01  &  2/7  	&  0.285714  	&  7  	& $16\sin^4(\pi/7)$ 	&  5.67040e$-$01  \\
8/11	&  0.727273  	&  11  & $16\cos^4(3\pi/22)$  &  1.09543e+01  &  3/11  &  0.272727  	&  22  	& $16\sin^4(3\pi/22)$   	&  4.76484e$-$01  \\
5/7  &  0.714286  	&  14  & $16\cos^4(\pi/7)$    &  1.05429e+01  &  4/15  &  0.266667  	&  15  	& $16\sin^4(2\pi/15)$ 	&  4.37898e$-$01  \\
7/10	&  0.7  		&  20  & $16\cos^4(3\pi/20)$  &  1.00842e+01  &  6/23  &  0.260870  	&  23  	& $16\sin^4(3\pi/23)$ 	&  4.03090e$-$01  \\
2/3  &  0.666667  	&  3   &  9  				&  9			  &  1/4  &  0.25	  	&  8  	& $2(3-2\sqrt{2})$ 		&  3.43146e$-$01  \\
7/11	&  0.636364  	&  22  & $16\cos^4(2\pi/11)$  &  8.01360e+00  &  4/17  &  0.235294  	&  17  	& $16\sin^4(2\pi/17)$  	&  2.72465e$-$01  \\
5/8  &  0.625  	&  16  & $16\cos^4(3\pi/16)$  &  7.64725e+00  &  2/9  	&  0.222222  	&  9  	& $16\sin^4(\pi/9)$ 	&  2.18941e$-$01  \\
8/13	&  0.615385  	&  13  & $16\cos^4(5\pi/26)$  &  7.33982e+00  &  4/19  &  0.210526  	&  19  	& $16\sin^4(2\pi/19)$  	&  1.77847e$-$01  \\
3/5	&  0.6  		&  10  & $(7+3\sqrt{5})/2$    &  6.85410e+00  &  1/5  	&  0.2  		&  10  	& $(7-3\sqrt{5})/2$ 	&  1.45898e$-$01  \\
10/17&  0.588235  	&  17  & $16\cos^4(7\pi/34)$  &  6.48887e+00  &  4/21  &  0.190476  	&  21  	& $16\sin^4(2\pi/21)$   	&  1.20772e$-$01  \\
7/12	&  0.583333  	&  24  & $16\cos^4(5\pi/24)$  &  6.33850e+00  &  2/11  &  0.181818  	&  11  	& $16\sin^4(\pi/11)$ 	&  1.00802e$-$01  \\
4/7	&  0.571429  	&  7   & $16\cos^4(3\pi/14)$  &  5.97823e+00  &  4/23  &  0.173913  	&  23  	& $16\sin^4(2\pi/23)$ 	&  8.47748e$-$02  \\
5/9	&  0.555556  	&  18  & $16\cos^4(2\pi/9)$   &  5.50980e+00  &  1/6  	&  0.166667  	&  12  	& $7-4\sqrt{3}$ 		&  7.17968e$-$02  \\
6/11	&  0.545455  	&  11  & $16\cos^4(5\pi/22)$  &  5.21953e+00  &  2/13  &  0.153846  	&  13  	& $16\sin^4(\pi/13)$ 	&  5.24813e$-$02  \\
8/15	&  0.533333  	&  15  & $16\cos^4(7\pi/30)$  &  4.87993e+00  &  1/7  	&  0.142857  	&  14  	& $16\sin^4(\pi/14)$	&  3.92287e$-$02  \\
10/19&  0.526316  	&  19  & $16\cos^4(9\pi/38)$  &  4.68791e+00  &  2/15  &  0.133333 	&  15  	& $16\sin^4(\pi/15)$ 	&  2.98976e$-$02  \\
1/2	&  1/2 		&  4   &     		4		&  4  		 &  1/8  	&  0.125  	&  16  	& $16\sin^4(\pi/16)$	&  2.31773e$-$02  \\
10/21&  0.476190  	&  21  &  $16\sin^4(5\pi/21)$ &  3.42450e+00  &  2/17  &  0.117647  	&  17  	& $16\sin^4(\pi/17)$ 	&  1.82400e$-$02  \\
8/17	&  0.470588  	&  17  & $16\sin^4(4\pi/17)$  &  3.29591e+00  &  1/9  	&  0.111111  	&  18  	& $16\sin^4(\pi/18)$ 	&  1.45479e$-$02  \\
6/13	&  0.461538  	&  13  & $16\sin^4(3\pi/13)$  &  3.09382e+00  &  2/19  &  0.105263  	&  19  	& $16\sin^4(\pi/19)$ 	&  1.17431e$-$02  \\
5/11	&  0.454545  	&  22  & $16\sin^4(5\pi/22)$  &  2.94250e+00  &  1/10  &  0.1  		&  20  	& $16\sin^4(\pi/20)$  	&  9.58186e$-$03  \\
4/9	&  0.444444  	&  9   &  $16\sin^4(2\pi/9)$  &  2.73143e+00  &  2/21  &  0.0952381  	&  21  	& $16\sin^4(\pi/21)$ 	&  7.89510e$-$03  \\
10/23&  0.434783  	&  23  &  $16\sin^4(5\pi/23)$ &  2.53793e+00  &  1/11  &  0.0909091  	&  22  	& $16\sin^4(\pi/22)$  	&  6.56328e$-$03  \\
3/7	&  0.428571  	&  14  & $16\sin^4(3\pi/14)$  &  2.41789e+00  &  2/23  &  0.0869565  	&  23  	& $16\sin^4(\pi/23)$ 	&  5.50051e$-$03  \\
8/19	&  0.421053  	&  19  &  $16\sin^4(4\pi/19)$ &  2.27717e+00  &  1/12  &  0.0833333  	&  24  	& $16\sin^4(\pi/24)$ 	&  4.64420e$-$03  \\
\end{tabular}
\end{ruledtabular}
\end{table*}

The critical fields $h_{\rm c}$ calculated from Eqs.~(\ref{Eqs:hcFan}) are listed for 72 values of $kd$ in Table~\ref{TabFanData} with both even and odd $n_\lambda\leq 24$.  We infer that from the list of analytic $h_{\rm c}$ expressions for the discrete values in the ranges $0< kd\leq \pi/2$ and $\pi/2 \leq kd \leq \pi$ in Table~\ref{TabFanData}, $h_{\rm c}$ can respectively be expressed for all cases as
\bse
\label{Eqs:hcVals}
\bea
h_{\rm c} &=& 16\sin^4\left(\frac{kd}{2}\right) \qquad(0 \leq kd \leq \pi/2),\label{smallhc}\\
h_{\rm c} &=& 16\cos^4\left(\frac{\pi-kd}{2}\right) \quad(\pi/2 \leq kd \leq \pi).\label{bighc}
\eea
\ese
Since Eqs.~(\ref{Eqs:hcVals}) apply to all discrete values of $kd$ in Table~\ref{TabFanData}, we suggest that the same formulas also apply to \mbox{incommensurate} (continuous) values of $kd$ in the respective ranges.  In the limit of small $kd$, Eq.~(\ref{smallhc}) gives
\be
h_{\rm c} = (kd)^4 \qquad (kd \to 0).
\label{Eq:hcSmallkd}
\ee
For such small values of~$kd$, the system is nearly ferromagnetic (see Fig.~\ref{Fig:J0Jz1Jz2PhaseDiag}). A result equivalent to Eq.~(\ref{Eq:hcSmallkd}) was obtained via a continuum model in Ref.~\cite{Enz1961}.

\begin{figure}
\includegraphics [width=3.4in]{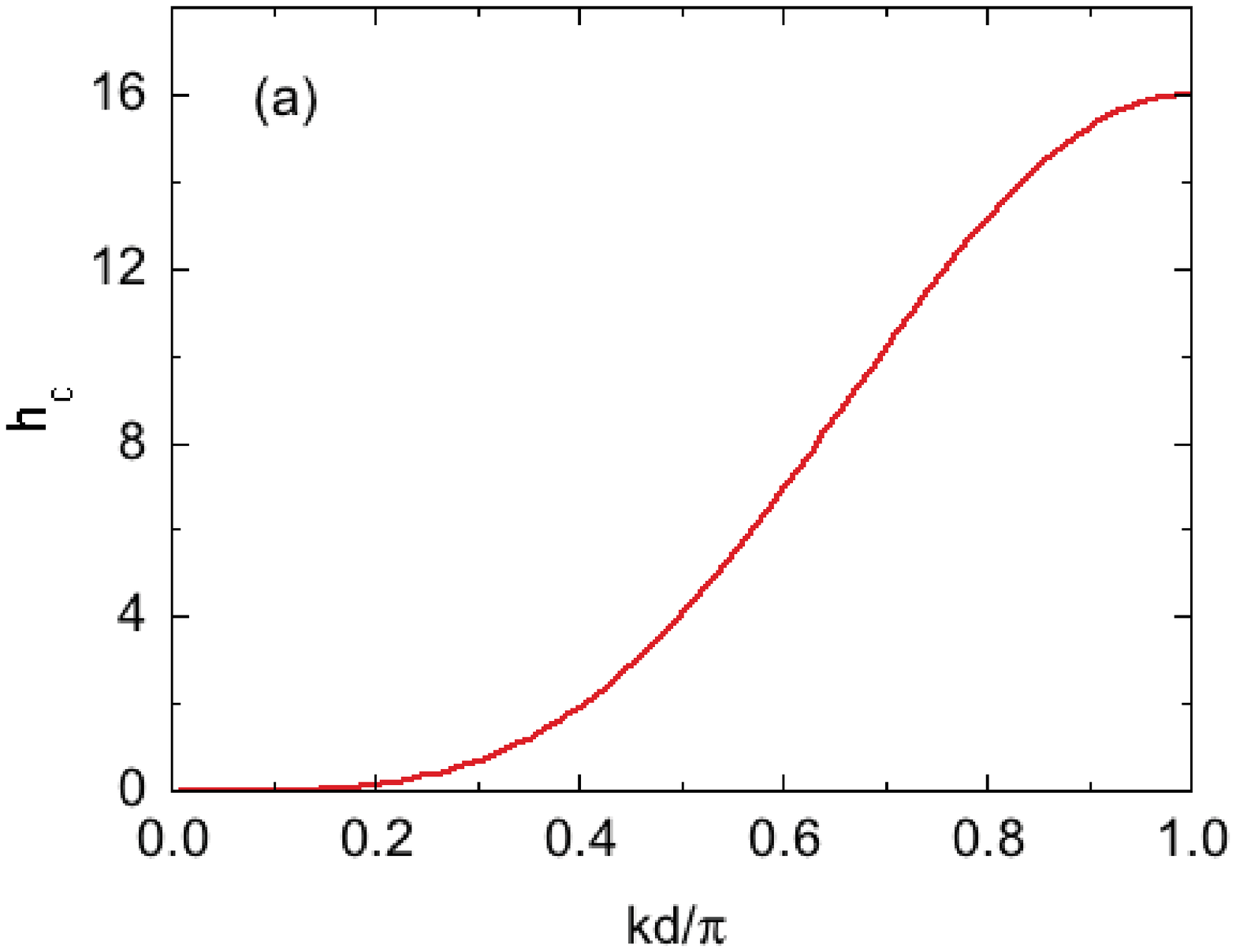}
\includegraphics [width=3.4in]{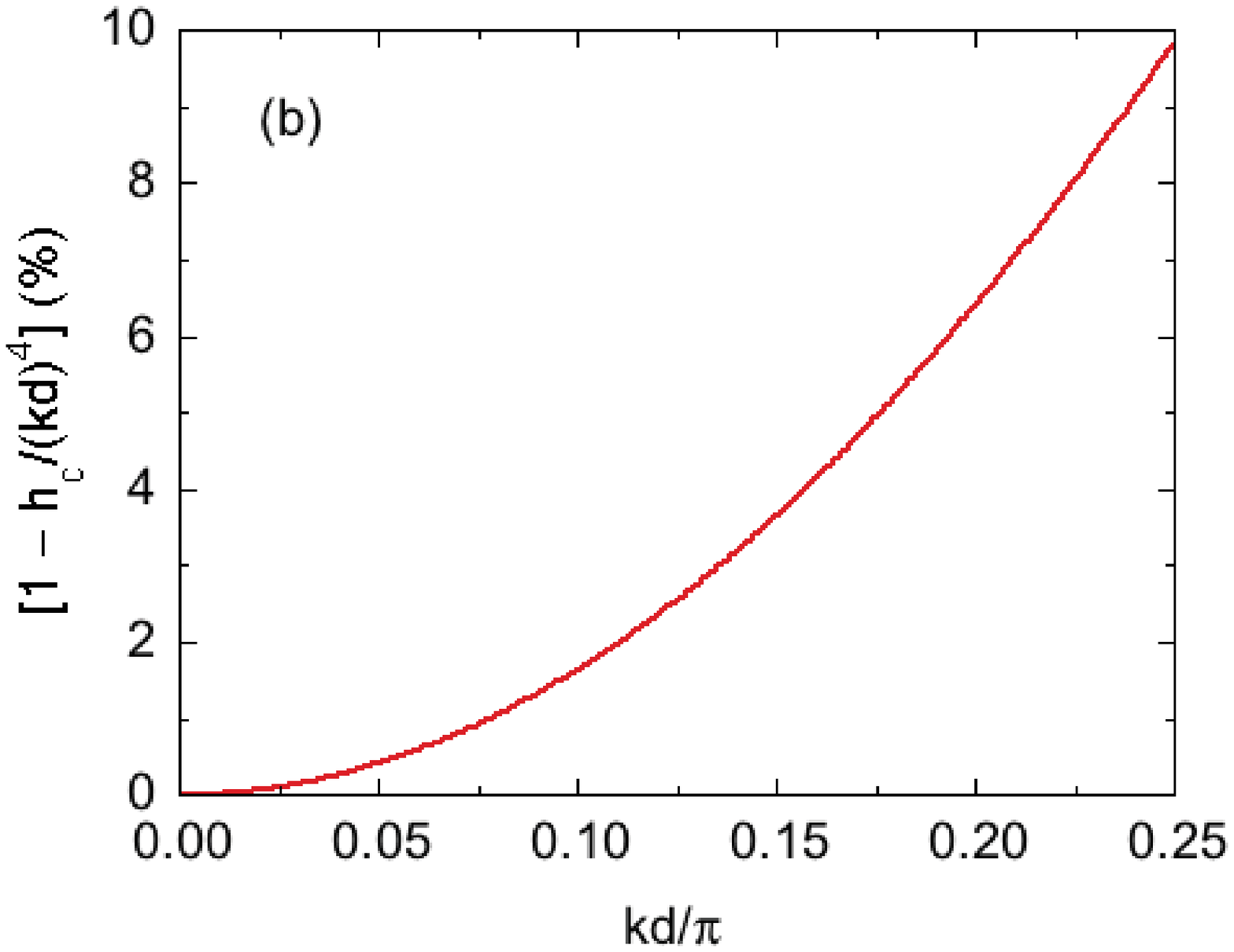}
\caption {(Color online) (a)~The critical field~$h_{\rm c}$ versus the turn angle~$kd$ between adjacent layers of a sinusoidal fan structure with $J_{12}$ set to $J_{12} = -4\cos(kd)$ for helices, in units of~$\pi$, for the range $0<kd\leq1$ described by Eqs.~(\ref{Eqs:hcVals}). (b)~The percentage difference between $h_{\rm c}$ and the limiting behavior $(kd)^4$ for the region $0<kd\leq 1/4$.}
\label{Fig:hc_fan_data}
\end{figure}

Shown in Fig.~\ref{Fig:hc_fan_data}(a) is a plot of $h_{\rm c}$ versus $kd/\pi$ over the full range $0\leq kd/\pi \leq 1$ according to the continuum Eqs.~(\ref{Eqs:hcVals}).  \mbox{Figure~\ref{Fig:hc_fan_data}(b)} shows the percentage difference between $h_{\rm c}$ and the limiting behavior $(kd)^4$ for $kd\to0$ in Eq.~(\ref{Eq:hcSmallkd}).  For example, the $h_{\rm c}$ value for the smallest value $kd = \pi/12$ in Table~\ref{TabFanData} is about 1.13\% larger than the limiting expression.

The dependence of $\phi_{\rm max}$ on $h_x$, $kd$ and~$n_\lambda$ is determined by setting the derivative of the energy with respect to $\phi_{\rm max}$ to zero and solving for $\phi_{\rm max}$ from the resultant expression.  For commensurate helices and fans, all possible $kd$~values are written as
\be
kd = \frac{2\pi m}{n_\lambda},
\ee
where~$m$ and $n_\lambda$ are positive integers with $2m \leq n_\lambda$ so that $kd \leq \pi$.  Here we solve for the critical behavior of $\phi_{\rm max}(h_x)$ for $h_x$ close to $h_{\rm c}$ by expanding the energy to fourth order in $\phi_{\rm max}$ and then setting the derivative of the energy with respect to $\phi_{\rm max}$ to zero.  Solving the resultant expression for $\phi_{\rm max}$ yields
\begin{widetext}
\be
\phi_{\rm max}^2(h_x) = \frac{8 \big[-6 + h_x + 8 \cos(2 \pi m/n_\lambda) - 2 \cos(4 \pi m/n_\lambda)\big]}{-22 + h_x + 28 \cos(2\pi m/n_\lambda) - 8 \cos(4\pi m/n_\lambda) + 4 \cos(6\pi m/n_\lambda)- 2 \cos(8\pi m/n_\lambda)} \quad (h_x\to h_{\rm c}^-).
\label{Eq:phimaxSqu}
\ee
\end{widetext}

As anticipated in Ref.~\cite{Nagamiya1962}, the critical behavior obtained from Eq.~(\ref{Eq:phimaxSqu}) is mean-field-like as expected for the present classical treatment, with
\bse
\be
\phi_{\rm max} = A(h_x)\left(1- \frac{h_x}{h_{\rm c}}\right)^{1/2} \qquad (h_x\to h_{\rm c}^-),
\label{Eq:CritPhiMaxA}
\ee
where the amplitude $A(h_x)$ in radians is given by
\be
A(h_x) = \sqrt{\frac{4}{3 + 2\cos(kd) + \cos(2kd)}}.
\label{Eq:Ahx}
\ee
\ese
To plot the field dependence of $\phi_{\rm max}$ from these expressions, one needs to first insert the appropriate $h_{\rm c}(m,n_\lambda)$ from Table~\ref{TabFanData} or Eqs.~(\ref{Eqs:hcVals}) into Eq.~(\ref{Eq:CritPhiMaxA}).

For $kd \ll1$ (close to the ferromagnetic limit, see Fig.~\ref{Fig:J0Jz1Jz2PhaseDiag}), a Taylor series expansion of Eq.~(\ref{Eq:Ahx}) to lowest order in $kd$ gives
\be
A(h_x) = \sqrt{\frac{2}{3}}\left[1 + \frac{(kd)^2}{4}\right] \quad (kd\ll1).
\label{Eq:AhxkdTo0}
\ee
For $m=1$, keeping only the first term in the expansion and combining this with Eq.~(\ref{Eq:CritPhiMaxA}) for $m=1$ gives the small-$kd$ continuum limit as
\be
\phi_{\rm max} = \sqrt{\frac{2}{3}}\left(1- \frac{h_x}{h_{\rm c}}\right)^{1/2} \quad (h_x\to h_{\rm c}^-,\ m=1),
\label{Eq:CritPhiMaxA2}
\ee
where $h_{\rm c}$ is given by Eq.~(\ref{Eq:hcSmallkd}).

\section{\label{Sec:HelixFieldDep} Field-Dependent Results: Helix Phases with Crossovers or Transitions to Fan Phases}

In general, the helix phase competes with the fan phase in high fields \cite{Nagamiya1962}.  However, in our treatment we minimize the energy of the helix with respect to all angles~$\phi_n$ independently, so there is no specification in the minimization about whether the system is in the helix or fan phase at a particular value of~$h_x$ or in some sort of transition between them.  One cannot avoid obtaining a fan phase if the energy minimization for a particular field gives a set of $\phi_n$ values corresponding to a fan, and correspondingly also for the helix phase.  However, as already noted, a fan phase obtained this way does not have an exact sinusoidal fan configuration except in the limit $h_x\to h_{\rm c}$.  This observation is explicitly illustrated later in Figs.~\ref{Fig:PhiMuxEnN11kd8PiOn11Helix}, \ref{Fig:PhiMuxEnN10kd3PiOn5Helix}, \ref{Fig:PhiMuxEnN11kd4PiOn11Helix}, and~\ref{Fig:PhiMuxEnN12kd2PiOn12Helix}.  Irrespective of the differences between the fan moment angles and those of the sinusoidal fans, the average moment per spin versus field appear to be identical for each~$kd$.

In the following two sections we discuss the structures and magnetizations versus transverse field of the two categories of helices with $J_{1}> 0$ (AFM) and $J_{1}<0$ (FM) with ${\pi/2 < kd \leq \pi}$ and ${0 < kd < \pi/2}$, respectively, according to Eq.~(\ref{Eq:Coskd2}).  Because we find that these properties vary nonmonotonically with $kd$, it is necessary to present the results for many $kd$ values to illustrate the variety and evolution of the results.  As part of these studies, we examined how the angles $\phi_n$ of the individual moments with respect to the $+x$ axis along which the field is aligned evolve with increasing field, as part of the energy minimization used to determine them.  Therefore, the subscript~$n$ in $\phi_n$ refers to the helix convention in Eqs.~(\ref{Eqs:phinHelixDefs}) and Fig.~\ref{Fig:HelixFanAngles} for all fields, even when the helix changes into a fan with increasing field.

\subsection{${\pi/2 < kd \leq \pi}$: AFM $J_{1}>0$}

\begin{figure}
\includegraphics [width=3.4in]{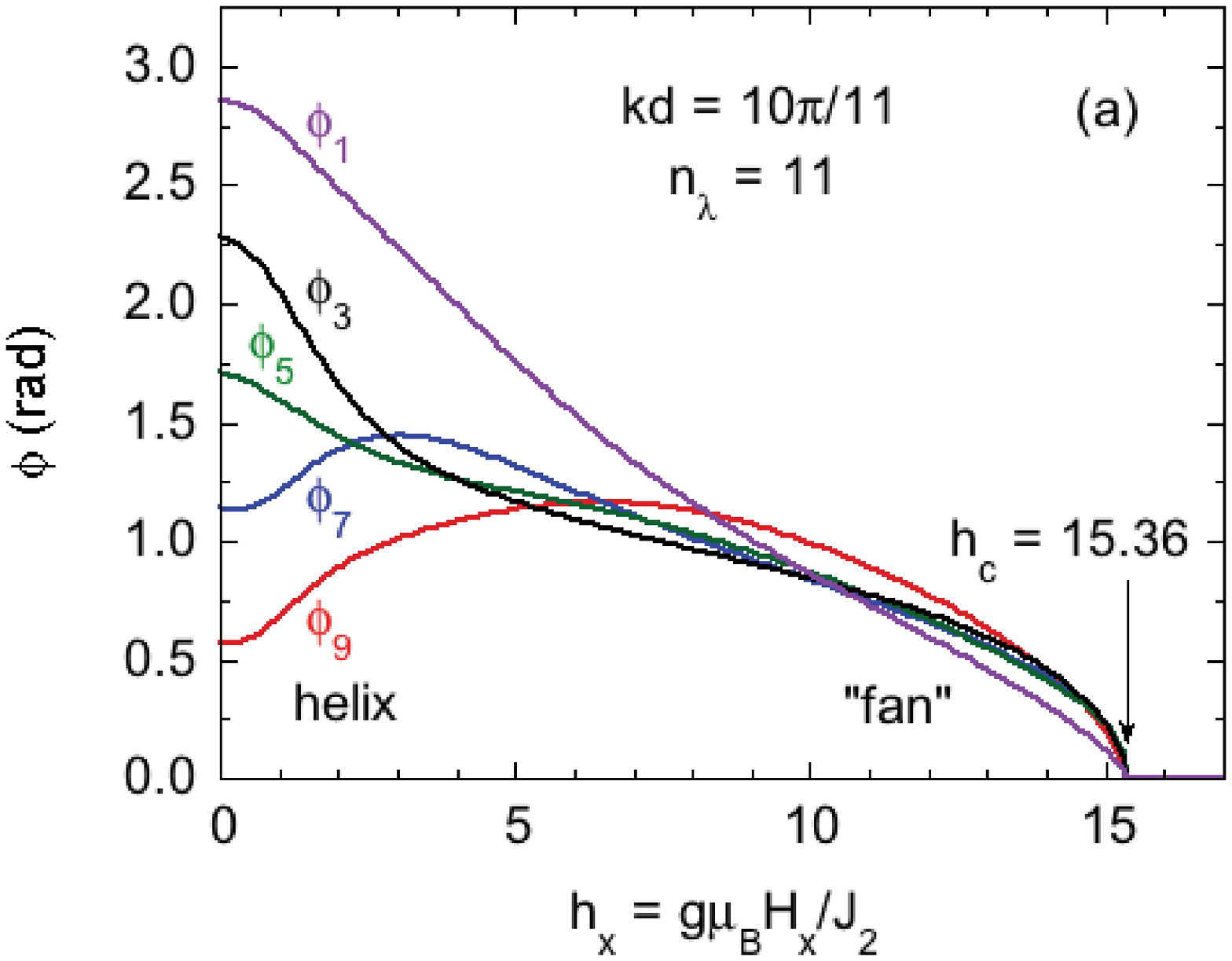}
\includegraphics [width=3.4in]{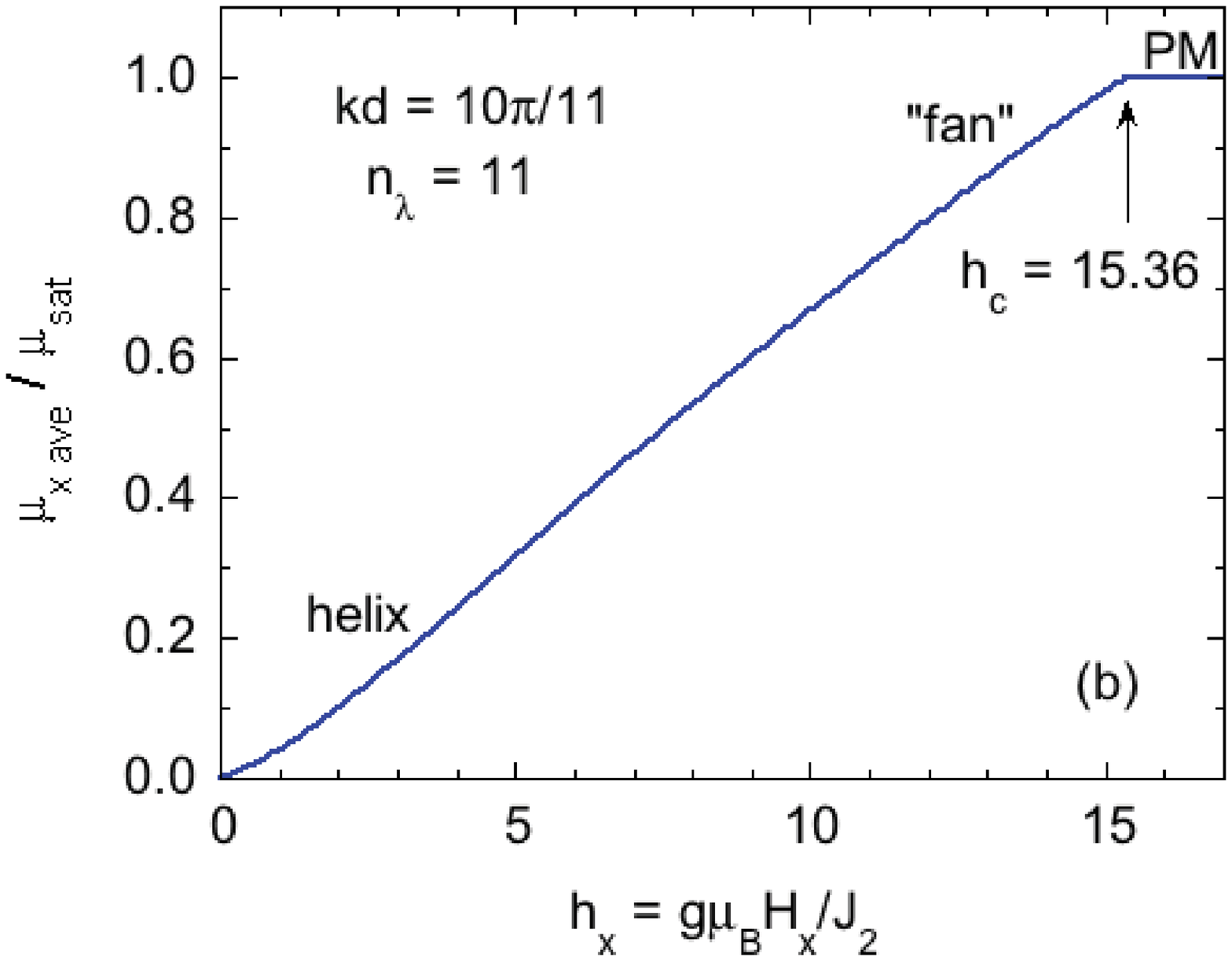}
\caption {(Color online) (a) The angles $\phi_9,\ \phi_7,\ \phi_5,\ \phi_3,$ and $\phi_1$ of the corresponding moments with respect to the $+x$ axis versus reduced in-plane field $h_x$ for a helix with turn angle $kd=10\pi/11$.  (b)~Average magnetic moment per spin in the field direction normalized by the moment magnitude, $\mu_{x{\rm ave}}/\mu_{\rm sat}$, versus~$h_x$.}
\label{Fig:PhiMuxEnN11kd10PiOn11Helix}
\end{figure}

\begin{figure}
\includegraphics [width=3.4in]{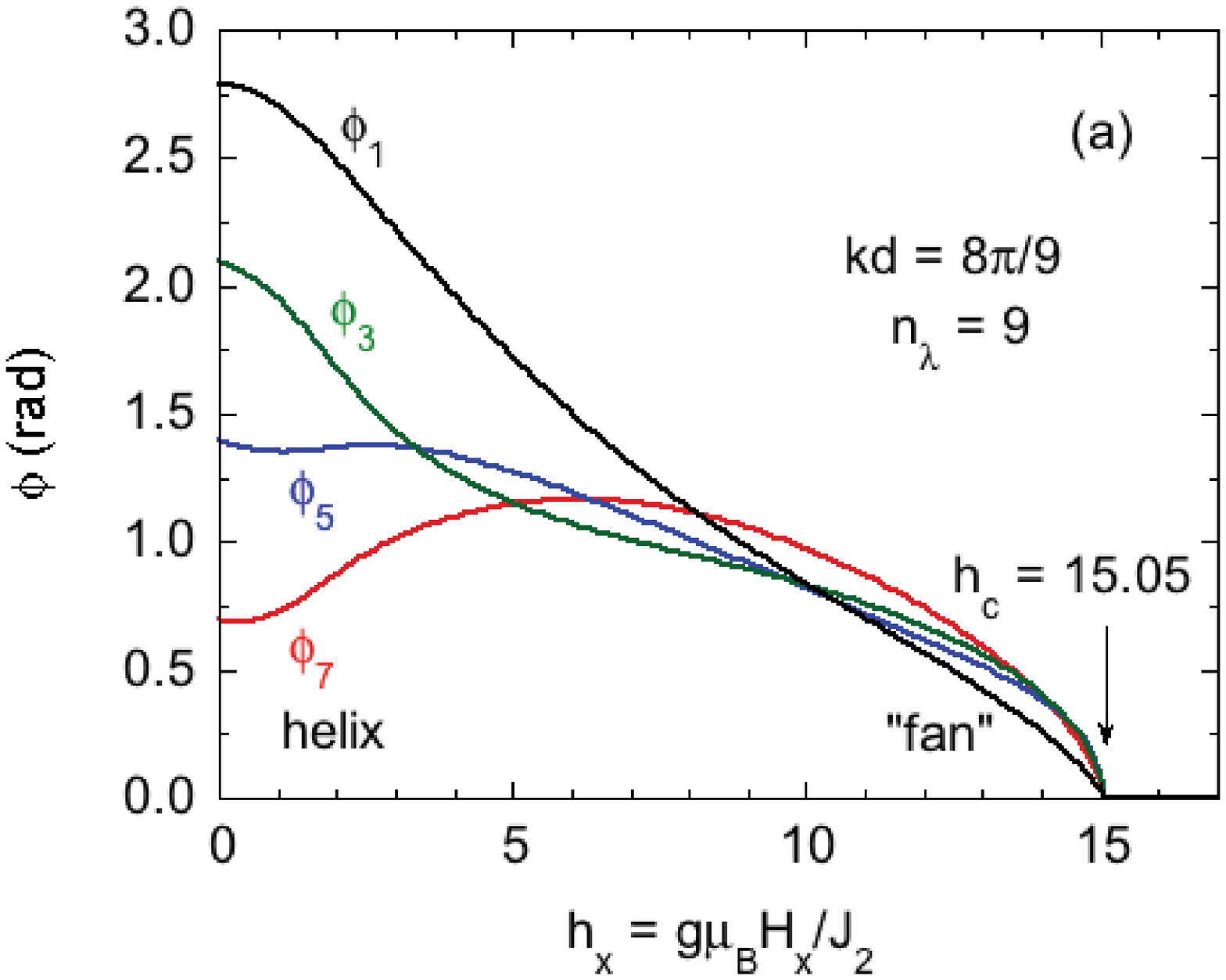}
\includegraphics [width=3.4in]{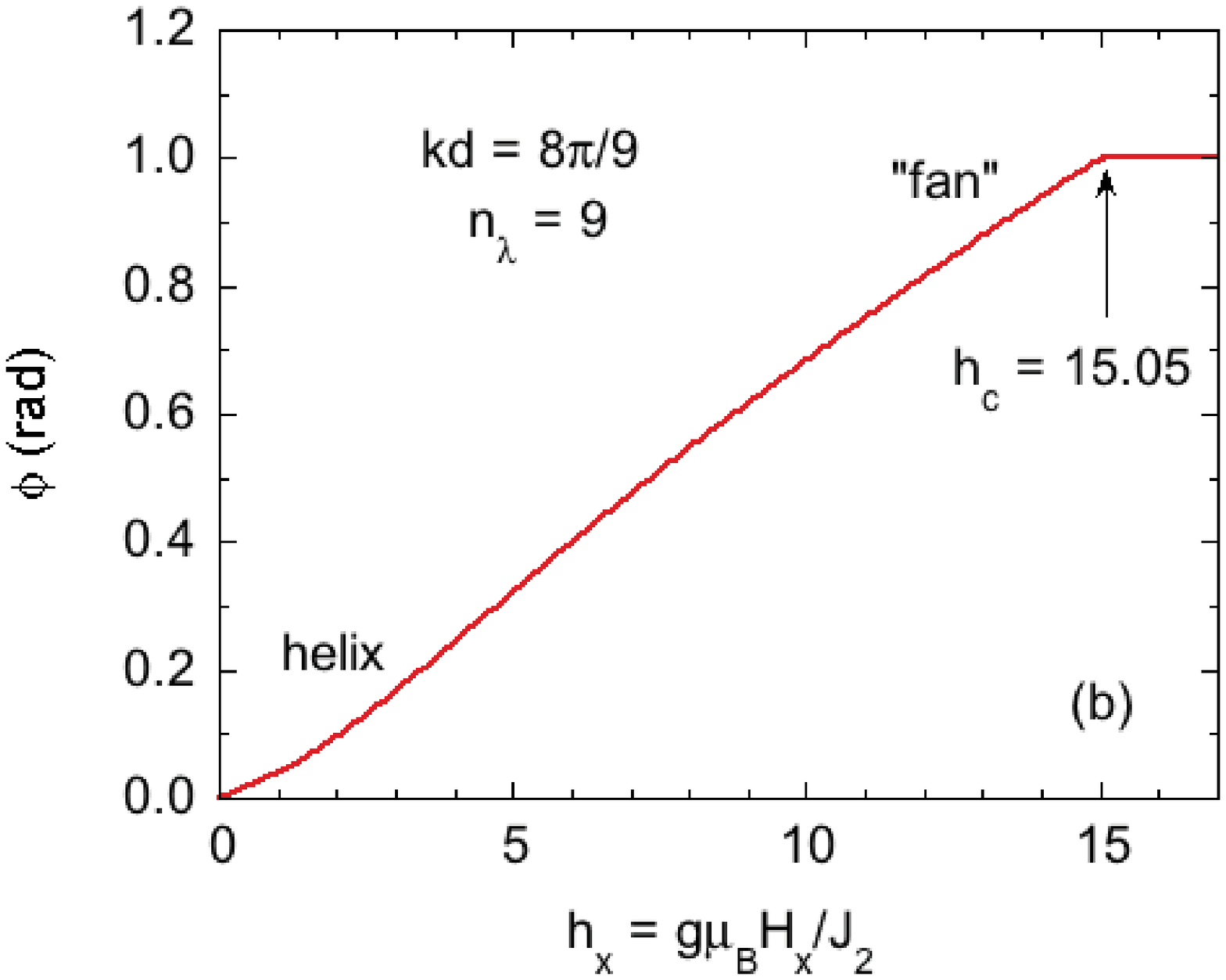}
\caption {(Color online) (a) The angles $\phi_7,\ \phi_5,\ \phi_3$, and $\phi_1$ of the corresponding moments with respect to the $+x$ axis versus reduced in-plane field $h_x$ for a helix with turn angle $kd=8\pi/9$.  (b)~Average magnetic moment per spin in the field direction normalized by the moment magnitude, $\mu_{x{\rm ave}}/\mu_{\rm sat}$, versus~$h_x$.}
\label{Fig:PhiMuxEnN9kd8PiOn9Helix}
\end{figure}

\begin{figure}
\includegraphics [width=3.3in]{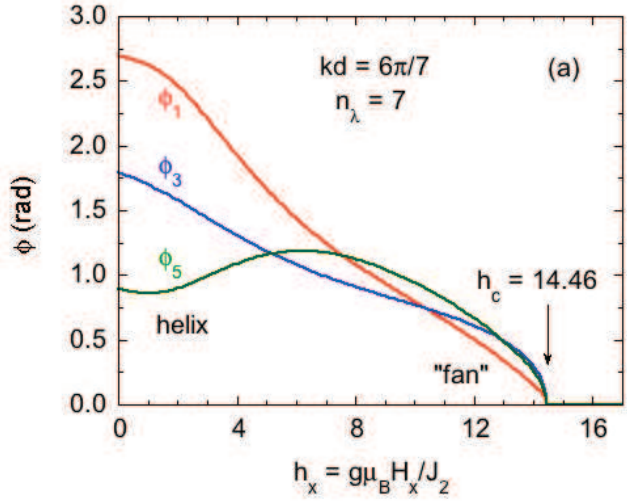}
\includegraphics [width=3.3in]{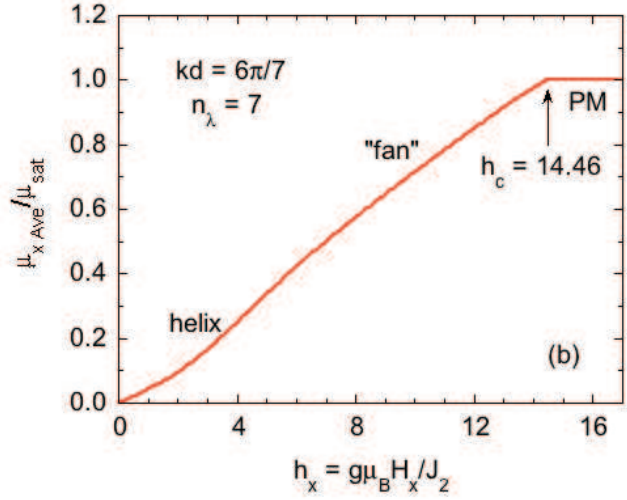}
\includegraphics [width=3.3in]{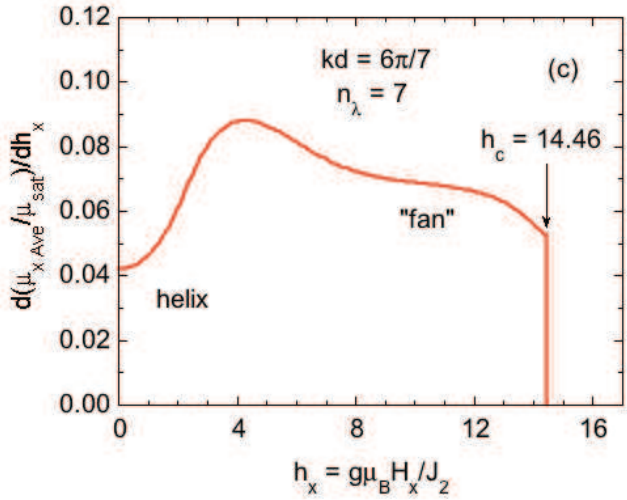}
\caption {(Color online) (a)~The angles $\phi_5$, $\phi_3$, and~$\phi_1$ of the corresponding moments with respect to the $+x$ axis versus reduced in-plane field~$h_x$ for a helix with turn angle $kd=6\pi/7$.  (b)~Average normalized magnetic moment per spin in the field direction, $\mu_{x{\rm ave}}/\mu_{\rm sat}$, versus~$h_x$.  (c)~Derivative of the average moment in~(b) with respect to~$h_x$, exhibiting the second-order ``fan'' to PM phase transition at the critical field~$h_{\rm c}$.}
\label{Fig:PhiMuxEnN7_kd6PiOn7}
\end{figure}

\begin{figure}
\includegraphics [width=3.4in]{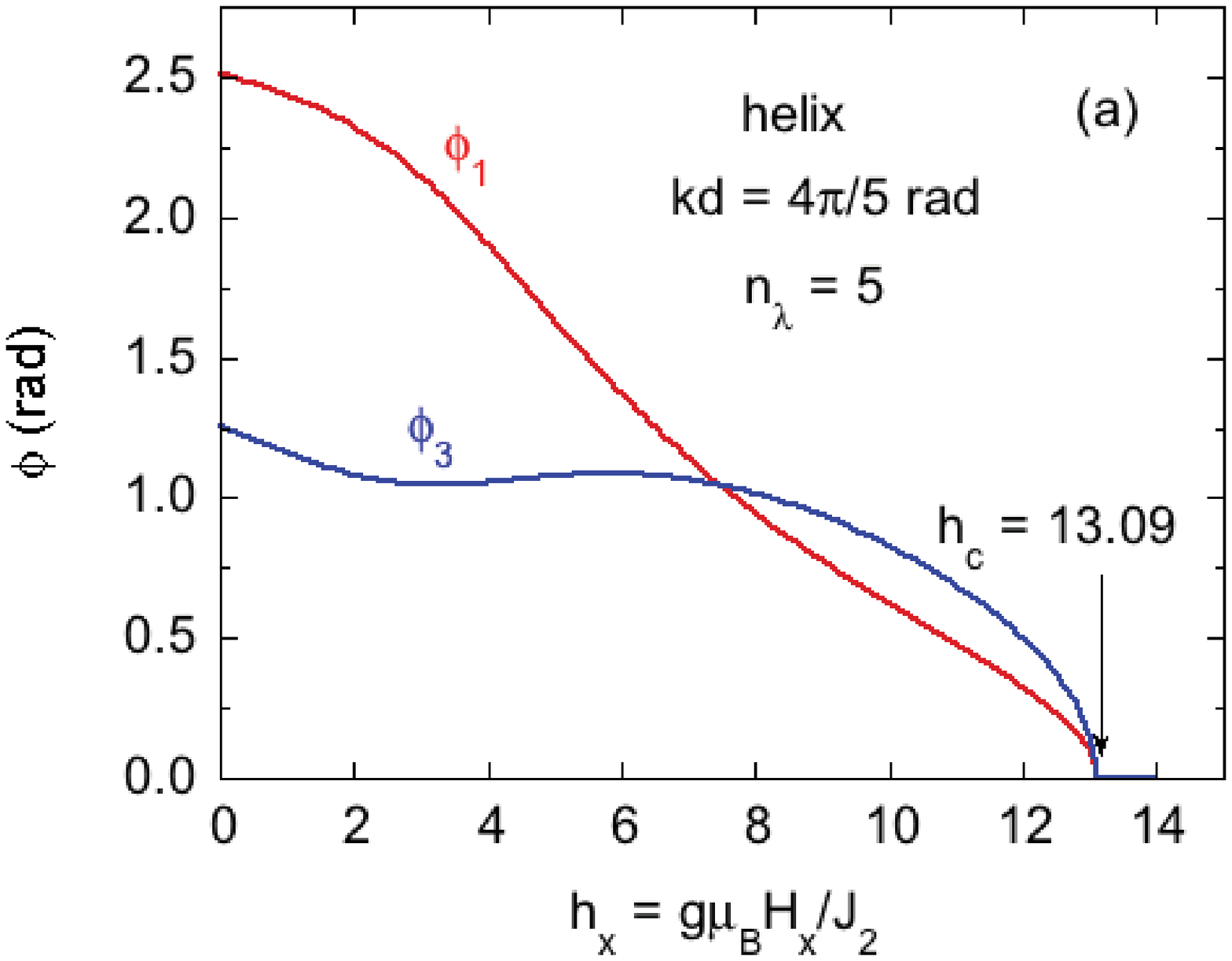}
\includegraphics [width=3.4in]{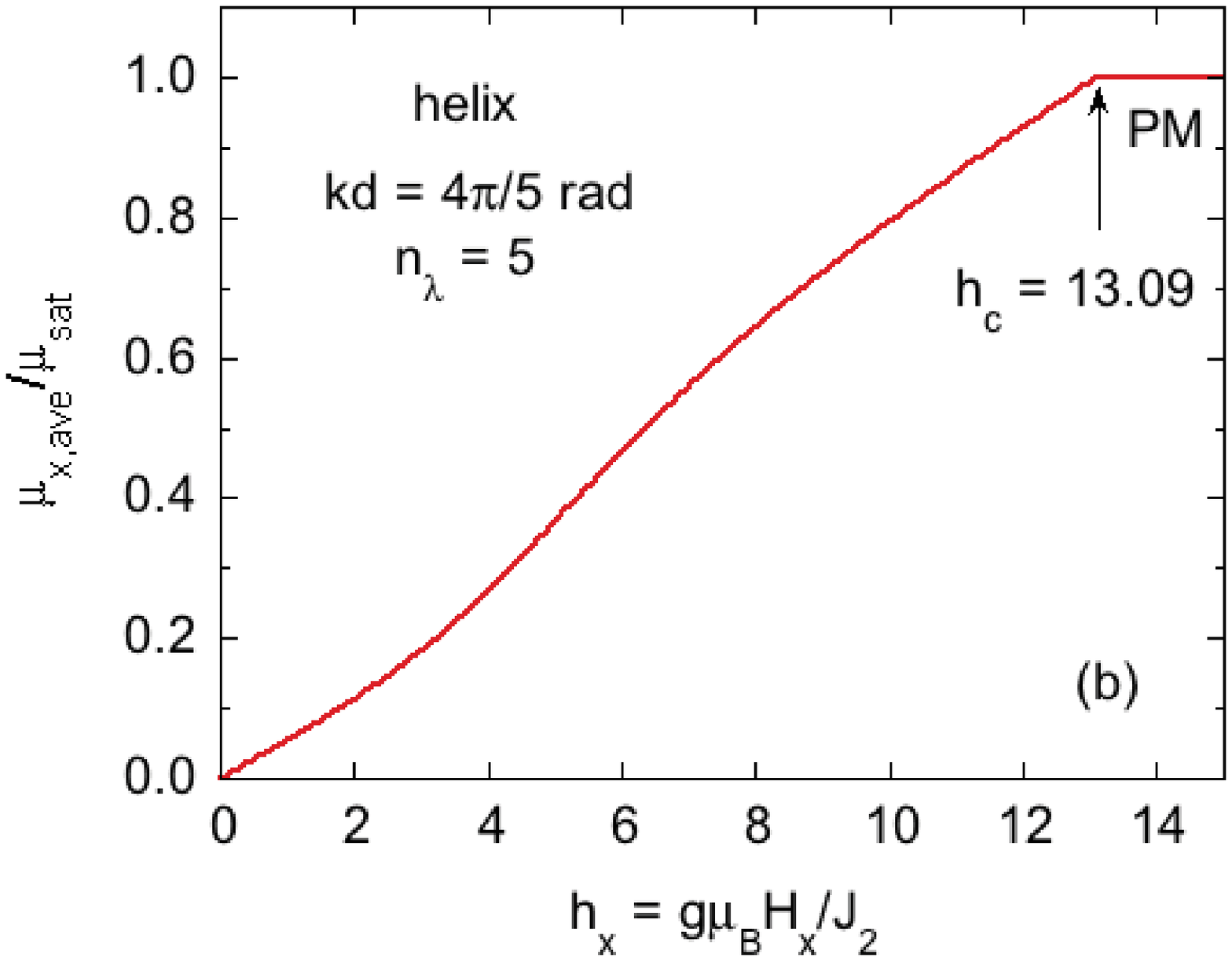}
\caption {(Color online) (a) The angles $\phi_3$ and $\phi_1$ of the corresponding moments with respect to the $+x$ axis versus reduced in-plane field~$h_x$ for a helix with turn angle $kd=4\pi/5$.  (b)~Average normalized magnetic moment per spin in the field direction $\mu_{x{\rm ave}}/\mu_{\rm sat}$ versus~$h_x$.}
\label{Fig:PhiMuxEnN5_kd4PiOn5}
\end{figure}

\begin{figure*}
\includegraphics [width=3.5in]{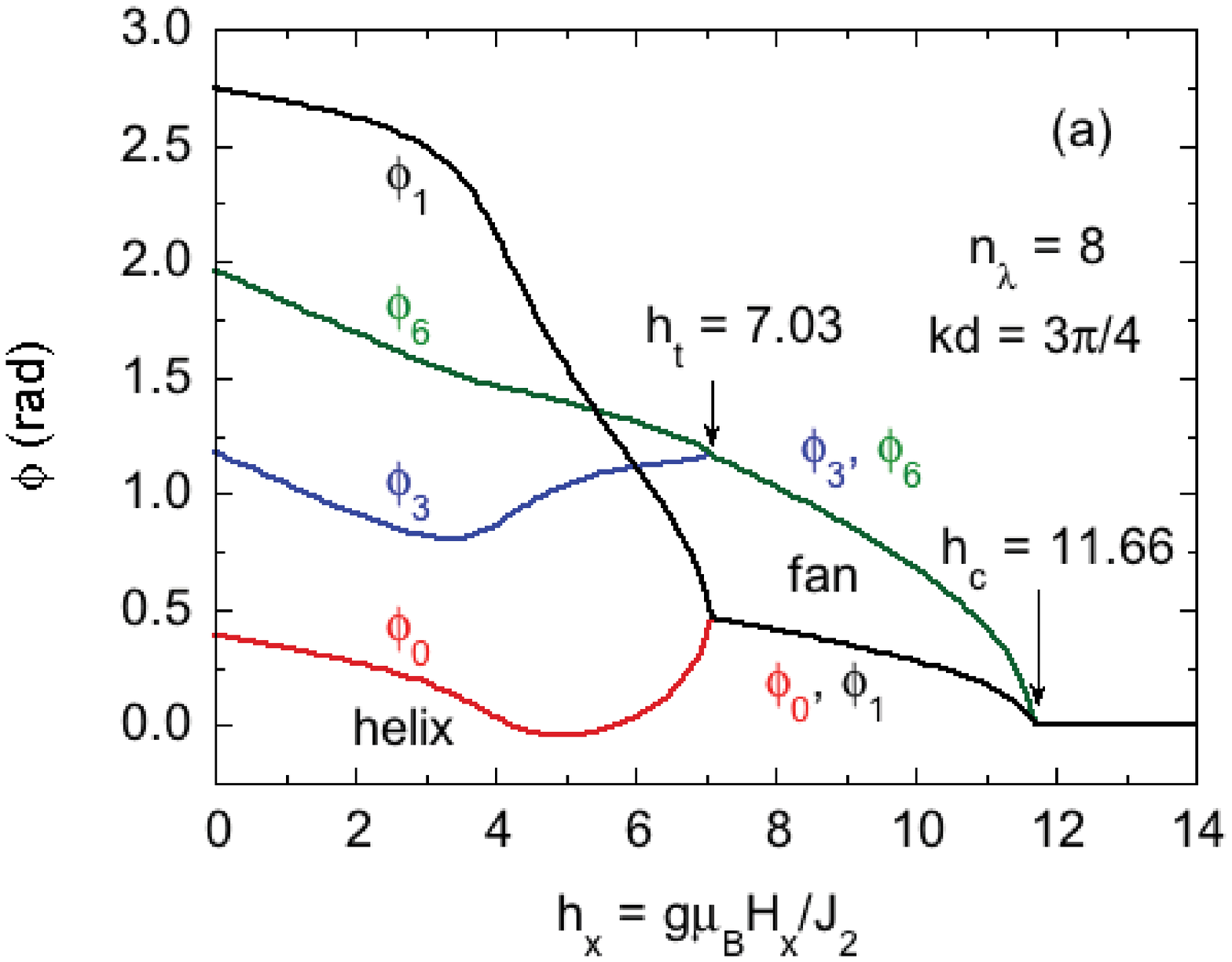}\includegraphics [width=3.5in]{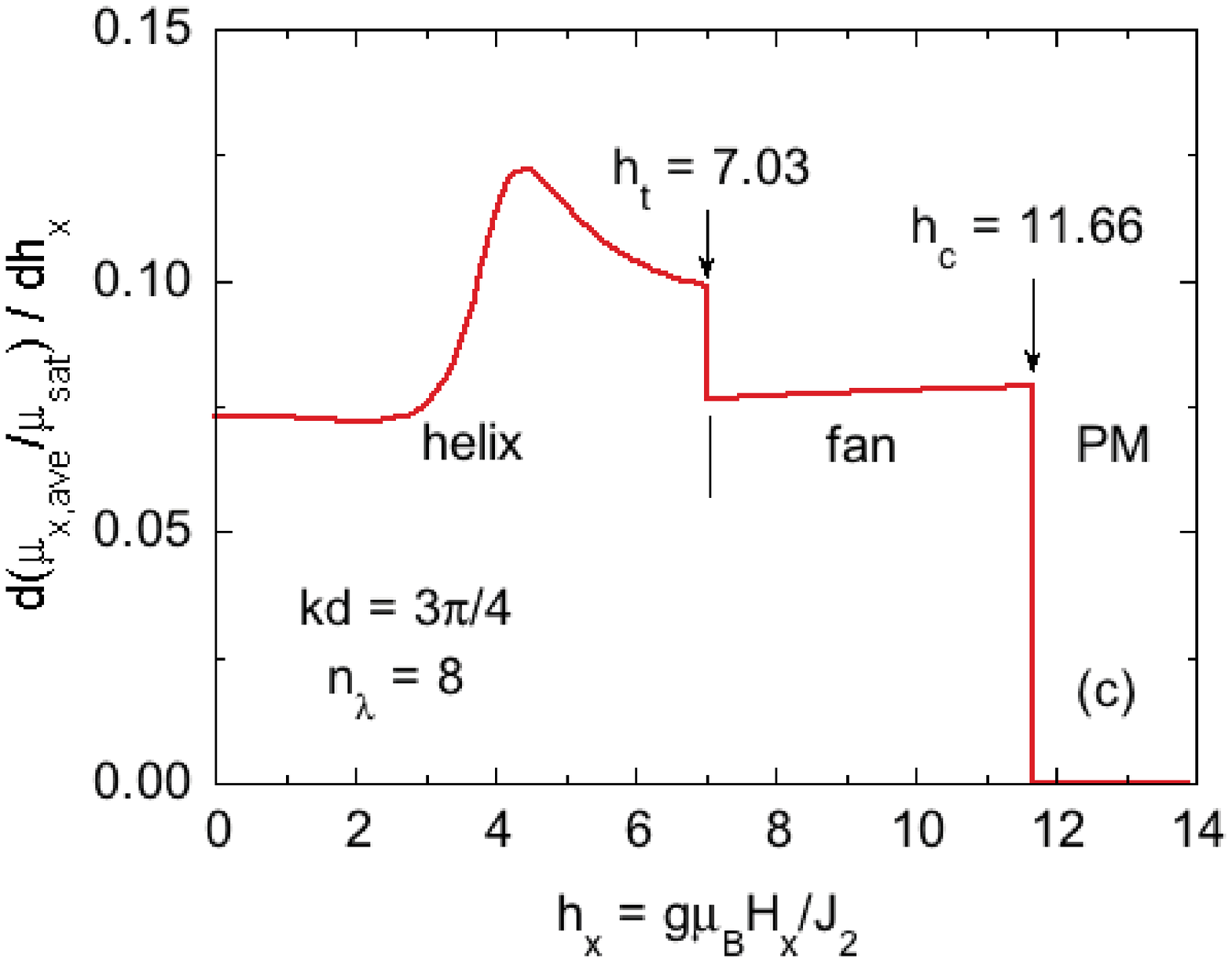}
\includegraphics [width=3.5in]{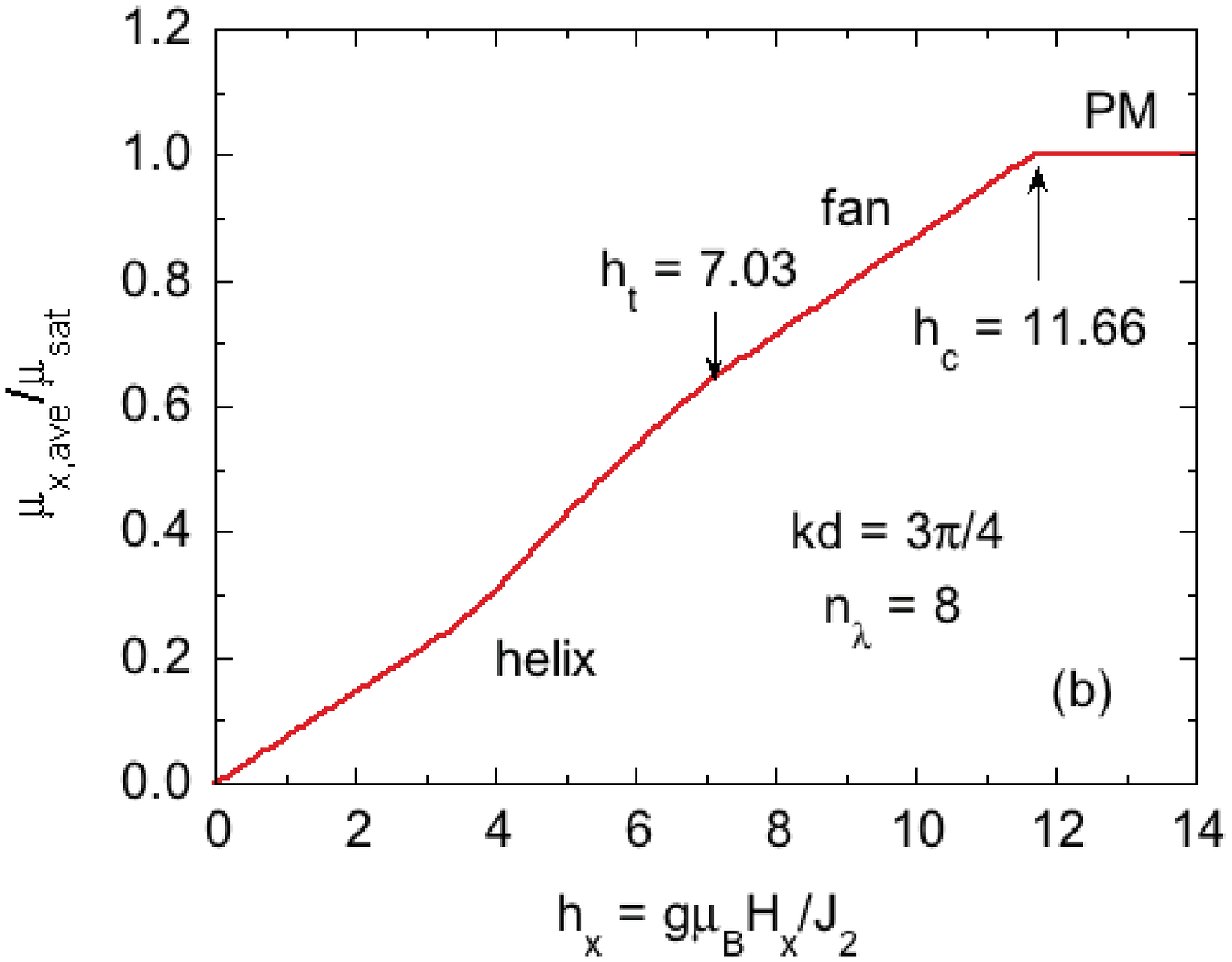}\includegraphics [width=3.5in]{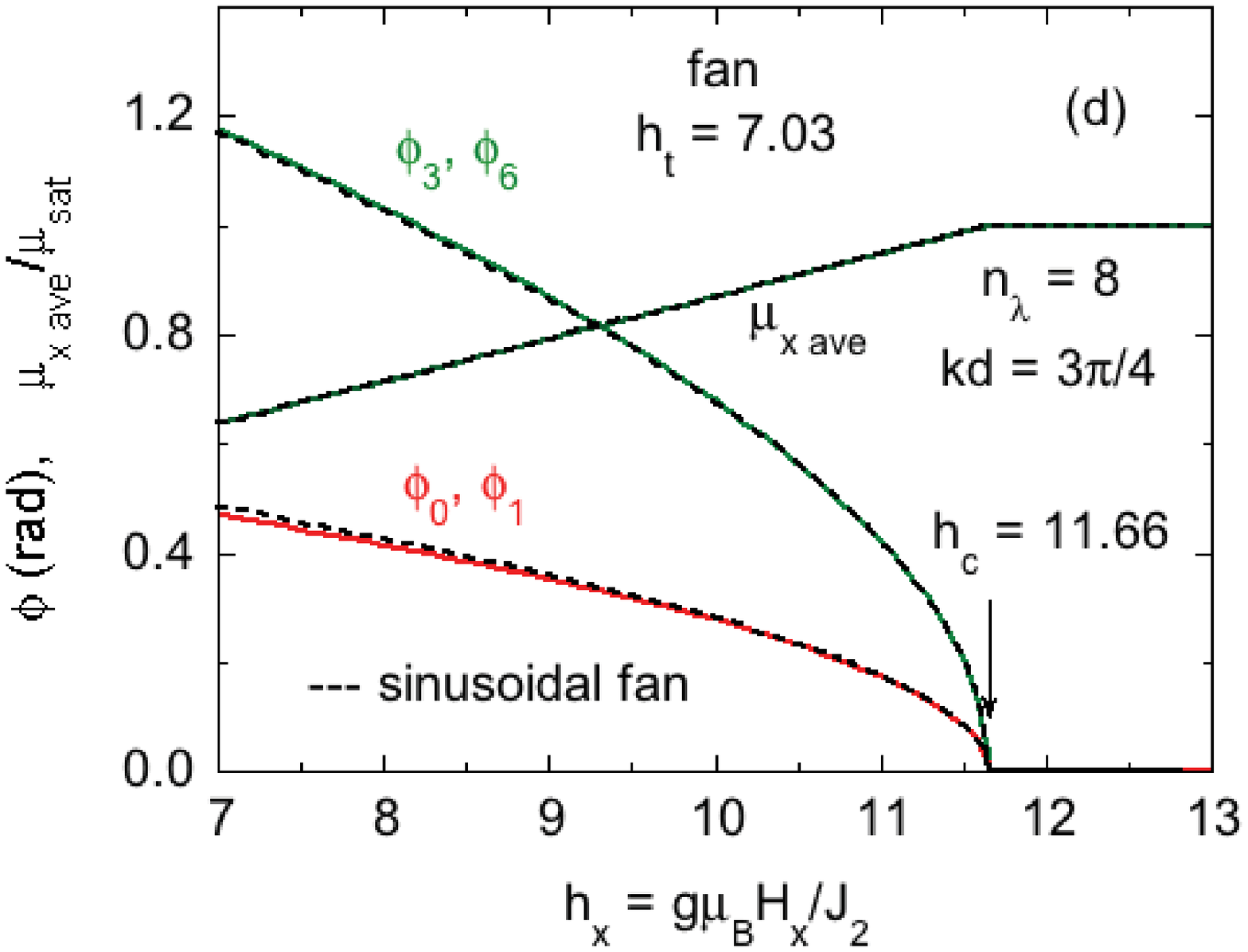}
\caption {(Color online) (a) The angles $\phi_0$, $\phi_3$, $\phi_6$, and~$\phi_1$ of the respective moments with respect to the $+x$ axis versus reduced in-plane field $h_x$ for a helix with turn angle $kd=3\pi/4$.  (b)~Average normalized magnetic moment per spin in the field direction, $\mu_{x{\rm ave}}/\mu_{\rm sat}$, versus~$h_x$.  (c) Derivative of the average moment in~(b) with respect to~$h_x$. (d)~Expanded plots of the $\phi_n$ and $\mu_{x{\rm ave}}/\mu_{\rm sat}$ versus~$h_x$ for the fan phase in~(a) and~(b), including corresponding data for the sinusoidal fan (dashed lines).}
\label{Fig:PhiMuxEnN8_kd3PiOn4}
\end{figure*}

\begin{figure}
\includegraphics [width=3.3in]{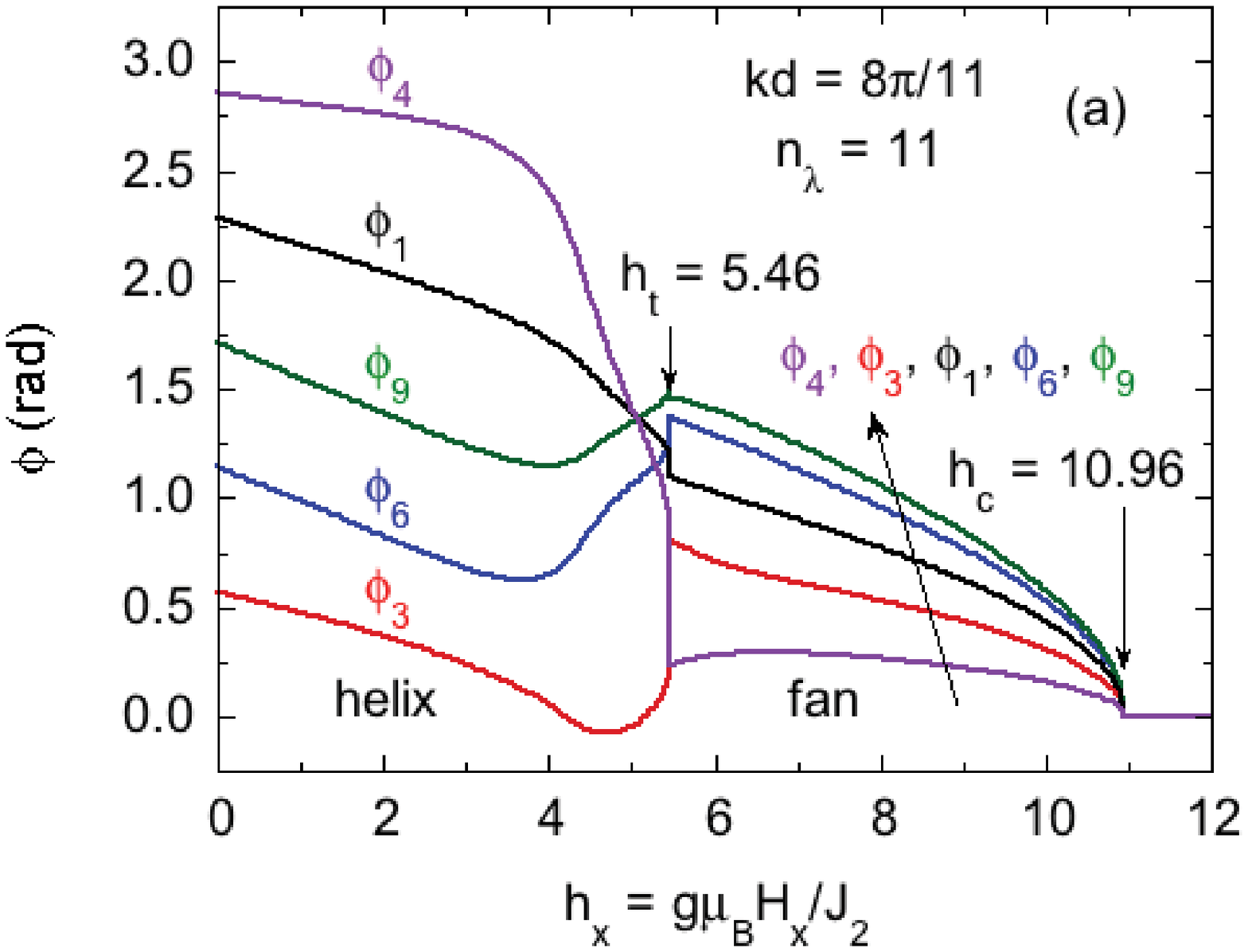}
\includegraphics [width=3.3in]{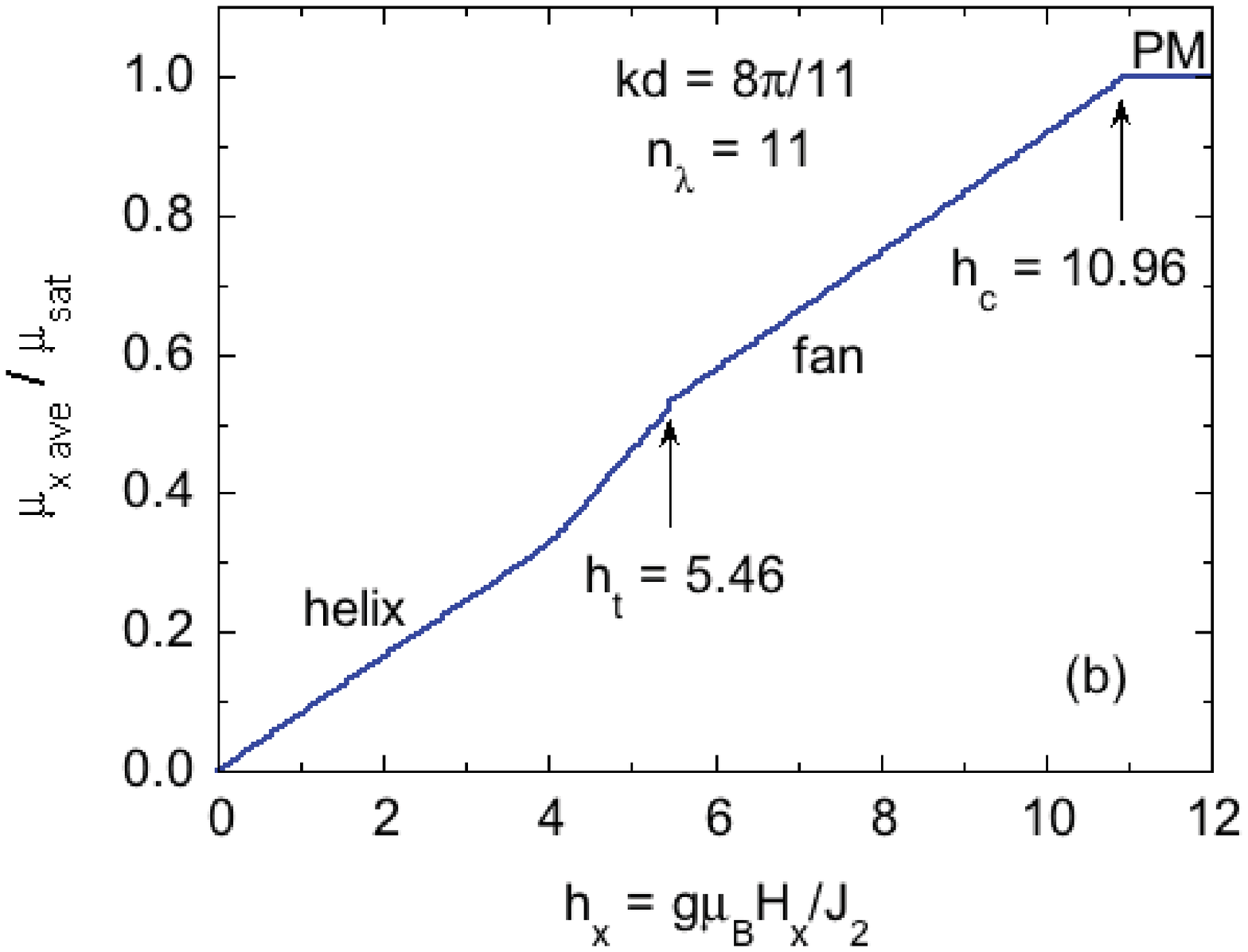}
\includegraphics [width=3.3in]{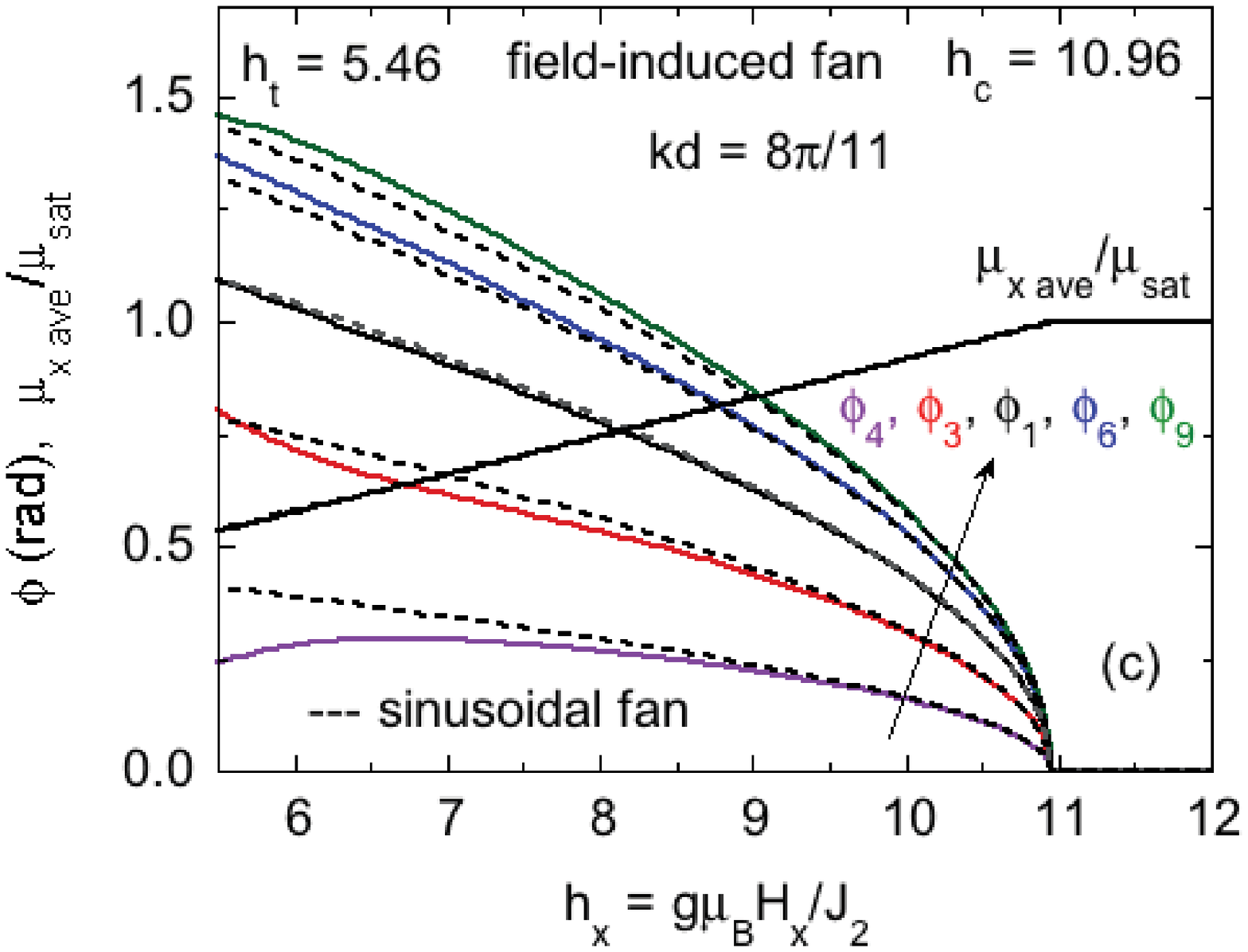}
\caption {(Color online) (a) The angles $\phi_3$, $\phi_6$, $\phi_9$, $\phi_1$, and~$\phi_4$ of the corresponding moments with respect to the $+x$ axis versus reduced in-plane field $h_x$ for a helix with turn angle $kd=8\pi/11$.  (b)~Average normalized magnetic moment per spin in the field direction, $\mu_{x{\rm ave}}/\mu_{\rm sat}$, versus~$h_x$, showing a weak first-order transition at field~$h_{\rm t}$. (c)~Expanded plots of the $\phi_n$ and $\mu_{x{\rm ave}}/\mu_{\rm sat}$ in the field-induced fan range (solid lines).  The predictions for a sinusoidal fan with the same~$kd$ and~$J_{12}$ are shown as black dashed lines.}
\label{Fig:PhiMuxEnN11kd8PiOn11Helix}
\end{figure}

\begin{figure}[h]
\includegraphics [width=3.3in]{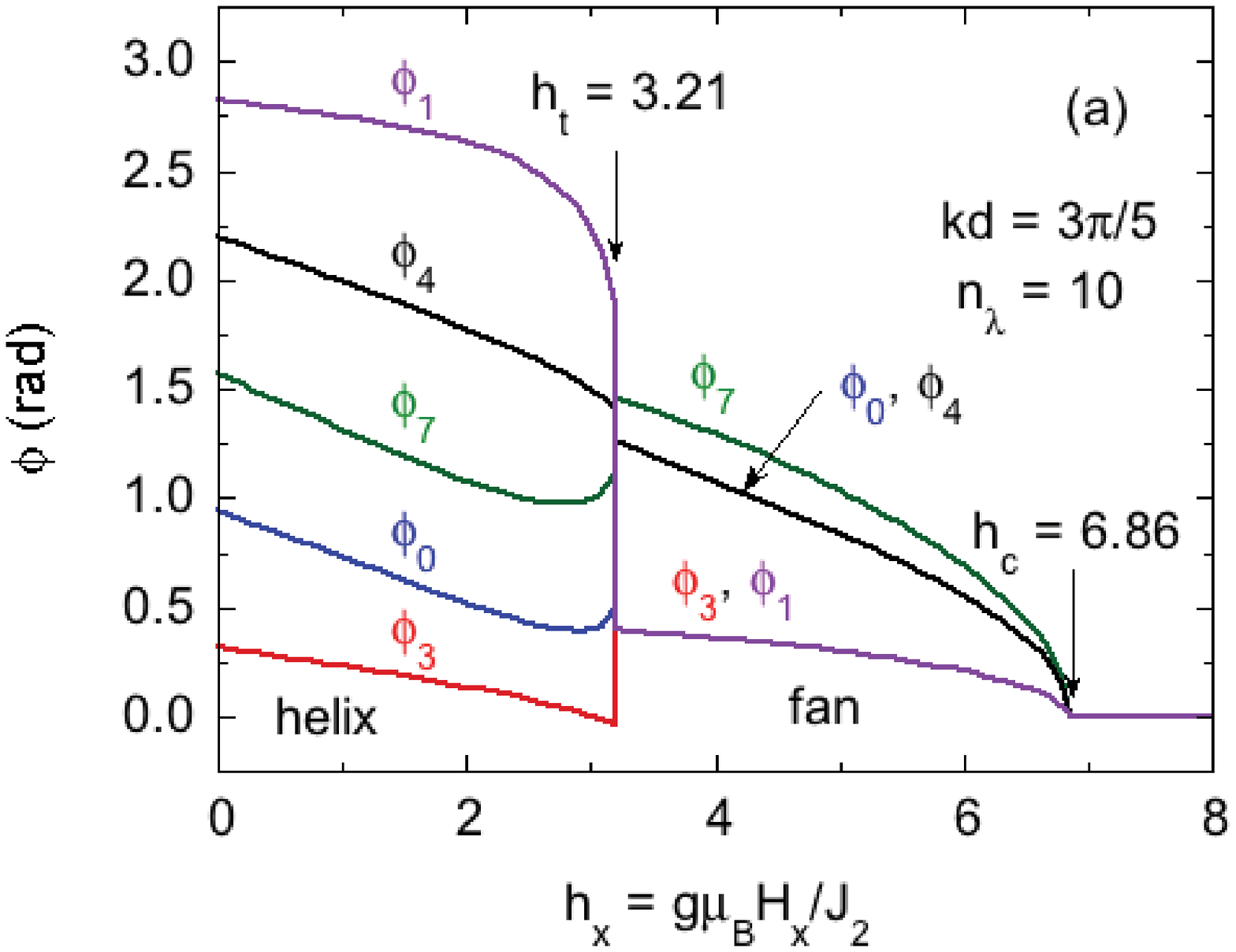}
\includegraphics [width=3.3in]{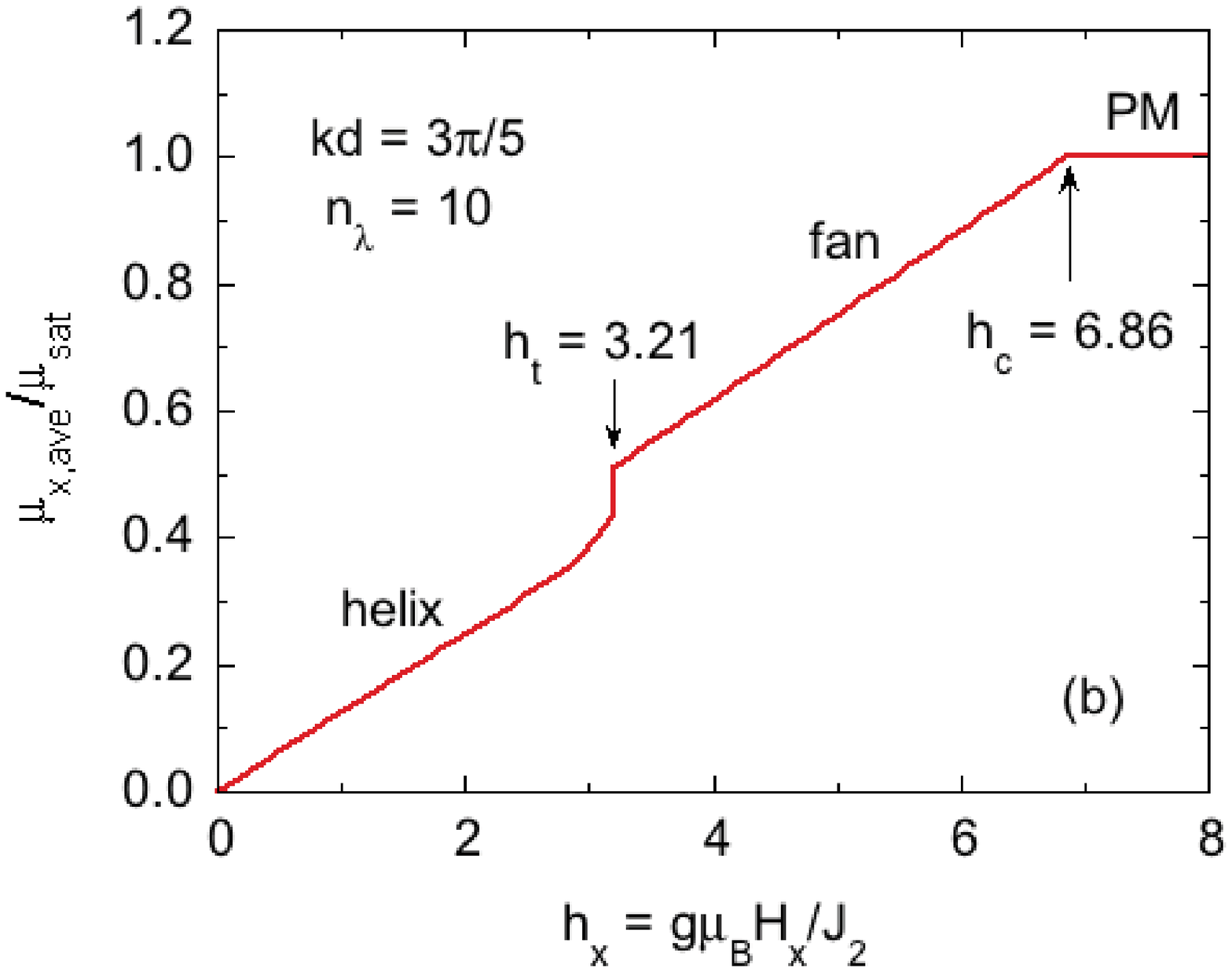}
\includegraphics [width=3.3in]{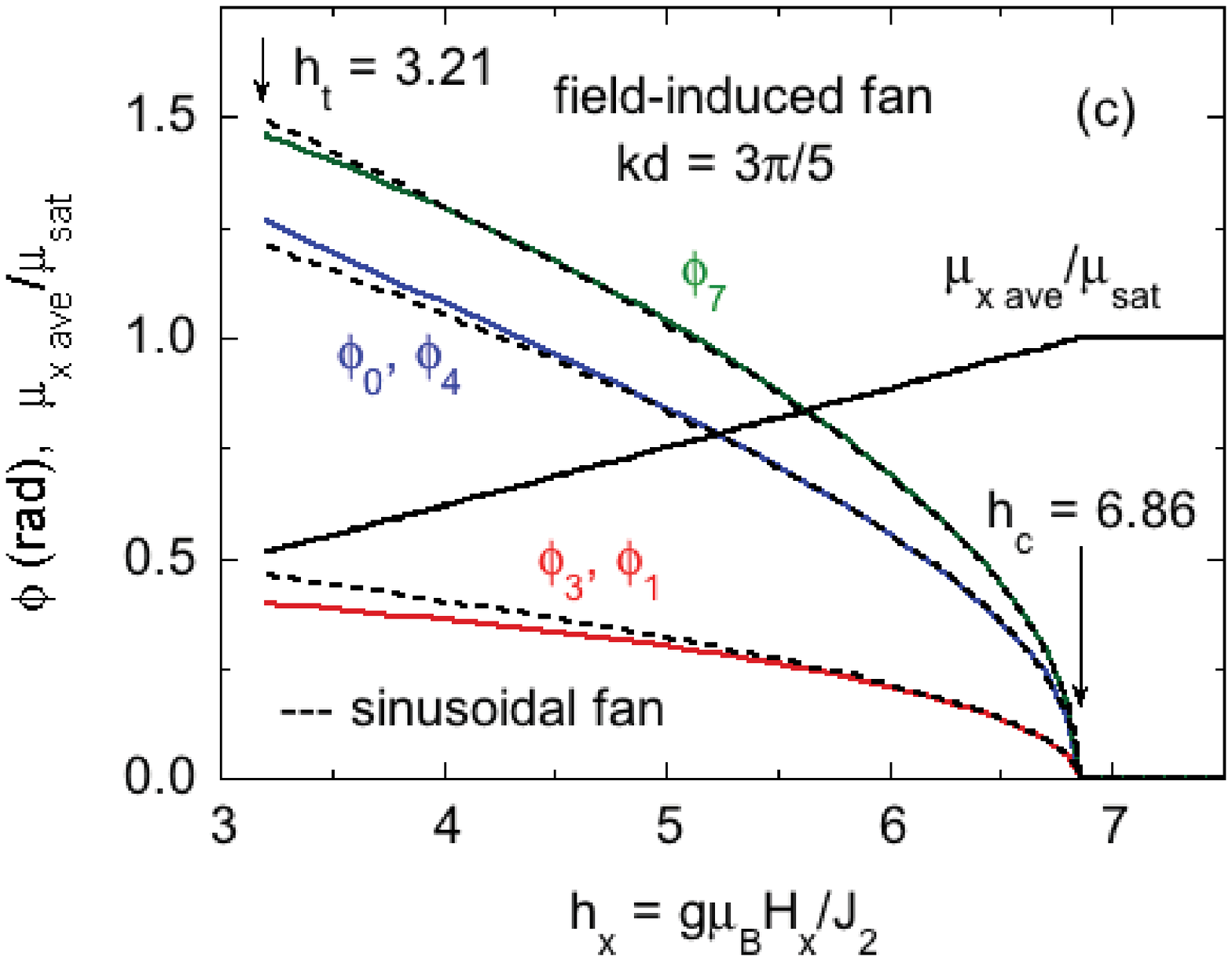}
\caption {(Color online) (a) The angles $\phi_3$, $\phi_0$, $\phi_7$, $\phi_4$, and~$\phi_1$ of the respective moments with respect to the $+x$ axis versus reduced in-plane field~$h_x$ for a helix with turn angle $kd=3\pi/5$.  (b)~Average normalized magnetic moment per spin in the field direction, $\mu_{x{\rm ave}}/\mu_{\rm sat}$, versus~$h_x$.  (c)~Expanded plots of the $\phi_n$ and $\mu_{x{\rm ave}}/\mu_{\rm sat}$ in the field-induced fan range (solid lines).  The predictions for a sinusoidal fan with the same~$kd$ and~$J_{12}$ are shown as black dashed lines.}
\label{Fig:PhiMuxEnN10kd3PiOn5Helix}
\end{figure}

\begin{figure}
\includegraphics [width=3.3in]{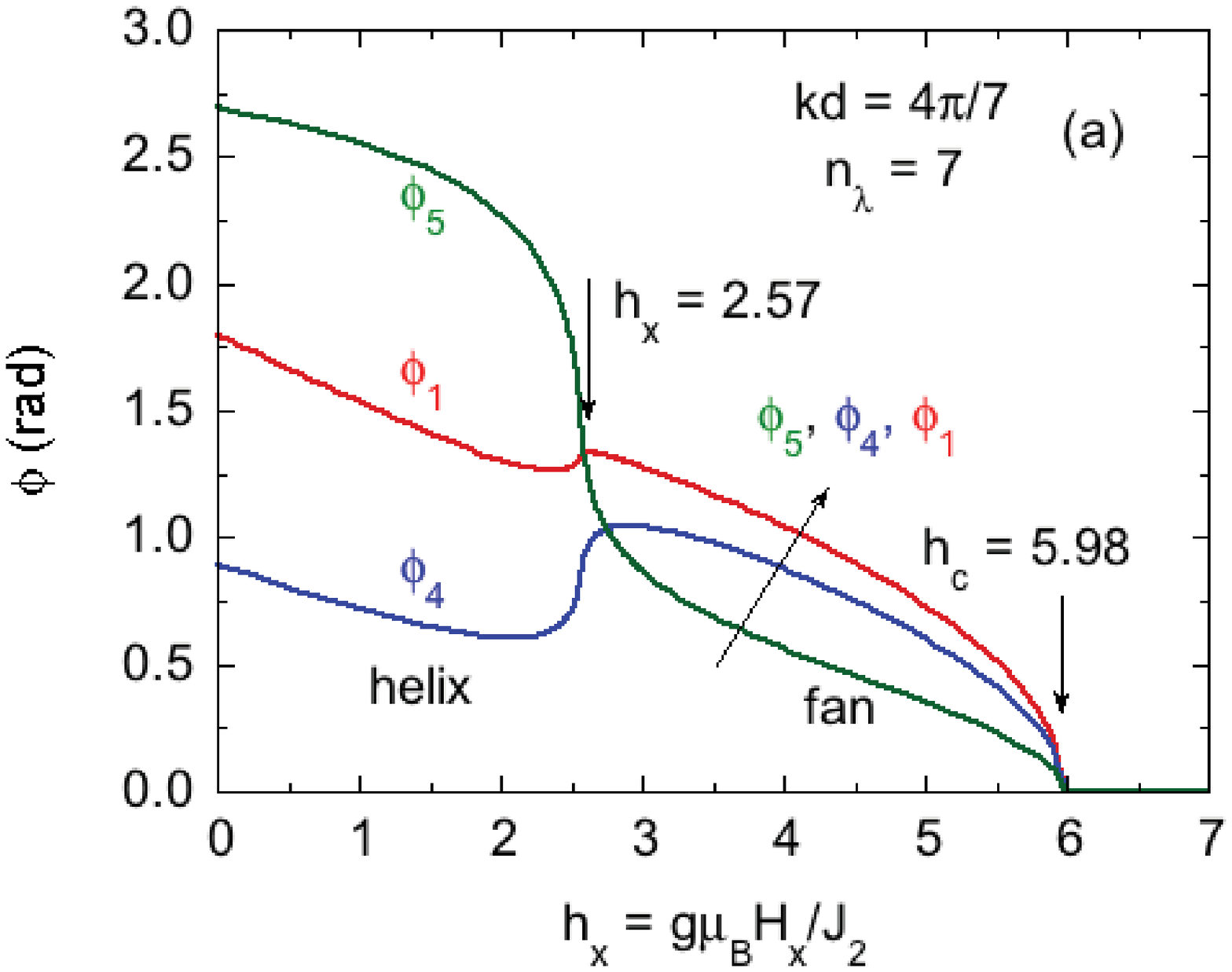}
\includegraphics [width=3.3in]{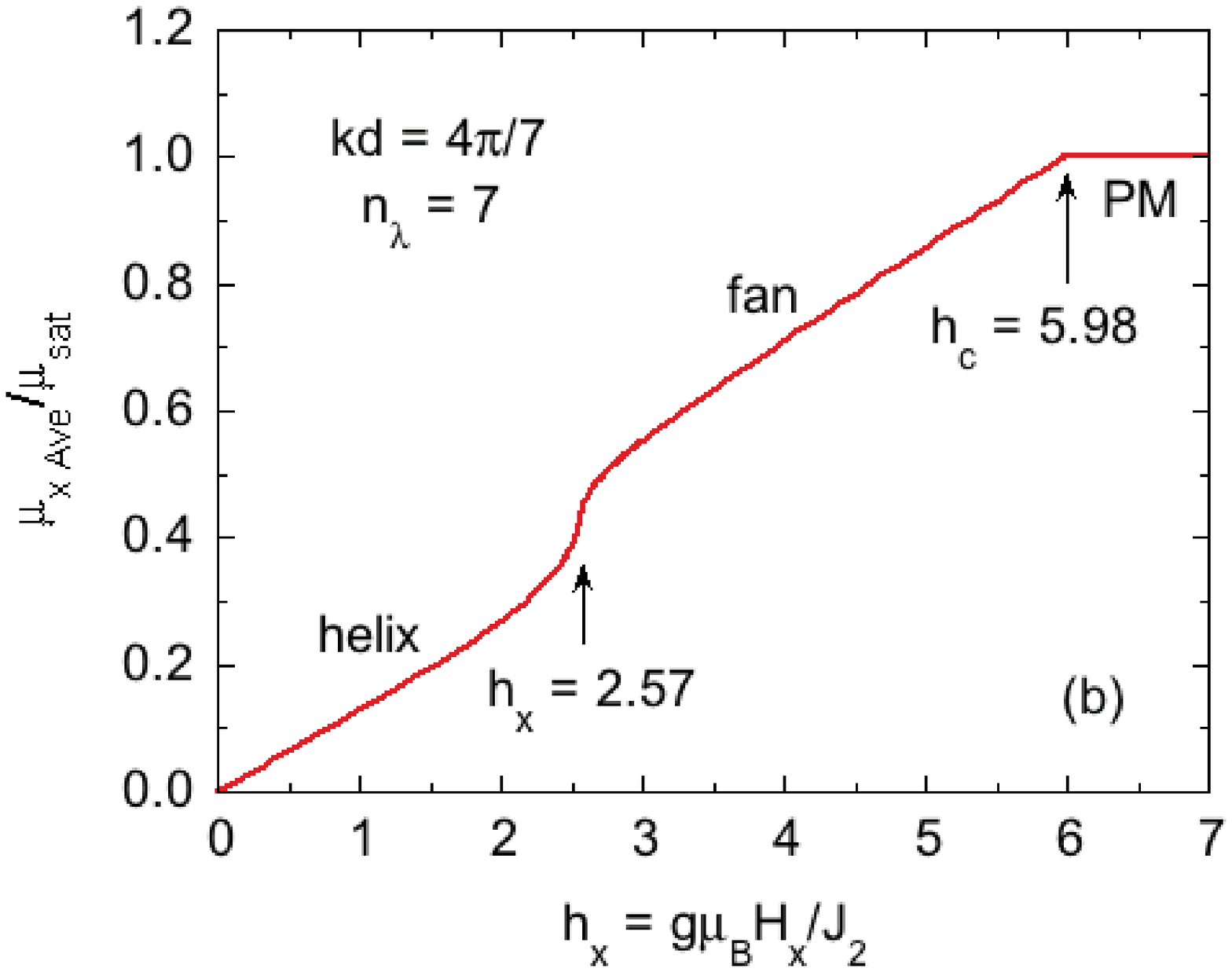}
\includegraphics [width=3.3in]{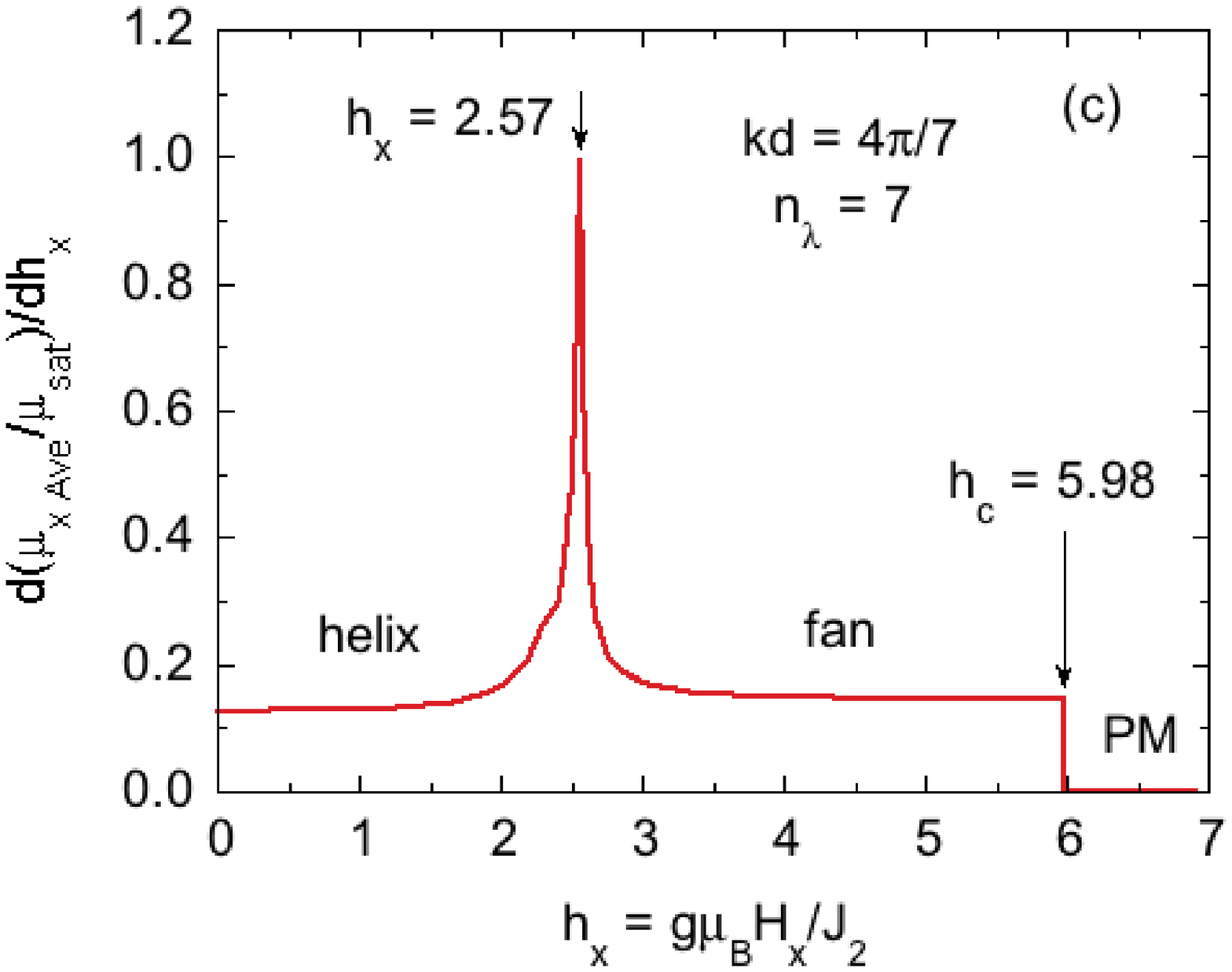}
\caption {(Color online) (a) The angles $\phi_4$, $\phi_1$, and~$\phi_5$ of the respective moments with respect to the $+x$ axis versus reduced in-plane field $h_x$ for a helix with turn angle $kd=4\pi/7$.  (b)~Average normalized magnetic moment per spin in the field direction, $\mu_{x{\rm ave}}/\mu_{\rm sat}$, versus~$h_x$.  (c) Derivative of the average moment in~(b) with respect to $h_x$, which shows a rapid smooth crossover between helix and fan phases at a field $h_{\rm X} = 2.57$ but no phase transition.}
\label{Fig:PhiMuxEnN7_kd4PiOn7}
\end{figure}

\begin{figure}
\includegraphics [width=3.3in]{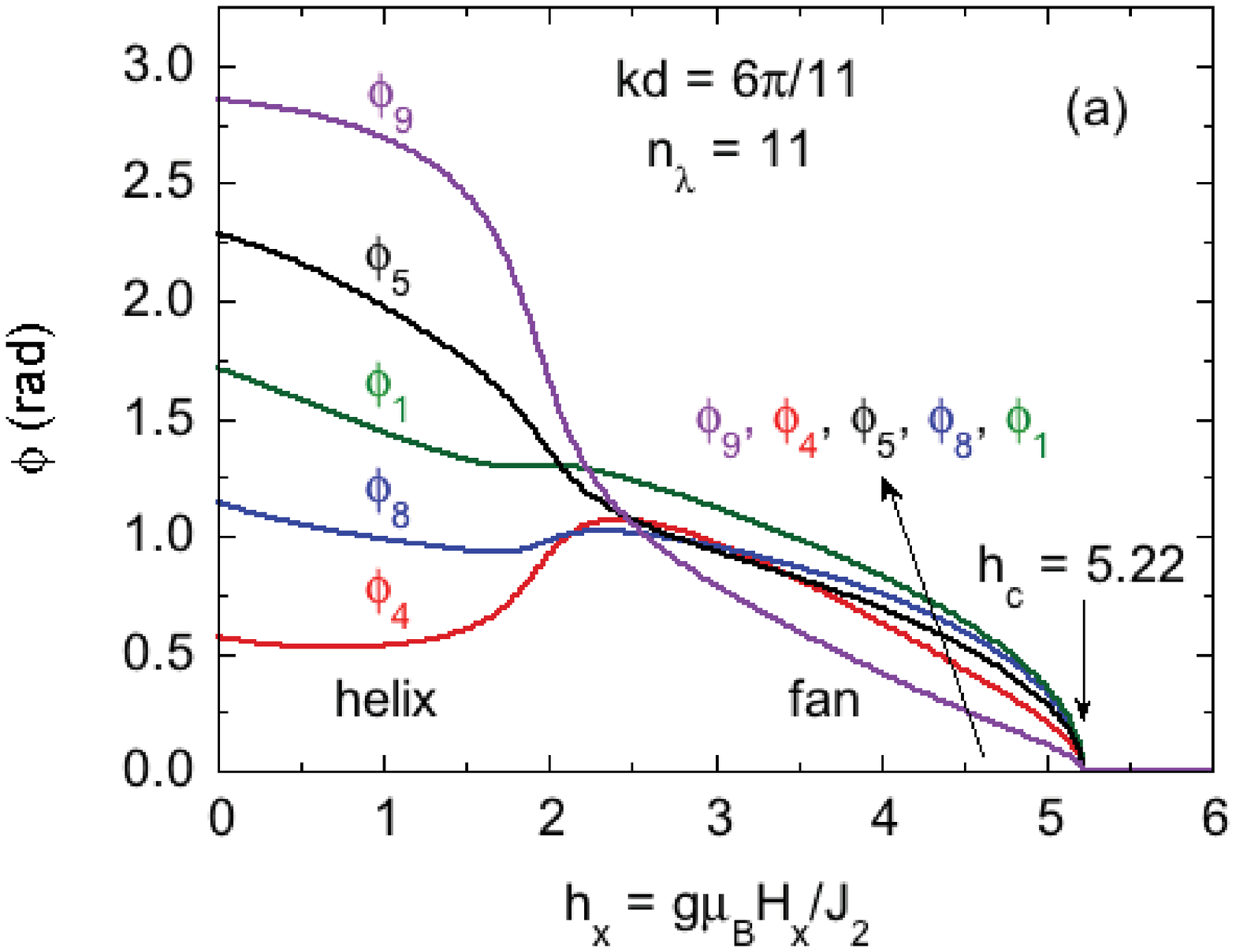}
\includegraphics [width=3.3in]{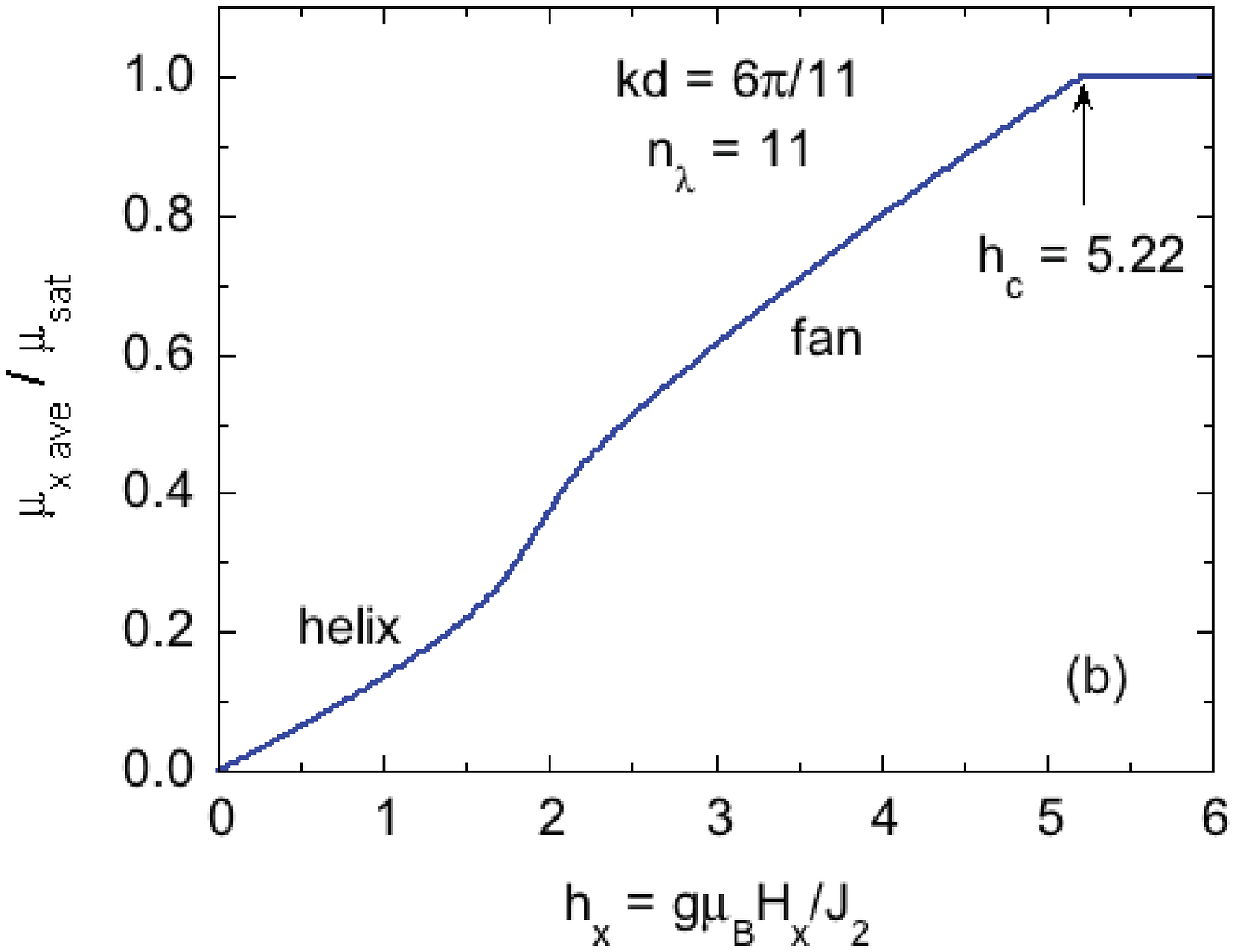}
\caption {(Color online) (a) The angles $\phi_4$, $\phi_8$, $\phi_1$, $\phi_5$, and $\phi_9$ of the corresponding moments with respect to the $+x$ axis versus reduced in-plane field $h_x$ for a helix with turn angle $kd=6\pi/11$.  (b)~Average normalized magnetic moment per spin in the field direction, $\mu_{x{\rm ave}}/\mu_{\rm sat}$, versus~$h_x$.}
\label{Fig:PhiMuxEnN11kd6PiOn11Helix}
\end{figure}

Figures~\ref{Fig:PhiMuxEnN11kd10PiOn11Helix}--\ref{Fig:PhiMuxEnN11kd6PiOn11Helix} show how the helix angles $\phi_n$ and the reduced average moment~$\mu_{x{\rm ave}}/\mu_{\rm sat}$ per spin versus reduced field~$h_x$ change as $kd$ is reduced from $10\pi/11$ to $6\pi/11$.  The dominant behavior is a smooth crossover in the ordering of the $\phi_n$ angles from their initial helical values to a distribution approximating a sinusoidal fan for $h_x\to h_{\rm c}$.  This smooth evoluation in $\phi_n(h_x)$ is accompanied by a smooth variation in $\mu_{x{\rm ave}}(h_x)$ as shown, which is not proportional to~$h_x$ but rather shows an S-shaped modulation of varying strength depending on~$kd$ that is strongest for $kd=4\pi/7$ where a first-order transition almost occurs.  We have previously shown a fit of the prediction for $kd = 6\pi/7$ in Fig.~\ref{Fig:PhiMuxEnN7_kd6PiOn7}(b) to the measured magnetization data for \ecp\ in Fig.~\ref{Fig:EuCo2P2_Mab_data_fit}, which also shows an S-shaped behavior.  The fit is not perfect in the S-shaped region, but it illustrates that a helix to fan transition in real materials need not be first order as often assumed previously but can be a smooth crossover instead, as suggested in Ref.~\cite{Carazza1991}.  

Several $kd$ values in the range ${\pi/2 < kd \leq \pi}$ were found to show interesting variations in the properties different from smooth crossovers from helix to fan with increasing field.  The data for $kd=3\pi/4$ in Fig.~\ref{Fig:PhiMuxEnN8_kd3PiOn4} exhibit a second-order transition at reduced field $h_{\rm t} = 7.03$ from a helix to fan structure with increasing $h_x$.  The second-order nature of the transition is clearly established from the field dependences of the $\phi_n$ in Fig.~\ref{Fig:PhiMuxEnN8_kd3PiOn4}(a).  It is also apparent from the discontinuity in $d\mu_{x{\rm ave}}/dh_x$ at $h_{\rm t}$ as illustrated in Fig.~\ref{Fig:PhiMuxEnN8_kd3PiOn4}(c).  For this $kd$, the variations of the $\phi_n(h_x)$ in the fan field range follow rather closely the prediction for the respective sinusoidal fan, as shown in Fig.~\ref{Fig:PhiMuxEnN8_kd3PiOn4}(d).

On the other hand, the $\phi_n(h_x)$ data for $kd = 8\pi/11$ in Fig.~\ref{Fig:PhiMuxEnN11kd8PiOn11Helix}(a) exhibit discontinuities with field at $h_{\rm t} = 5.46$ indicative of a first-order transition.  The magnetization data in Fig.~\ref{Fig:PhiMuxEnN11kd8PiOn11Helix}(b) show a small discontinuity at $h_{\rm t}$, reflecting a weak first-order transition.  Expanded plots of the $\phi_n(h_x)$ in the fan region are shown in Fig.~\ref{Fig:PhiMuxEnN11kd8PiOn11Helix}(c).  Except for the region $h_x\to h_{\rm c}$, the data are not well described by sinusoidal fan angles as shown by the dashed black curves.  A stronger first-order transition is found in the magnetization versus field for $kd=3\pi/5$ in Fig.~\ref{Fig:PhiMuxEnN10kd3PiOn5Helix}(b), where again the expanded plots of the $\phi_n(h_x)$ data in Fig.~\ref{Fig:PhiMuxEnN10kd3PiOn5Helix}(c) are not well described by the sinusoidal fan model except for $h_x\to h_{\rm c}$.

Finally, Fig.~\ref{Fig:PhiMuxEnN7_kd4PiOn7} for $kd = 4\pi/7$ demonstrates that the behaviors of $\phi_n$ and $\mu_{x{\rm ave}}$ with field do not vary monotonically with $kd$.  In particular, instead of first-order transitions found for the previous two $kd$ values, the $\phi_n$ values now vary smoothly with $h_x$ indicating a smooth but distinct crossover at a field $h_{\rm X} = 2.57$ between the helix and fan phases as shown in Fig.~\ref{Fig:PhiMuxEnN7_kd4PiOn7}(a).  This behavior is reflected in the data for $\mu_{x{\rm ave}}(h_x)$ in Figs.~\ref{Fig:PhiMuxEnN7_kd4PiOn7}(b) and~\ref{Fig:PhiMuxEnN7_kd4PiOn7}(c).  These data show that $\mu_{x{\rm ave}}(h_x)$ almost undergoes a first-order transition at~$h_{\rm X}$.  On the other hand, the next data set for $kd=6\pi/11$ in Fig.~\ref{Fig:PhiMuxEnN11kd6PiOn11Helix} again show smooth crossover behaviors more characteristic of the data for $kd = 10\pi/11$ to~$4\pi/5$ in Figs.~\ref{Fig:PhiMuxEnN11kd10PiOn11Helix}--\ref{Fig:PhiMuxEnN5_kd4PiOn5}.

\clearpage

\clearpage

\subsection{${0 < kd < \pi/2}$: FM $J_{1}<0$}

\begin{figure}
\includegraphics [width=3.3in]{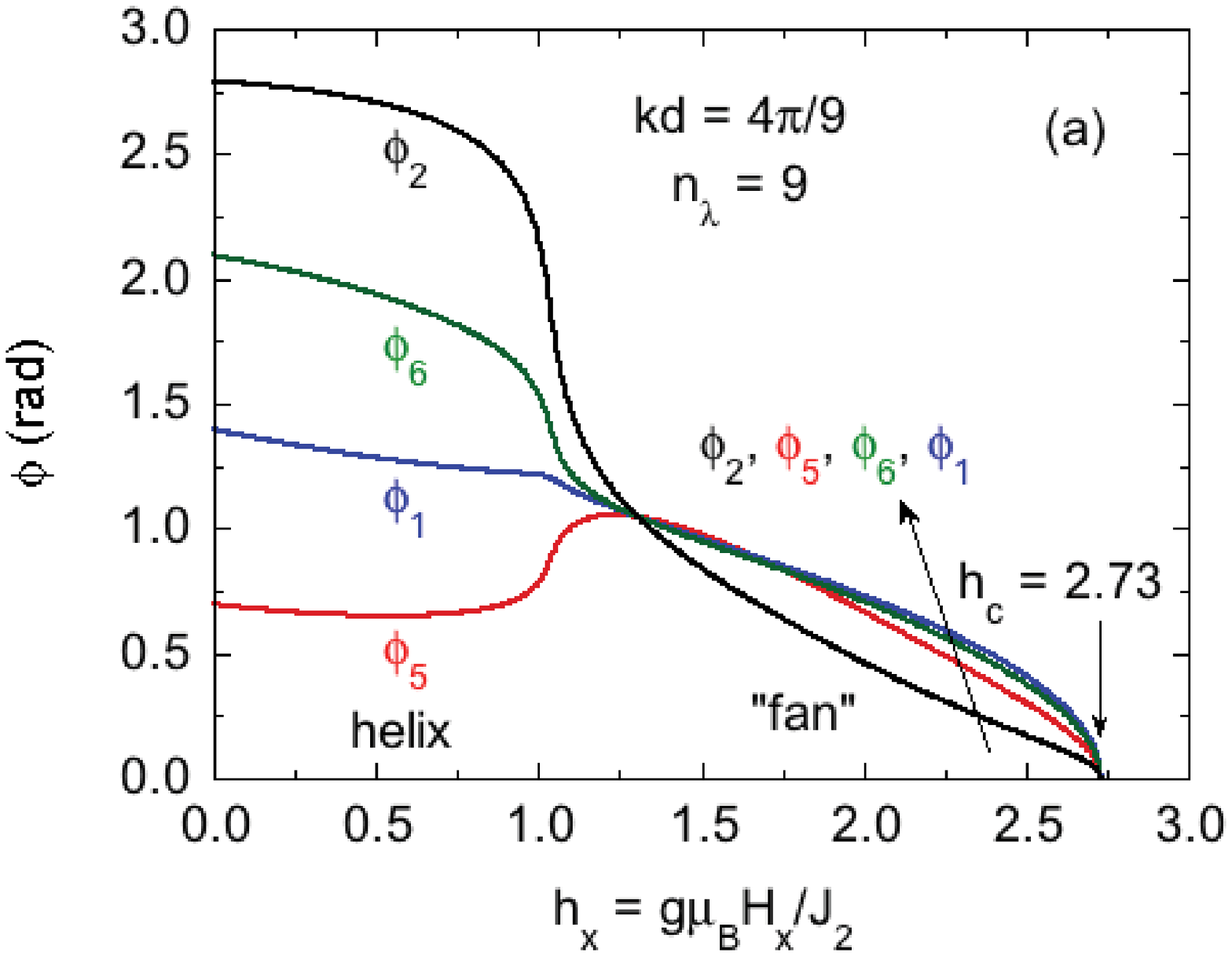}
\includegraphics [width=3.4in]{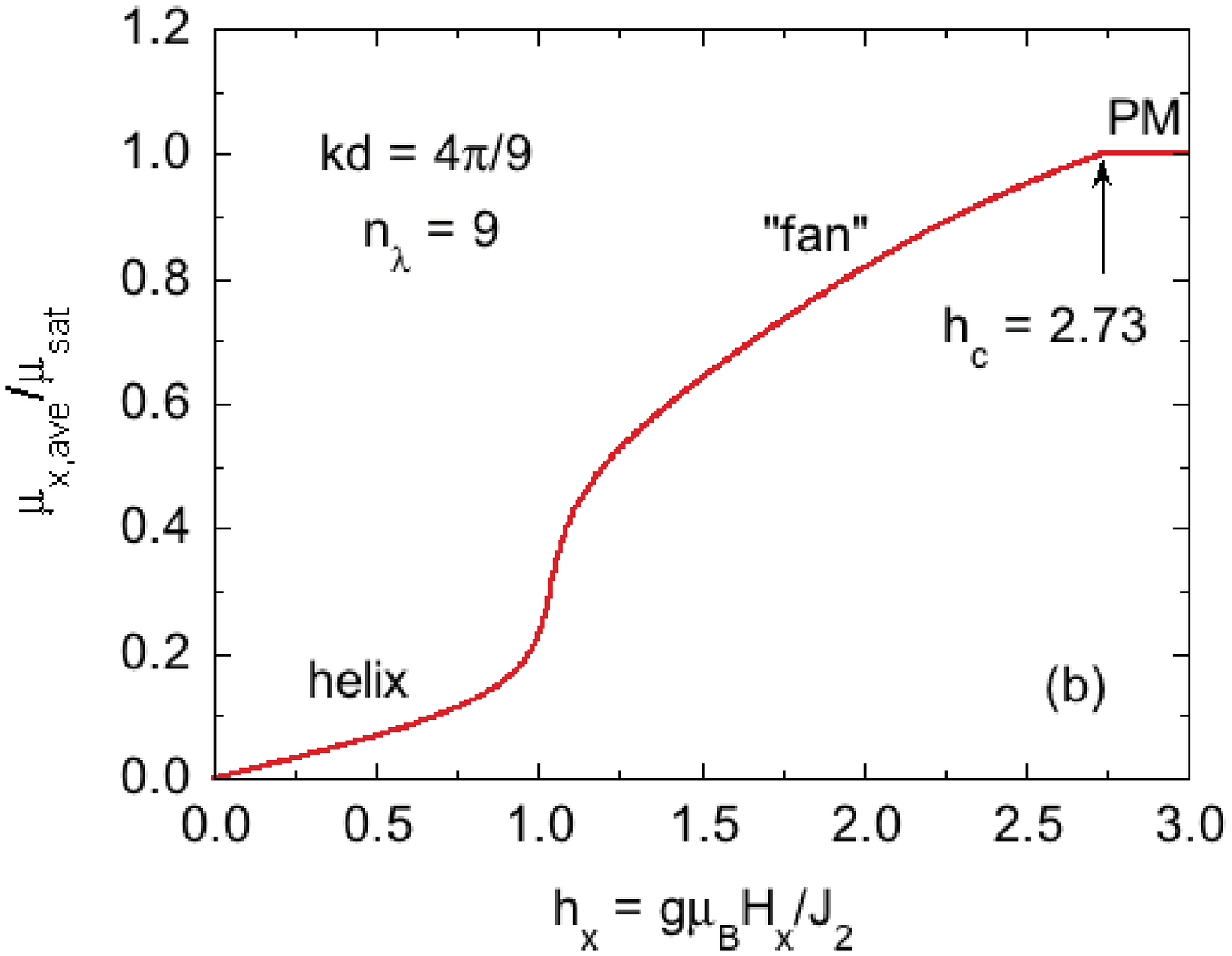}
\caption {(Color online) (a) The angles $\phi_5$, $\phi_1$, $\phi_6$, and $\phi_2$ of the corresponding moments with respect to the $+x$ axis versus reduced in-plane field $h_x$ for a helix with turn angle $kd=4\pi/9$.  (b)~Average normalized magnetic moment per spin in the field direction, $\mu_{x{\rm ave}}/\mu_{\rm sat}$, versus~$h_x$.}
\label{Fig:PhiMuxEnN9kd4PiOn9Helix}
\end{figure}

When moving into the regime of FM (negative) values of~$J_{1}$, with decreasing~$kd$ we again find a smooth crossover between helix and fan phases as revealed for $kd=4\pi/9$ in Fig.~\ref{Fig:PhiMuxEnN9kd4PiOn9Helix}.  However, this crossover results in a stronger S-shape to the $\mu_{x{\rm ave}}(h_x)$ data than found above in the region $\pi/2 < kd \leq \pi$, as shown in Fig.~\ref{Fig:PhiMuxEnN9kd4PiOn9Helix}(b).  It is clear from Fig.~\ref{Fig:PhiMuxEnN9kd4PiOn9Helix}(a) that the fan angles are not sinusoidal except for $h_x\to h_{\rm c}$.

\begin{figure}
\includegraphics [width=3.4in]{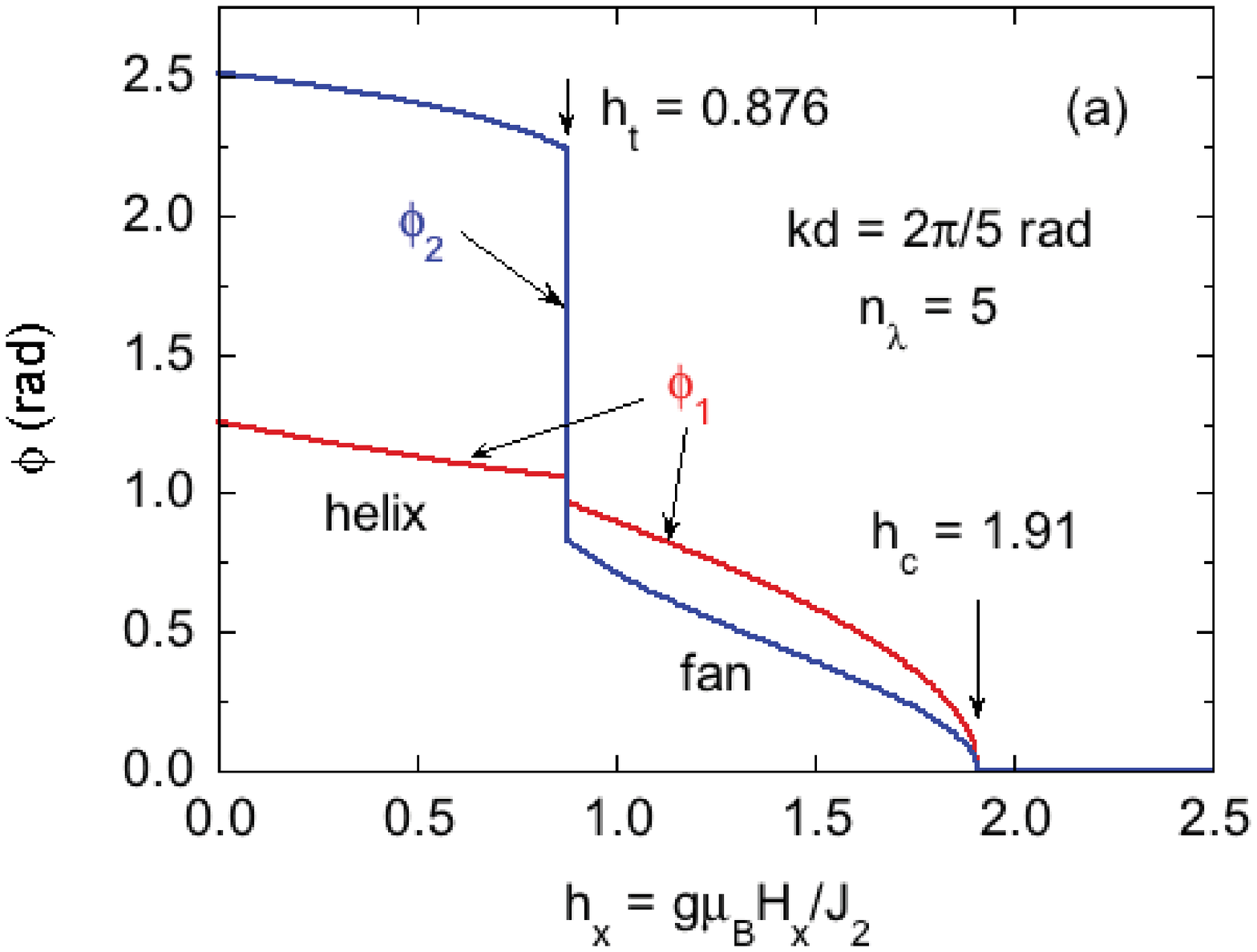}
\includegraphics [width=3.4in]{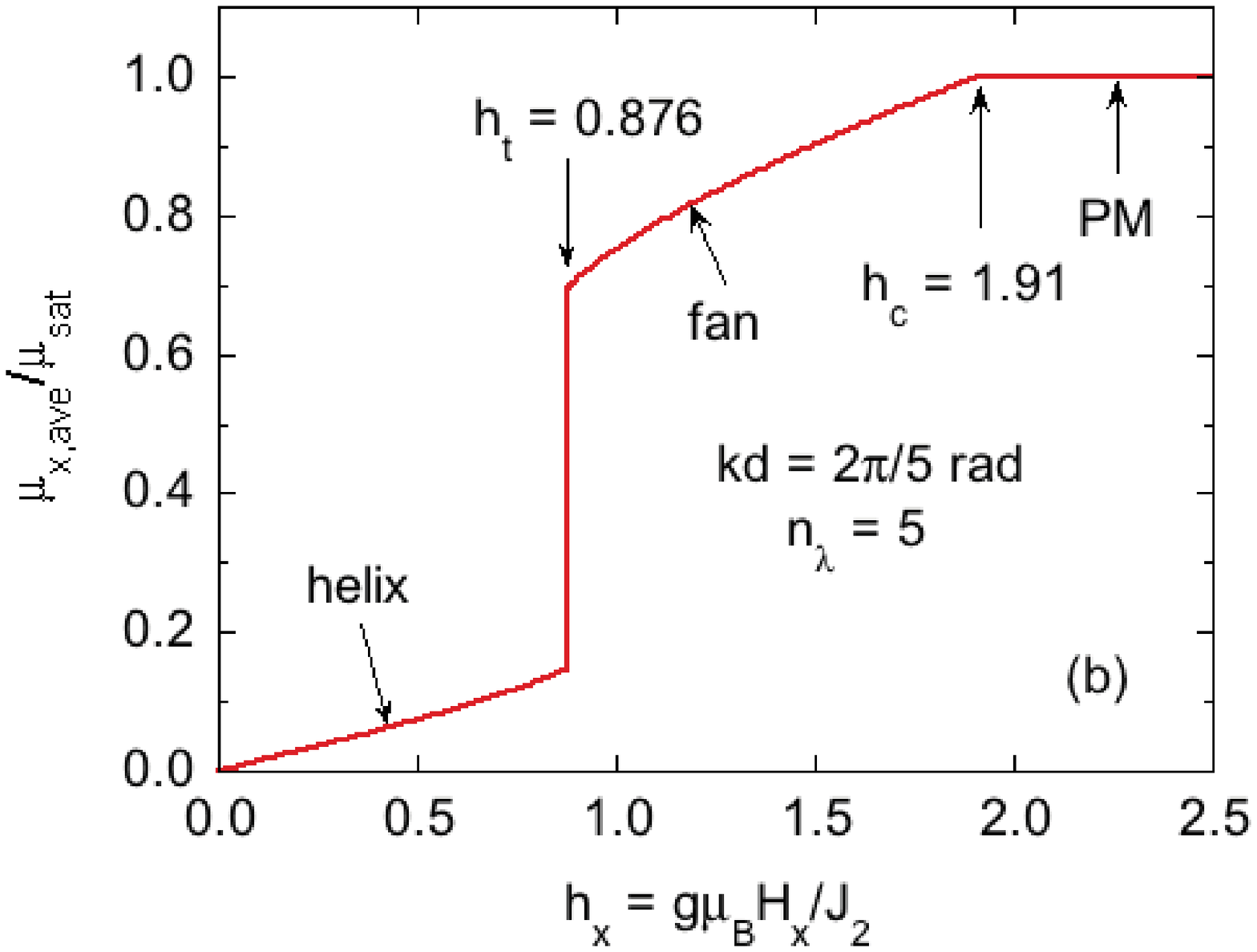}
\caption {(Color online) (a) The angles $\phi_1$ and $\phi_2$ of moments $\vec{\mu}_1$ and $\vec{\mu}_2$ with respect to the $+x$ axis versus reduced in-plane field $h_x$ for a helix with turn angle $kd=2\pi/5$.  (b)~Average normalized magnetic moment per spin in the field direction, $\mu_{x{\rm ave}}/\mu_{\rm sat}$, versus~$h_x$.}
\label{Fig:MH_T0_n5_Helix_Fan}
\end{figure}

\begin{figure}
\includegraphics [width=3.4in]{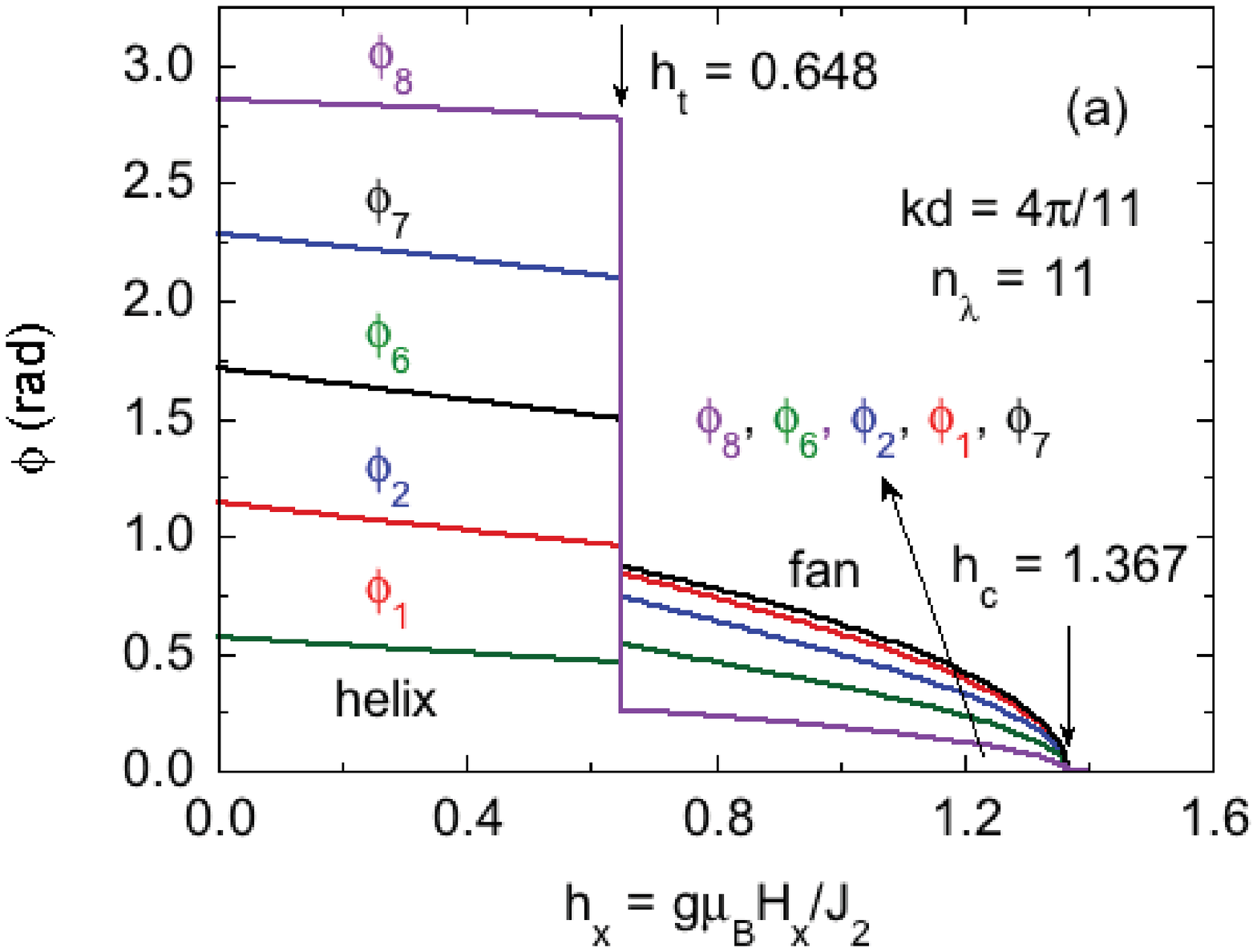}
\includegraphics [width=3.4in]{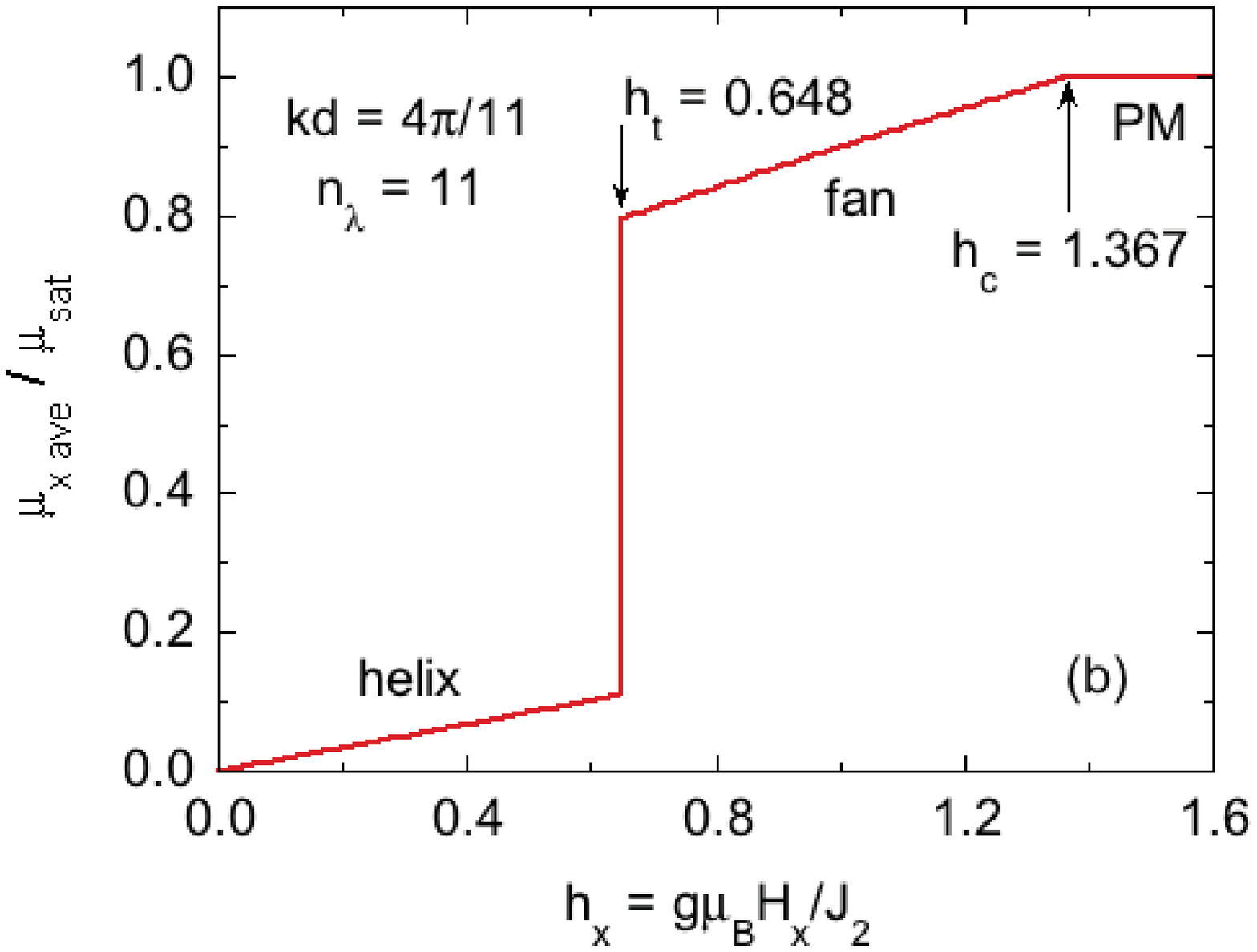}
\includegraphics [width=3.4in]{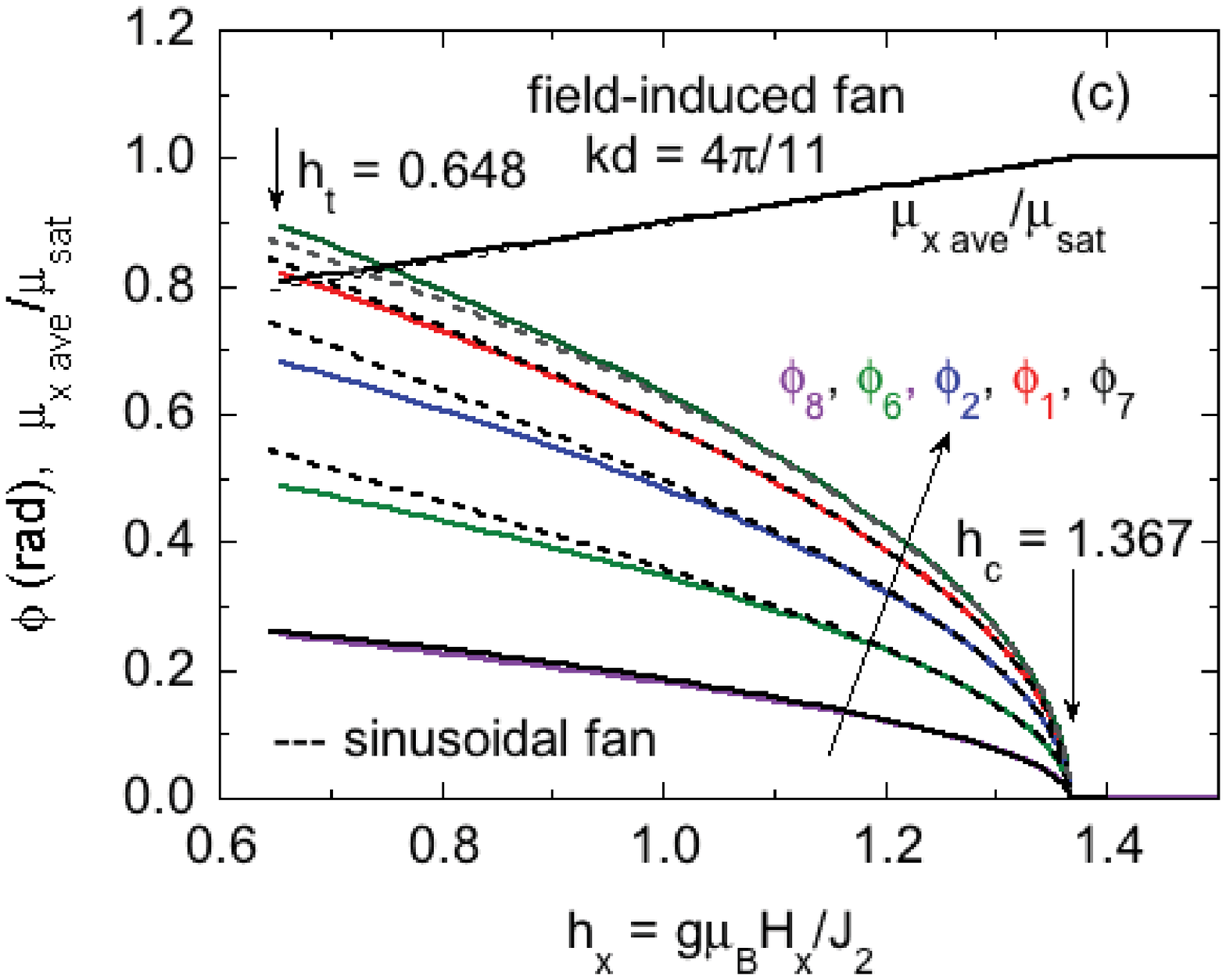}
\caption {(Color online) (a) The angles $\phi_1$, $\phi_2$, $\phi_6$, $\phi_7$, and $\phi_8$ of the respective moments with respect to the $+x$ axis versus reduced in-plane field $h_x$ for a helix with turn angle $kd=4\pi/11$.  (b)~Average normalized magnetic moment per spin in the field direction, $\mu_{x{\rm ave}}/\mu_{\rm sat}$, versus reduced field~$h_x$.  (c)~Expanded plots of the moment angles in~(a) for the field-induced fan angles versus~$h_x$ (solid lines), together with the predictions for the corresponding sinusoidal fan (dashed lines).}
\label{Fig:PhiMuxEnN11kd4PiOn11Helix}
\end{figure}

\begin{figure}
\includegraphics [width=3.3in]{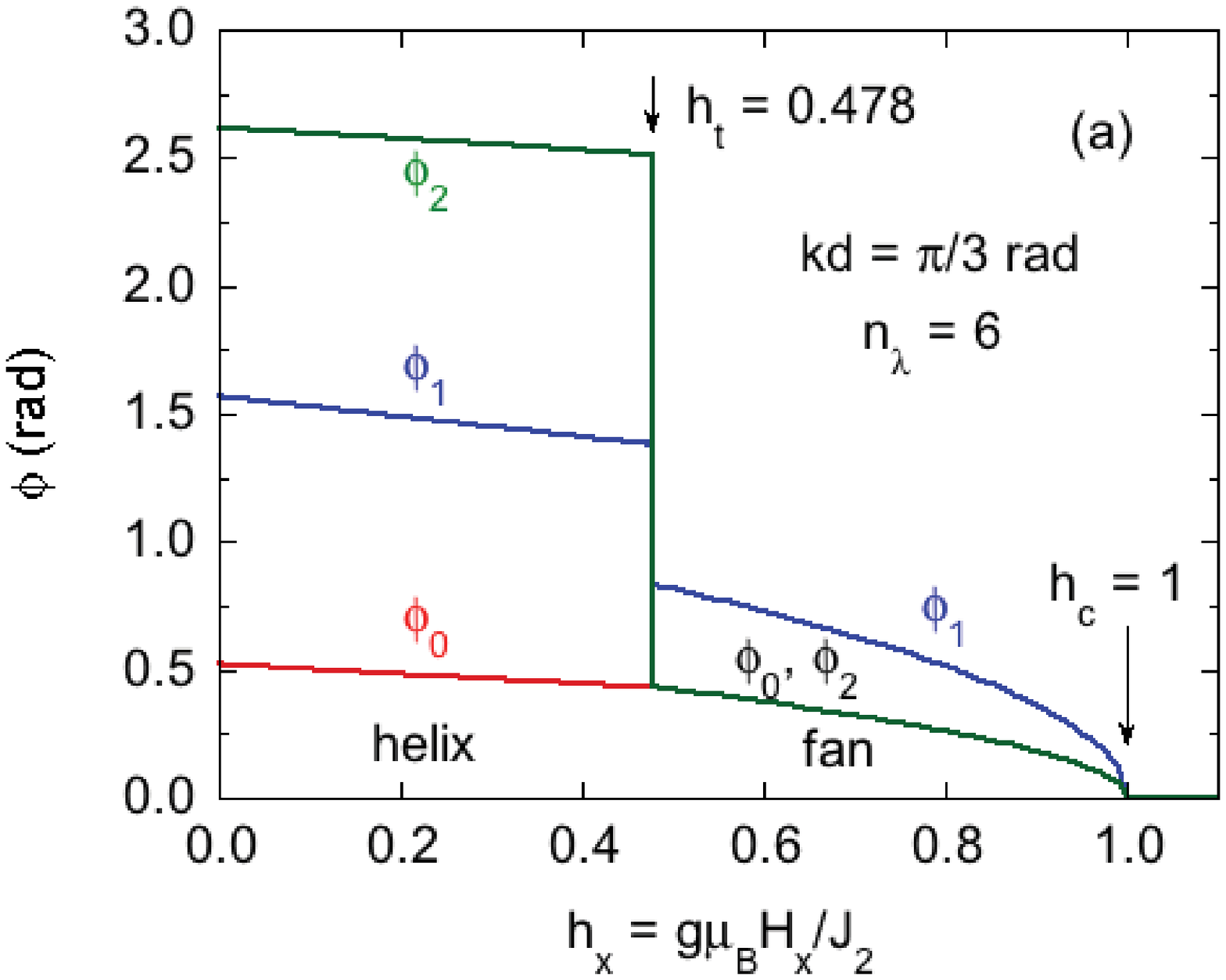}
\includegraphics [width=3.3in]{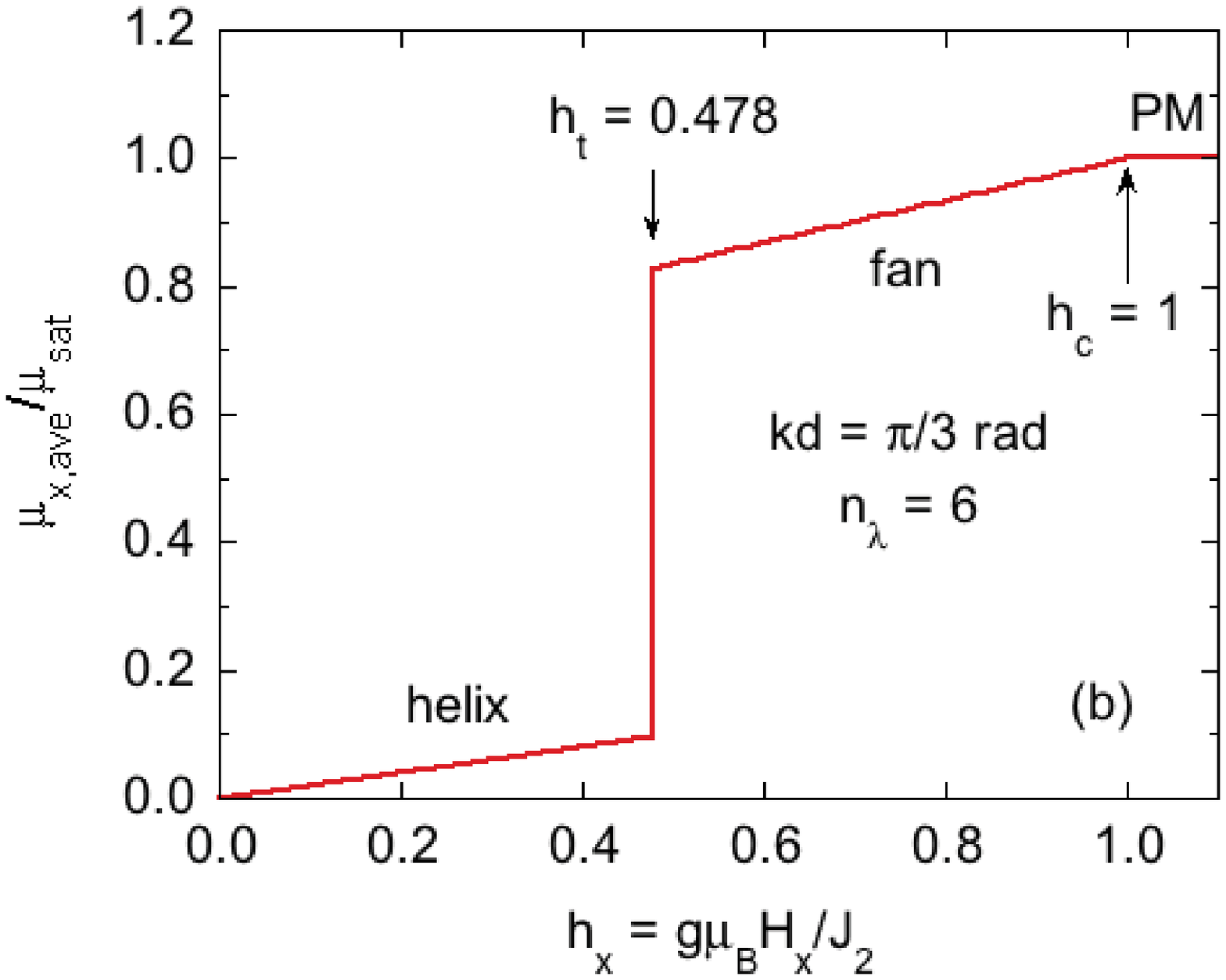}
\caption {(Color online) (a) The angles $\phi_0$, $\phi_1$, and $\phi_2$ of the corresponding moments with respect to the $+x$ axis versus reduced in-plane field $h_x$ for a helix with turn angle $kd=\pi/3$.  (b)~Average normalized magnetic moment per spin in the field direction, $\mu_{x{\rm ave}}/\mu_{\rm sat}$, versus reduced field $h_x$.}
\label{Fig:MH_T0_n6_Helix_Fan}
\end{figure}

\begin{figure}
\includegraphics [width=3.3in]{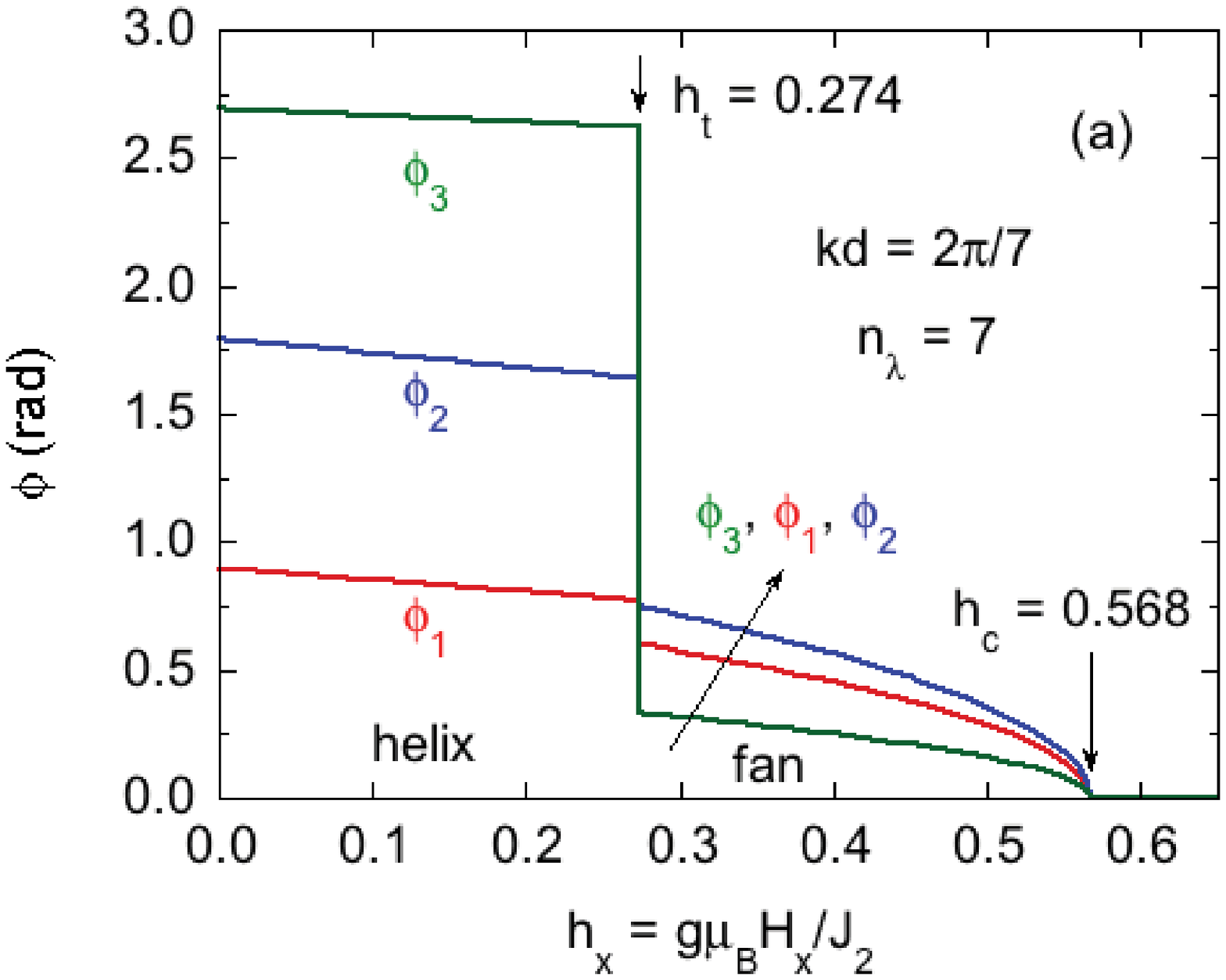}
\includegraphics [width=3.3in]{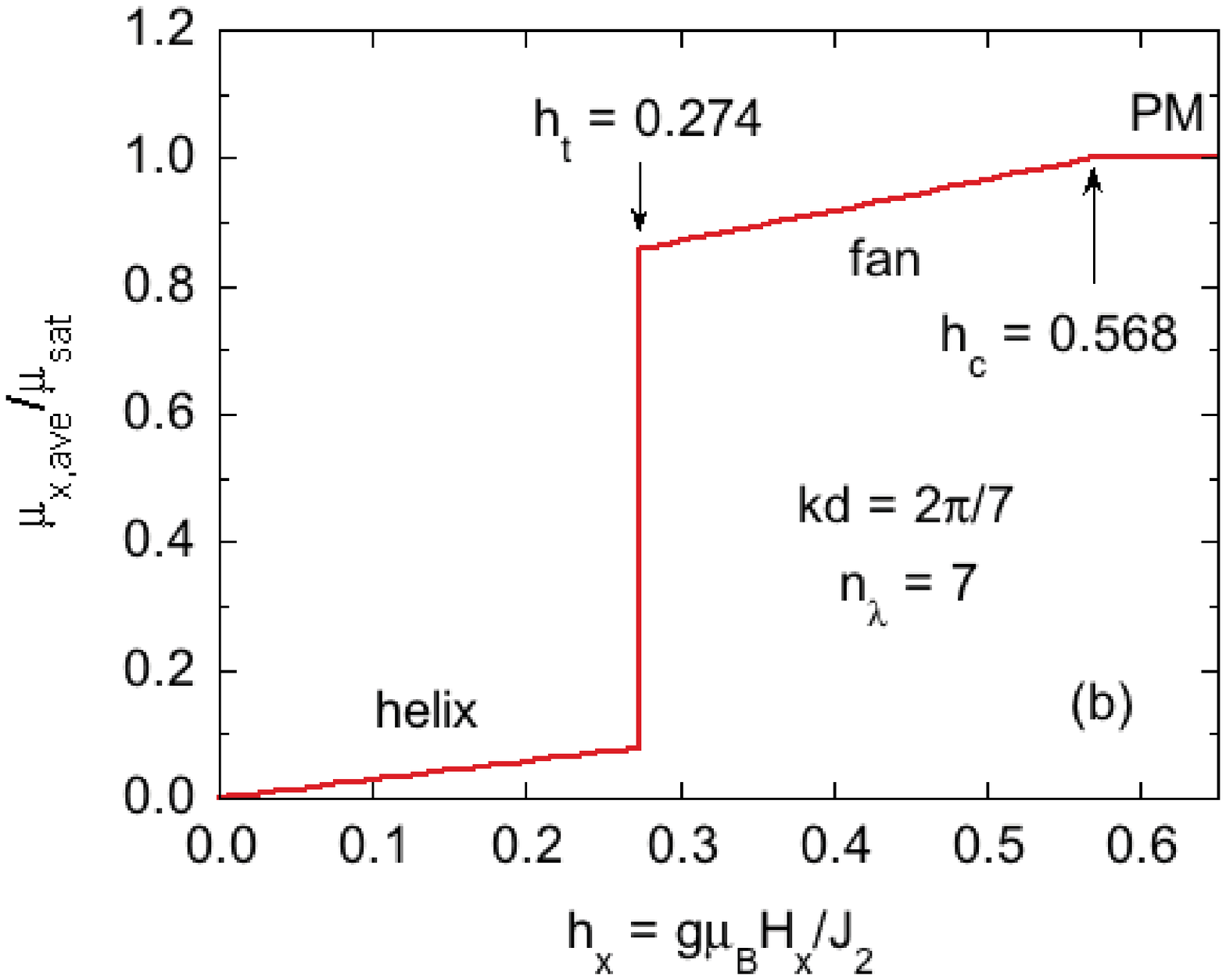}
\caption {(Color online) (a) The angles $\phi_1$, $\phi_2$, and $\phi_3$ of the corresponding moments with respect to the $+x$ axis versus reduced in-plane field $h_x$ for a helix with turn angle $kd=2\pi/7$.  (b)~Average normalized magnetic moment per spin in the field direction, $\mu_{x{\rm ave}}/\mu_{\rm sat}$, versus reduced field $h_x$.}
\label{Fig:MH_T0_n7_Helix_Fan}
\end{figure}

\begin{figure}
\includegraphics [width=3.3in]{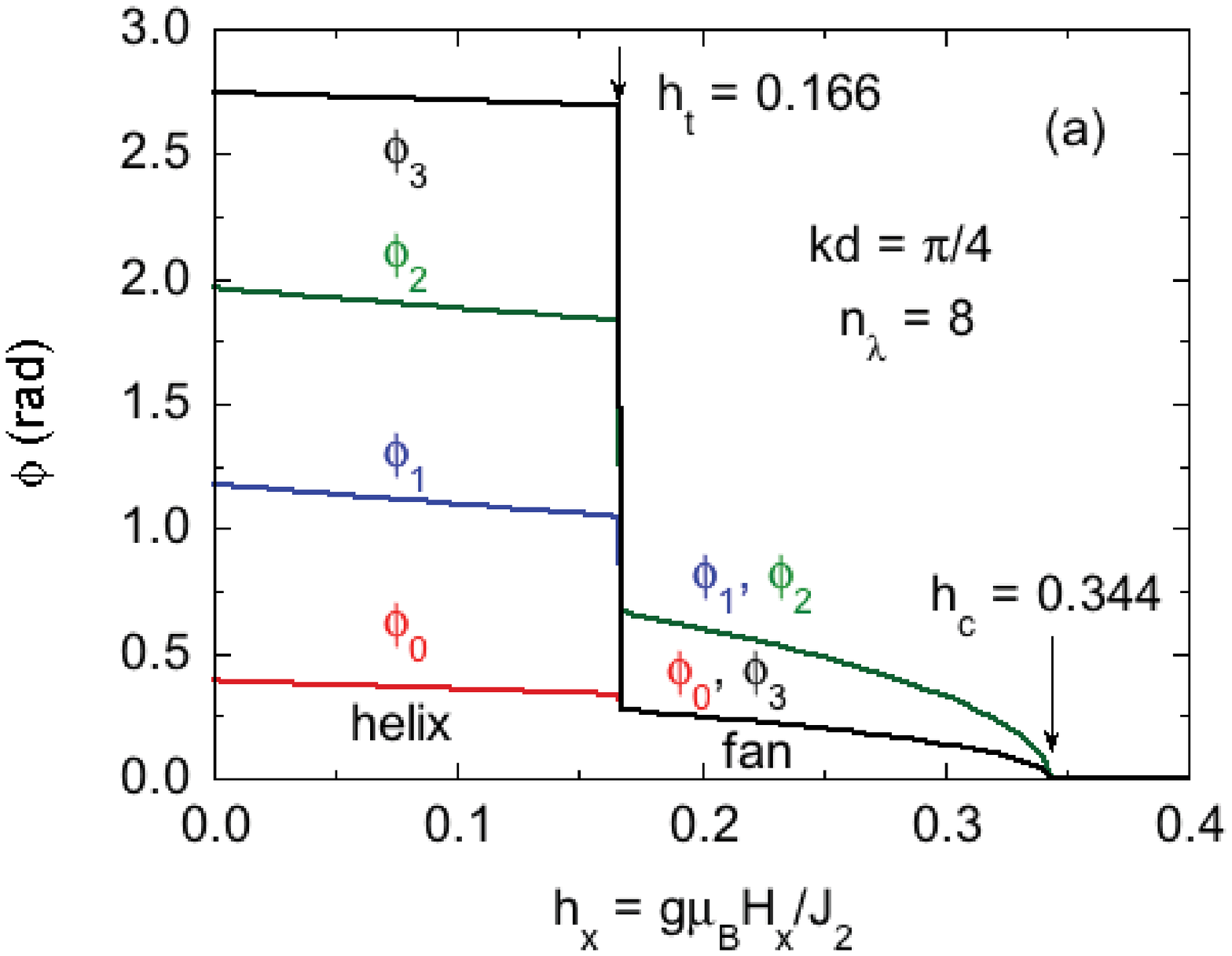}
\includegraphics [width=3.3in]{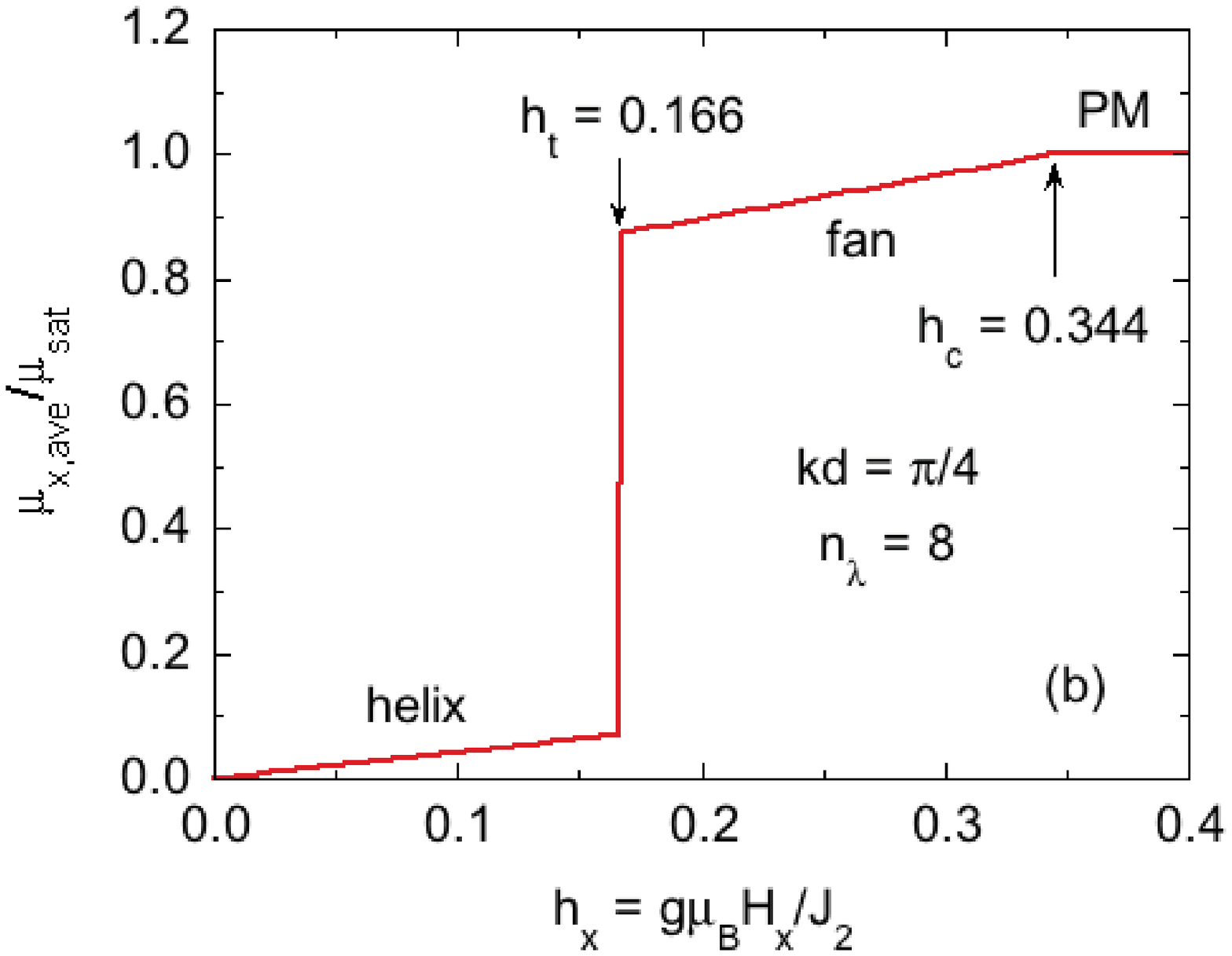}
\caption {(Color online) (a) The angles $\phi_0$, $\phi_1$, $\phi_2$, and $\phi_3$ of the corresponding moments with respect to the $+x$ axis versus reduced in-plane field $h_x$ for a helix with turn angle $kd=\pi/4$.  (b)~Average normalized magnetic moment per spin in the field direction, $\mu_{x{\rm ave}}/\mu_{\rm sat}$, versus~$h_x$.}
\label{Fig:MH_T0_n8_Helix_Fan}
\end{figure}

\begin{figure}
\includegraphics [width=3.3in]{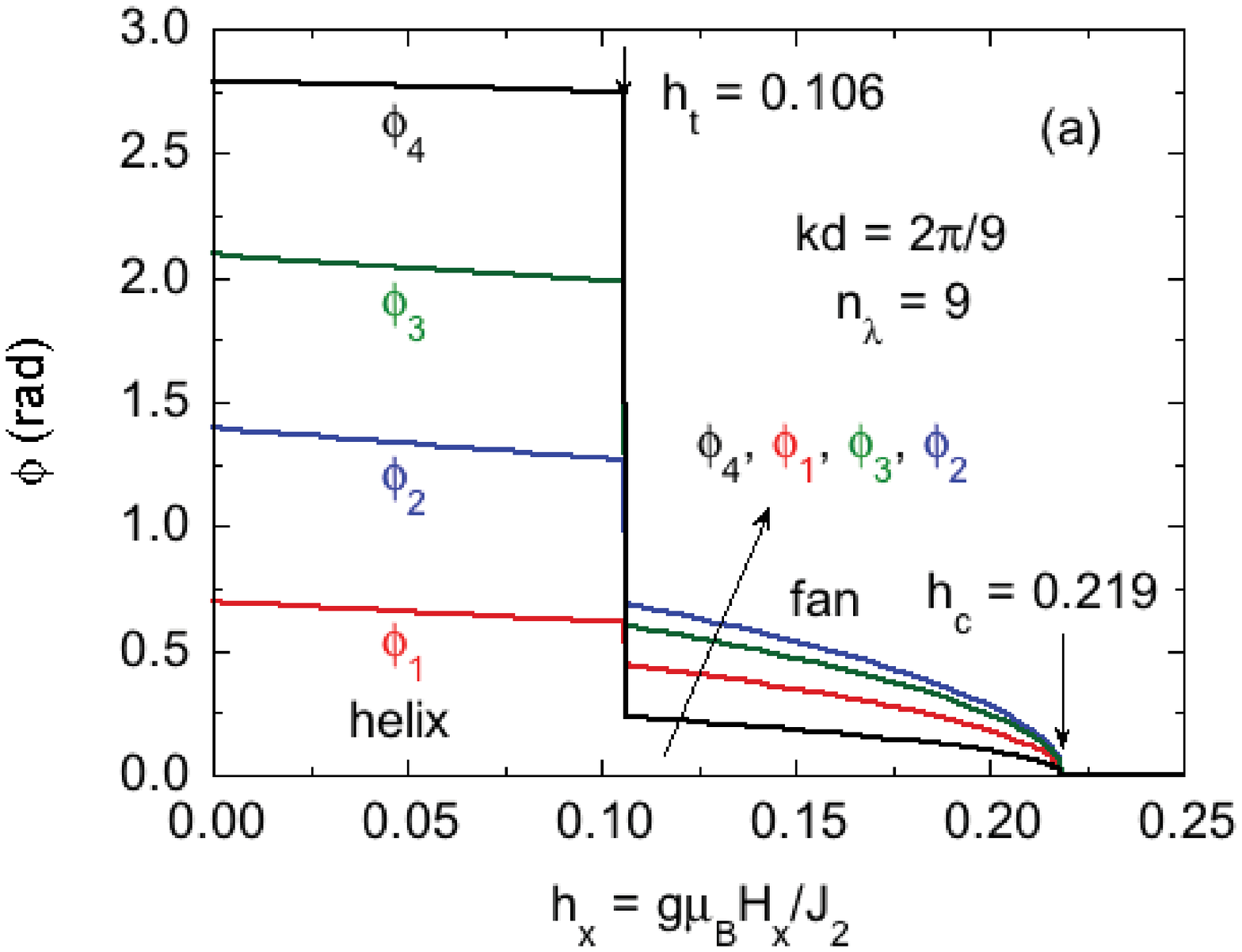}
\includegraphics [width=3.4in]{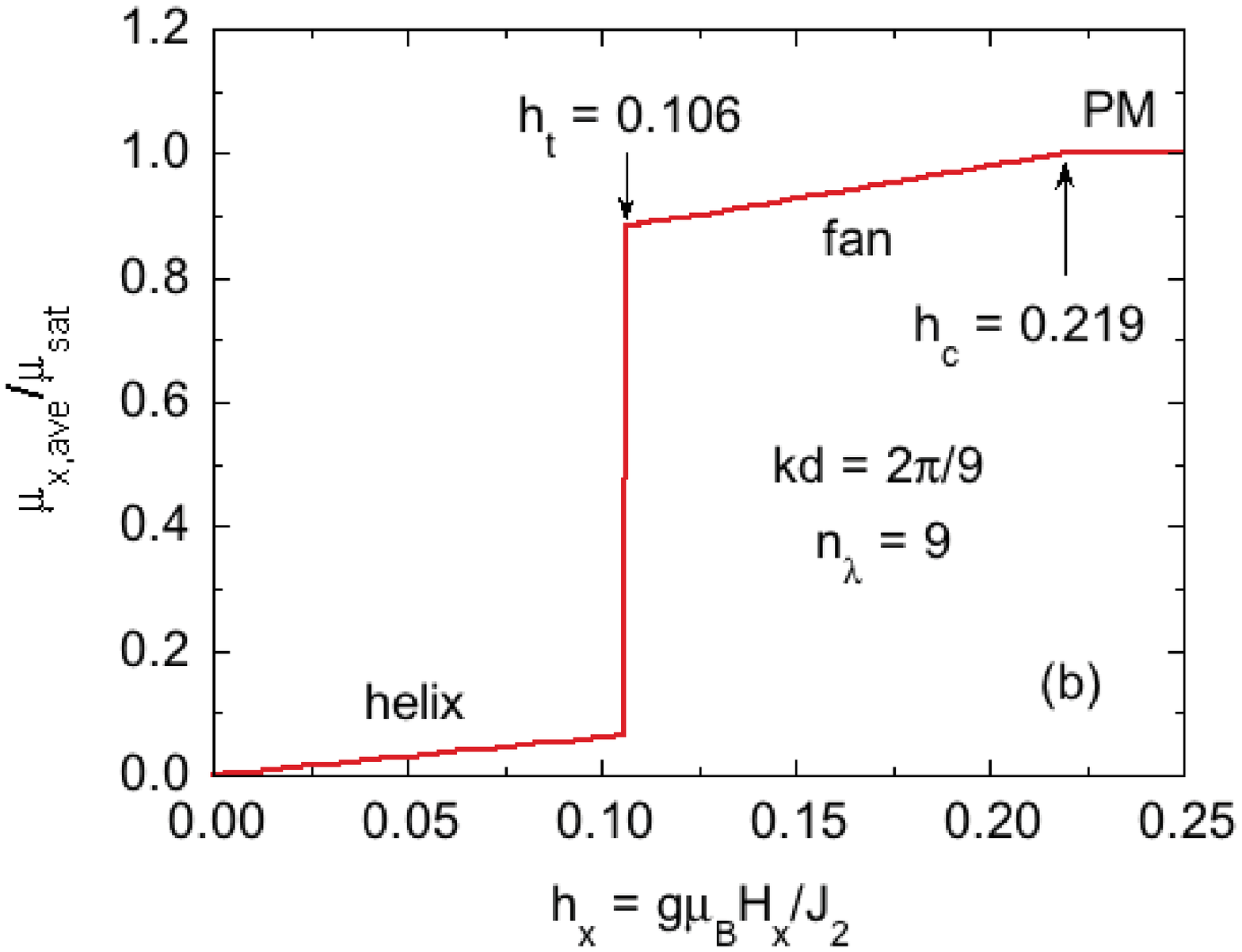}
\caption {(Color online) (a) The angles $\phi_1$--$\phi_4$ of the corresponding moments with respect to the $+x$ axis versus reduced in-plane field~$h_x$ for a helix with turn angle $kd=2\pi/9$.  (b)~Average normalized magnetic moment per spin in the field direction, $\mu_{x{\rm ave}}/\mu_{\rm sat}$, versus~$h_x$.}
\label{Fig:MH_T0_n9_Helix_Fan}
\end{figure}

\begin{figure}
\includegraphics [width=3.3in]{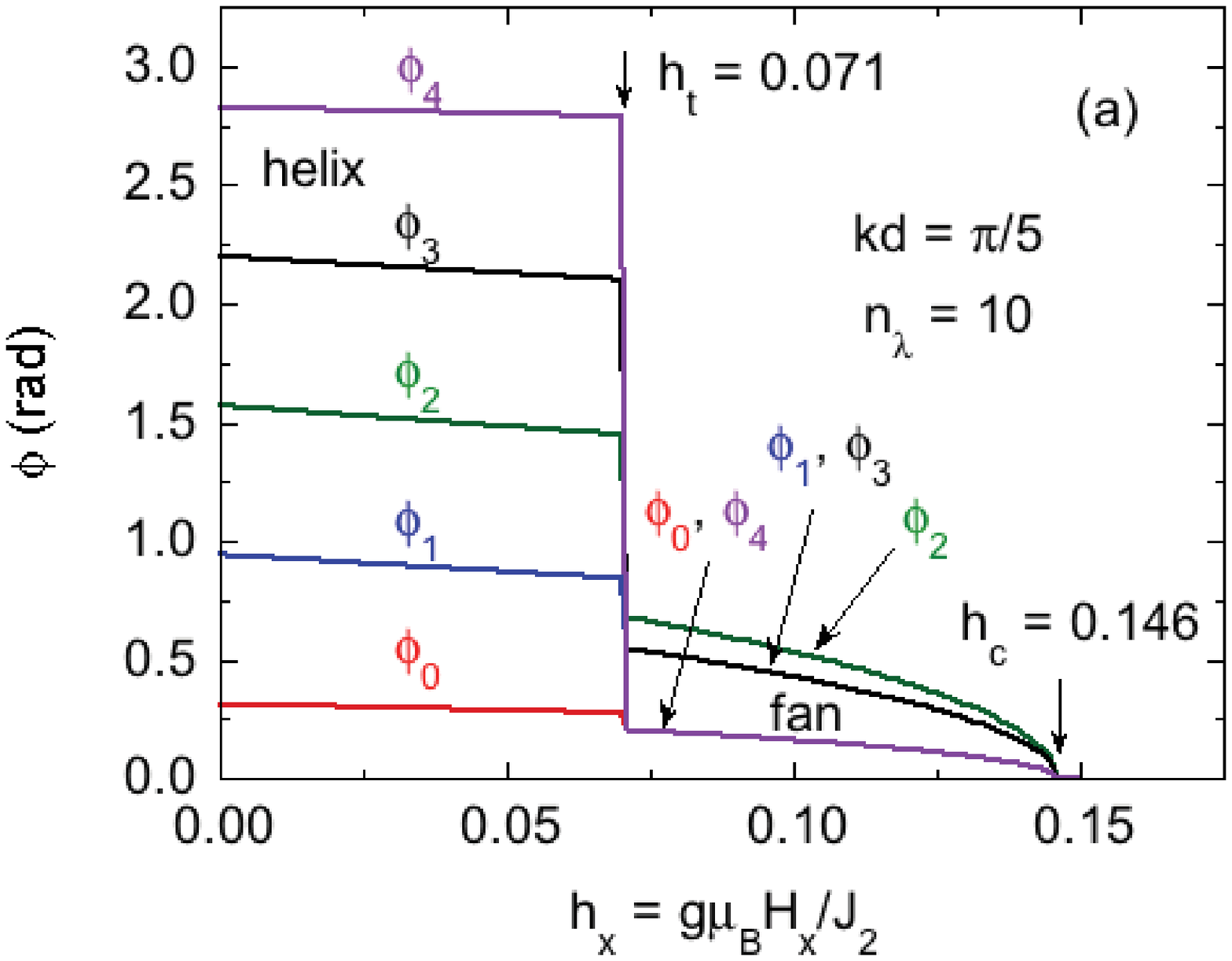}
\includegraphics [width=3.3in]{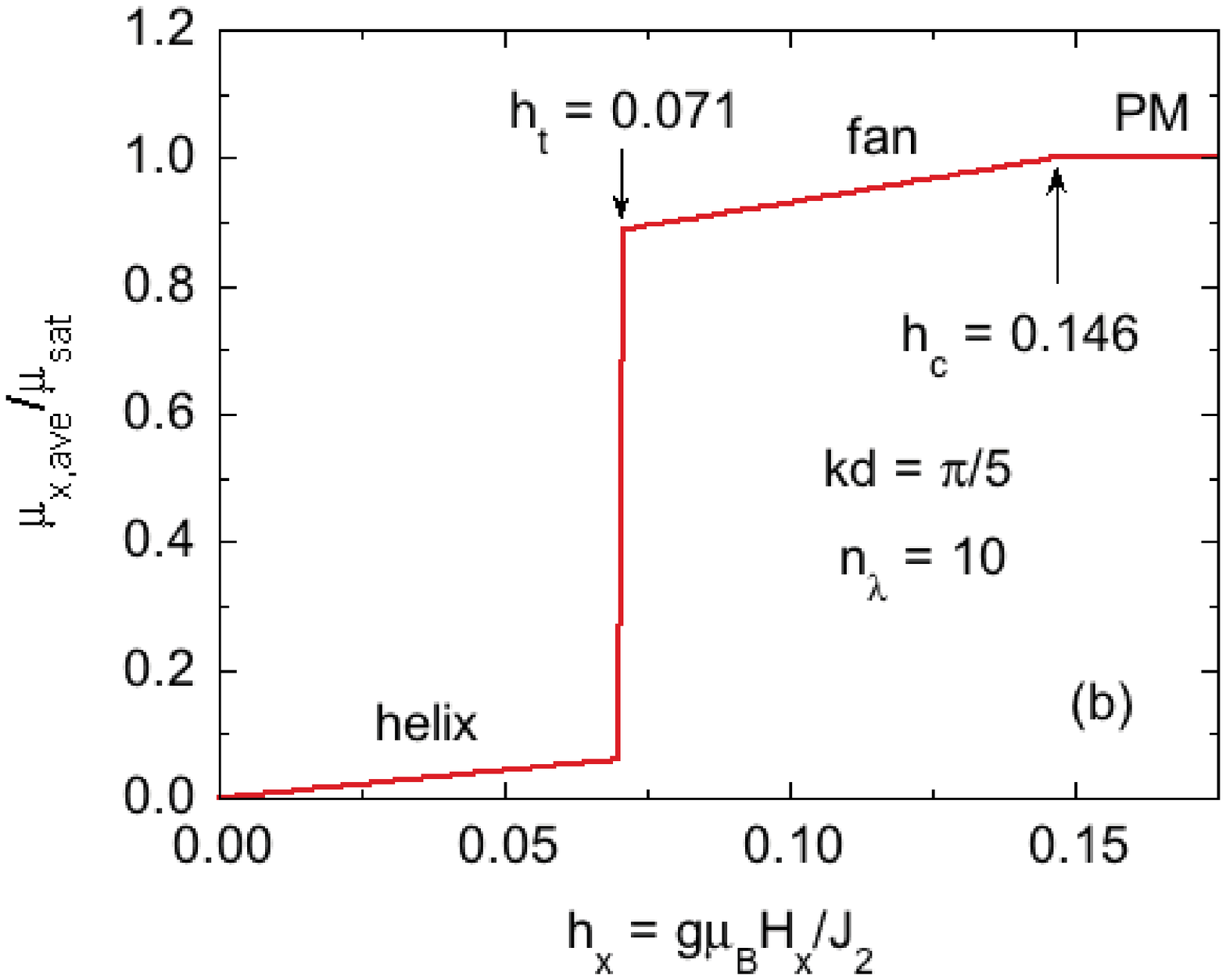}
\caption {(Color online) (a) The angles $\phi_0$ to $\phi_4$ of the corresponding moments with respect to the $+x$ axis versus reduced in-plane field $h_x$ for a helix with turn angle $kd=\pi/5$.  (b)~Average normalized magnetic moment per spin in the field direction, $\mu_{x{\rm ave}}/\mu_{\rm sat}$, versus~$h_x$.}
\label{Fig:MH_T0_n10_Helix_Fan}
\end{figure}

\begin{figure}
\includegraphics [width=3.3in]{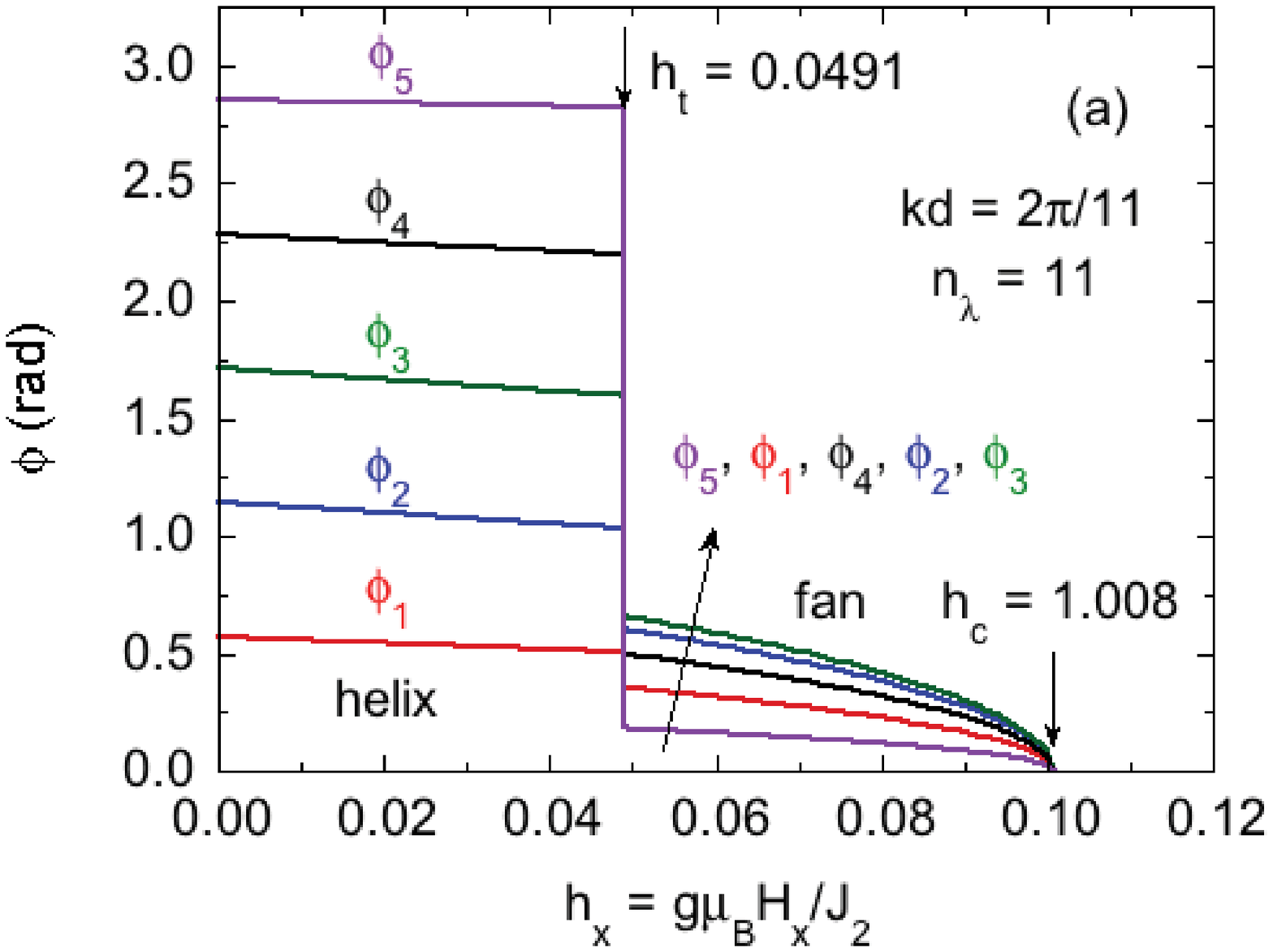}
\includegraphics [width=3.3in]{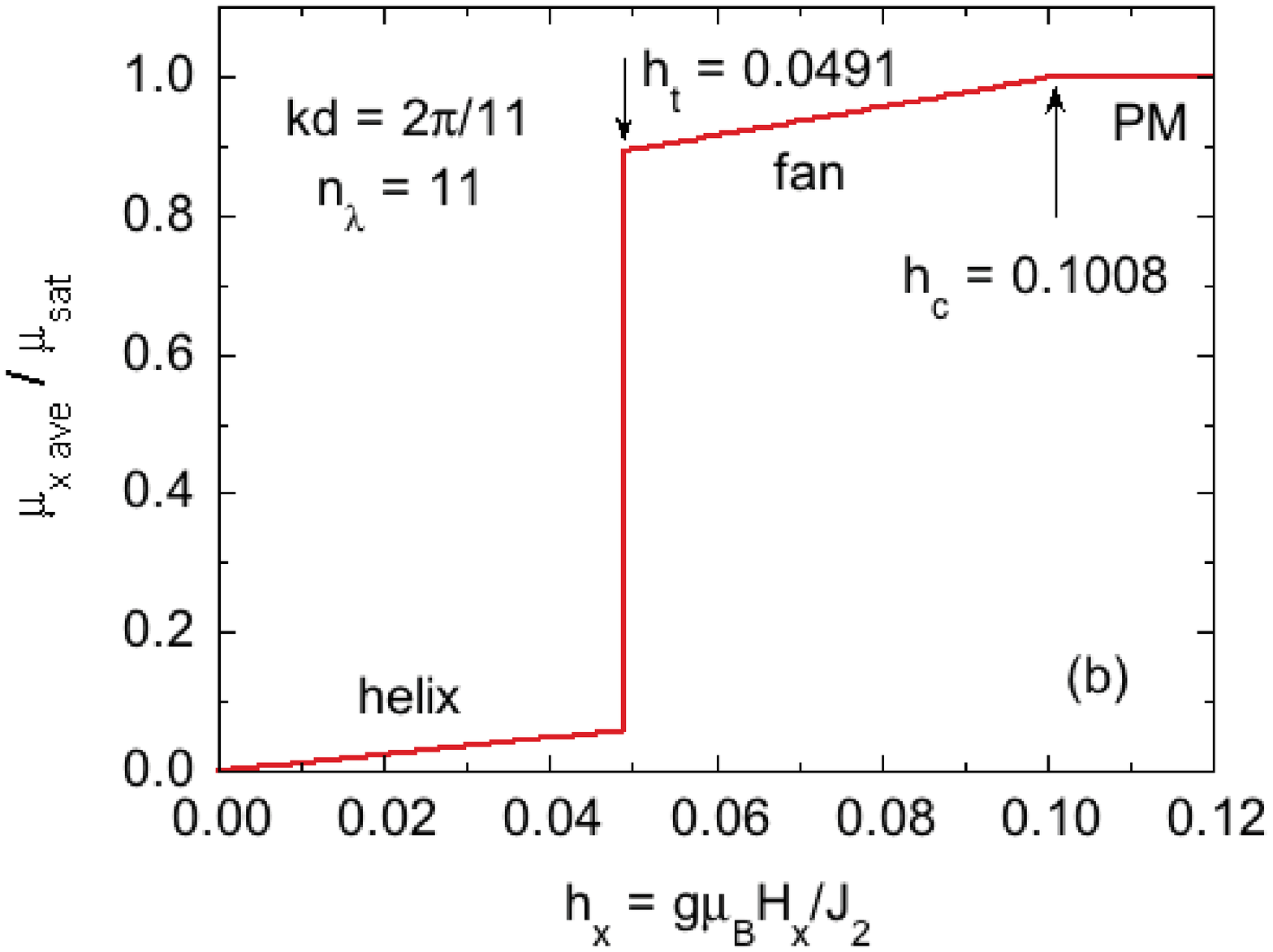}
\caption {(Color online) (a) The angles $\phi_1$ to $\phi_5$ of the correspondin moments with respect to the $+x$ axis versus reduced in-plane field $h_x$ for a helix with turn angle $kd=2\pi/11$.  (b)~Average normalized magnetic moment per spin in the field direction, $\mu_{x{\rm ave}}/\mu_{\rm sat}$, versus~$h_x$.}
\label{Fig:PhiMuxEnN11kd2PiOn11Helix}
\end{figure}

\begin{figure}[t]
\includegraphics [width=3.4in]{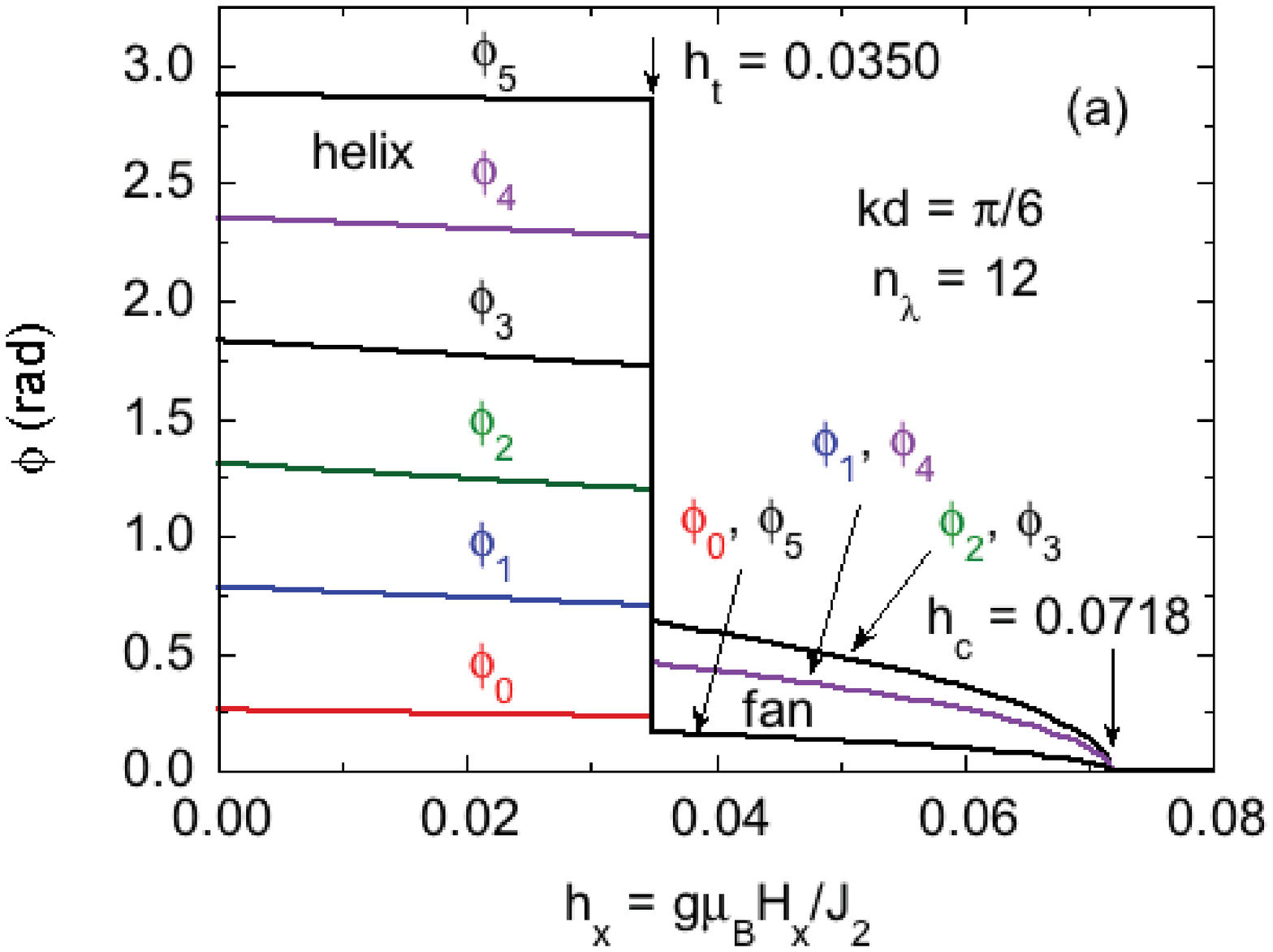}
\includegraphics [width=3.4in]{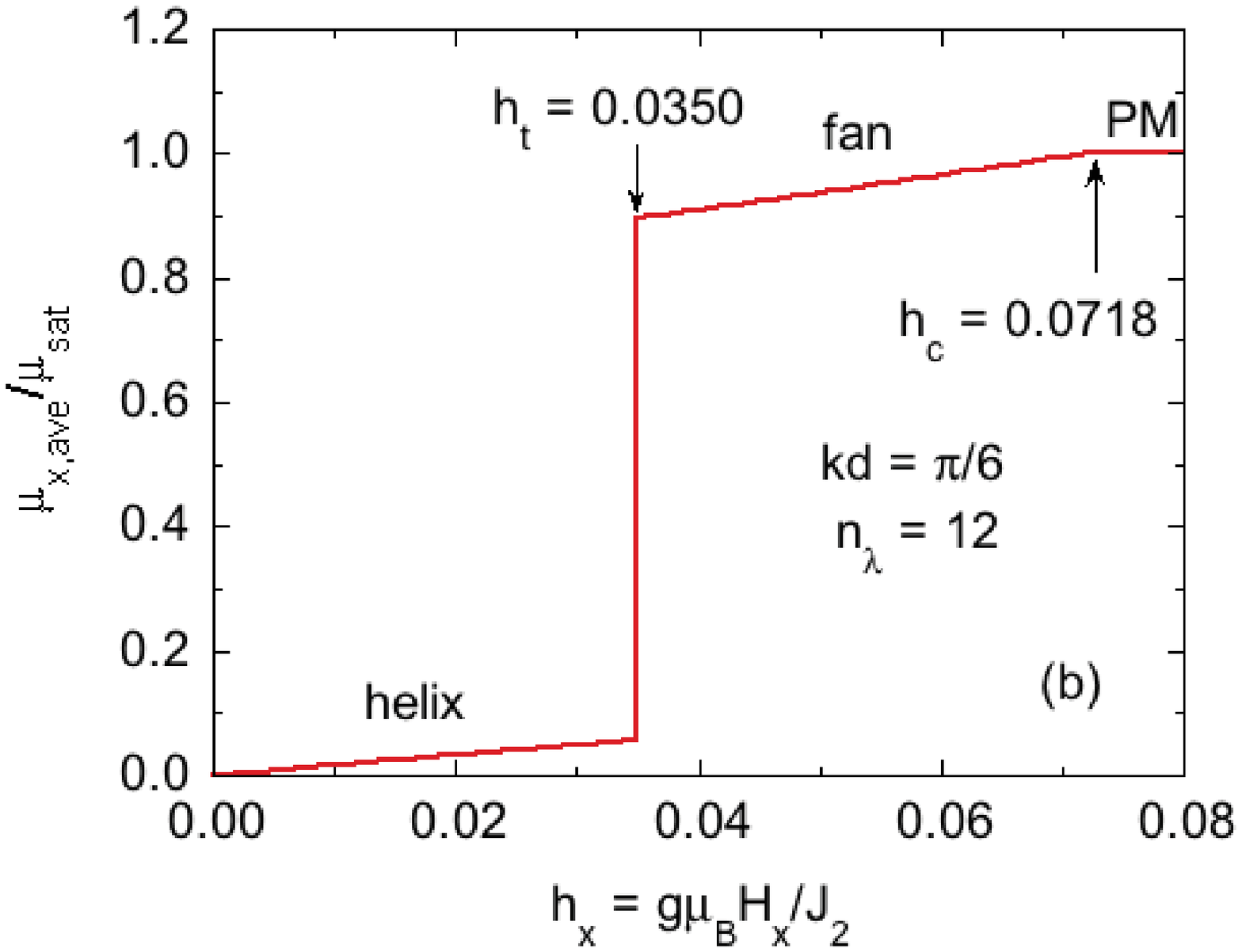}
\includegraphics [width=3.4in]{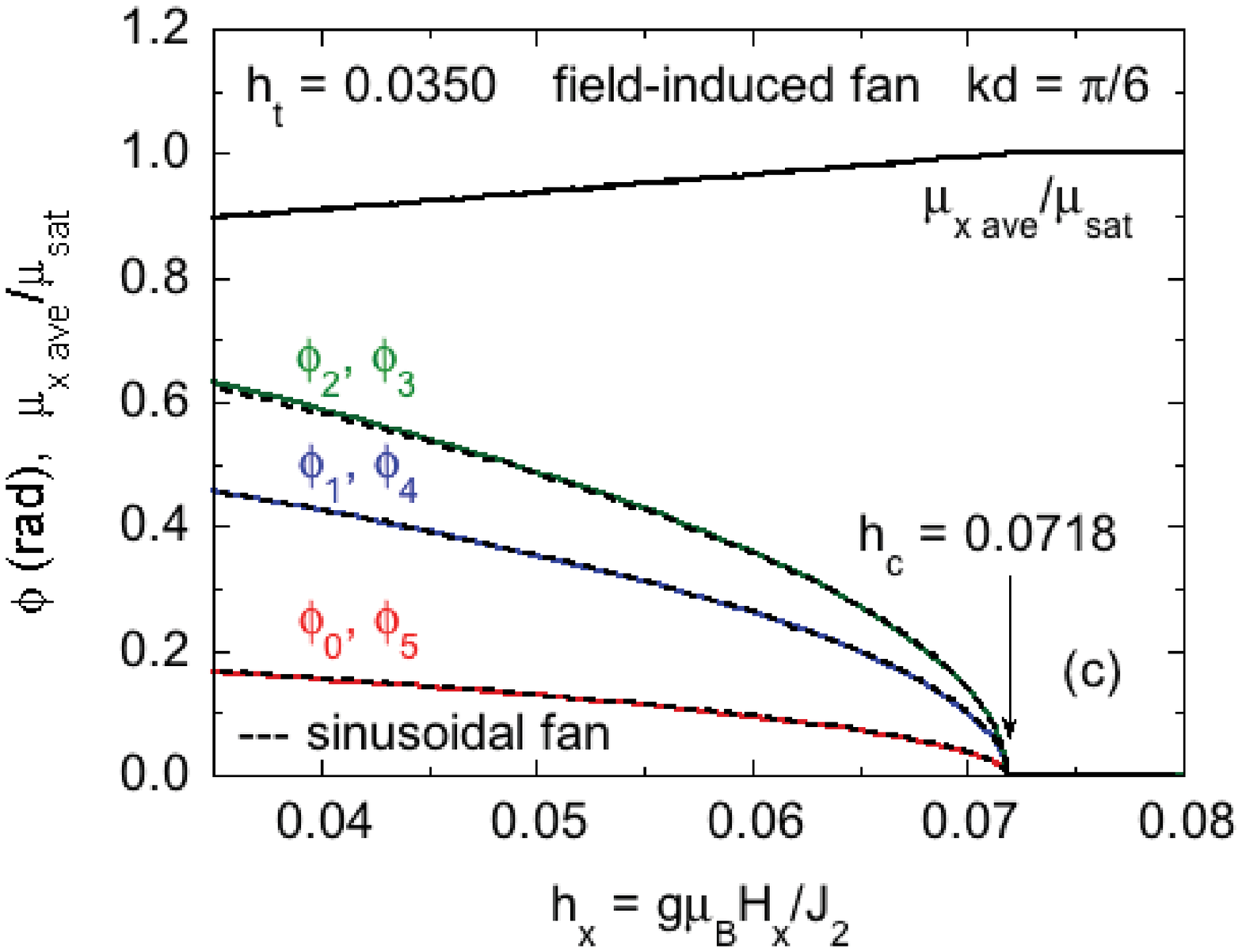}
\caption {(Color online) (a) The angles $\phi_0$ to $\phi_5$ of of the respective moments with respect to the $+x$ axis versus reduced in-plane field $h_x$ for a helix with turn angle $kd=\pi/6$.  (b)~Average normalized magnetic moment per spin in the field direction, $\mu_{x{\rm ave}}/\mu_{\rm sat}$, versus~$h_x$.  (c)~Expanded plots of the angles and average moment in the field-induced fan regime in (a) and~(b) (solid lines).  Also shown are the predictions for a sinusoidal fan with the same $kd$ and~$J_{12}$ (dashed lines).  }
\label{Fig:PhiMuxEnN12kd2PiOn12Helix}
\end{figure}

On the other hand, Figs.~\ref{Fig:MH_T0_n5_Helix_Fan}--\ref{Fig:PhiMuxEnN12kd2PiOn12Helix} for discrete values $kd = 2\pi/5$ to~$\pi/6$ exhibit strongly first-order transitions at~$h_{\rm t}$.  In all such cases, the ratio $h_{\rm t}/h_{\rm c}\sim 0.5$, even though $h_{\rm c}$ decreases by more than a factor of 20 from 1.91 to 0.0718 over this~$kd$ range.  Furthermore, the $\phi_n(h_x>h_{\rm t})$ data are approximated by sinusoidal fans, as shown  for $kd = 4\pi/11$ and~$\pi/6$ in Figs.~\ref{Fig:PhiMuxEnN11kd4PiOn11Helix} and~\ref{Fig:PhiMuxEnN12kd2PiOn12Helix}, respectively, where again the sinusoidal fan model is most accurate for $h_x\to h_{\rm c}$.

The discontinuity in $\mu_{x{\rm ave}}/\mu_{\rm sat}$ at~$h_x = h_{\rm t}$ increases strongly with decreasing~$kd$ from 0.547 for $kd=2\pi/5$ in Fig.~\ref{Fig:MH_T0_n5_Helix_Fan} to 0.839 for~$kd = \pi/6$ in Fig.~\ref{Fig:PhiMuxEnN12kd2PiOn12Helix}.  This monotonic increase in the discontinuity with decreasing~$kd$ is consistent with the value for the continuum model with $kd\to0$ in Fig.~\ref{Fig:muxVShx_Enz}, for which the discontinuity has a value of~0.875.

\clearpage

\section{\label{HelixDatSumm} Summary of Helix Data with Fixed Turn Angle \lowercase{kd} }

A summary of our data for the helix phase and helix to fan transitions where $kd$ is independent of field for the full range $0 < kd \leq \pi$ is given in Table~\ref{Tab:Helix/FanData}.  The data include the first- or second-order (as noted) reduced transition field~$h_{\rm t}$ and the critical field $h_{\rm c}$ at which the normalized average magnetization per spin $\mu_{x{\rm ave}}/\mu_{\rm sat}$ saturates to the value of unity obtained from energy minimization.  If the data for a given value of $kd$ exhibit only a crossover from helix to fan with increasing~$h_x$, the $h_{\rm t}$ column entry reads~``none''.  Also included for each value of~$kd$ in the table are the critical field for the sinusoidal fan by itself obtained from Eqs.~(\ref{Eqs:hcVals}), the ratio~$h_{\rm t}/h_{\rm c}$, and the initial reduced susceptibility $\chi_x(h_x\to 0) \equiv \mu_{x{\rm ave}}/h_x$ of the helix.

Table~\ref{Tab:Helix/FanData} shows that $h_{\rm c}$ determined from energy minimization, where the $\phi_n$ were free to vary independently to obtain the minimum energy, and the value $h_{\rm c\,Fan}$ for the sinusoidal fan with the same~$kd$ and $J_{12}$ given by the value $J_{12} = -4\cos(kd)$ for the helix, are identical.  This indicates that the field-induced fan structure of the helix approaches a sinusoidal fan with the same~$kd$ and~$J_{12}$ for $h_x\to h_{\rm c}$, as previously inferred \cite{Nagamiya1962}. 

\begin{figure}
\includegraphics [width=3.4in]{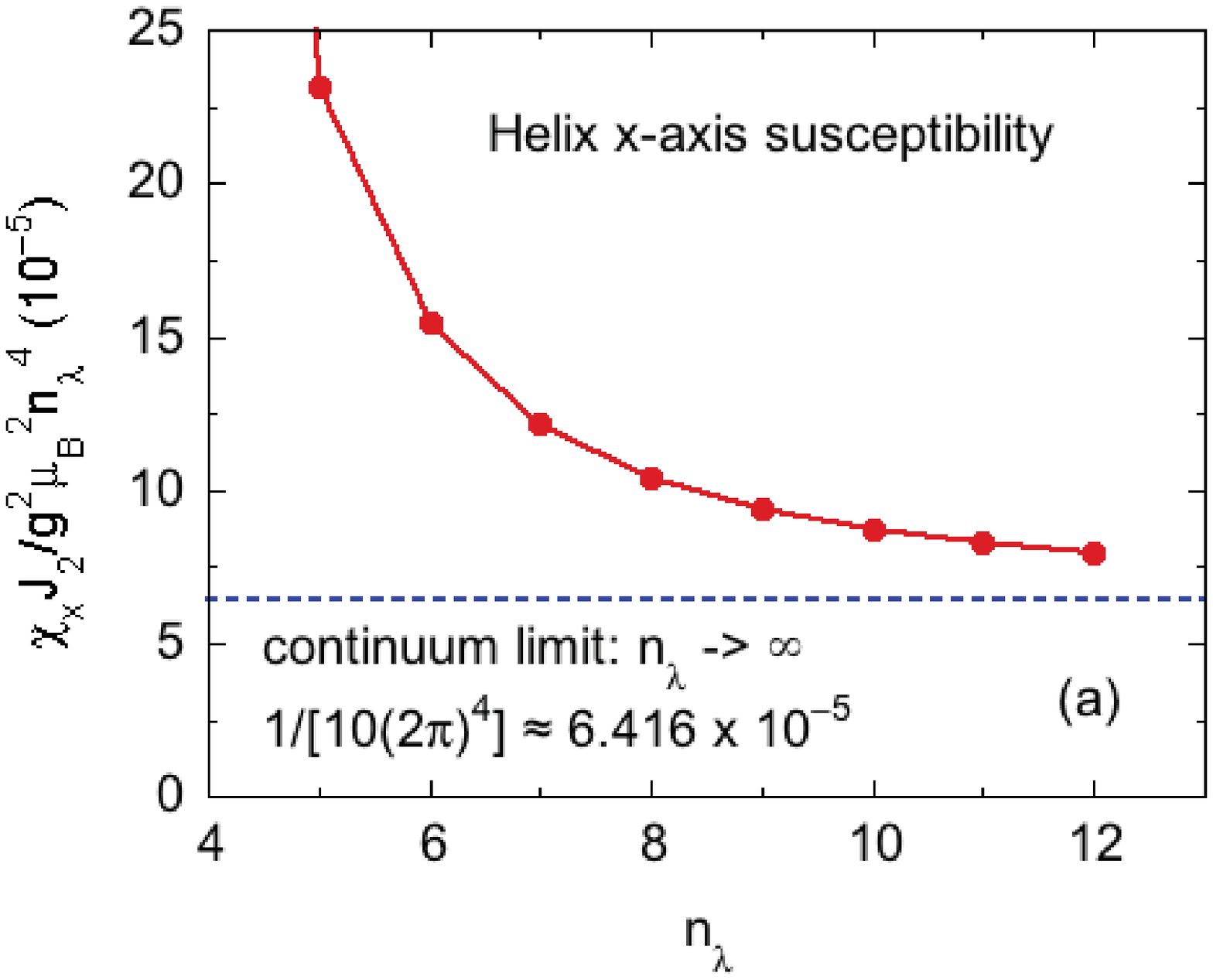}
\includegraphics [width=3.4in]{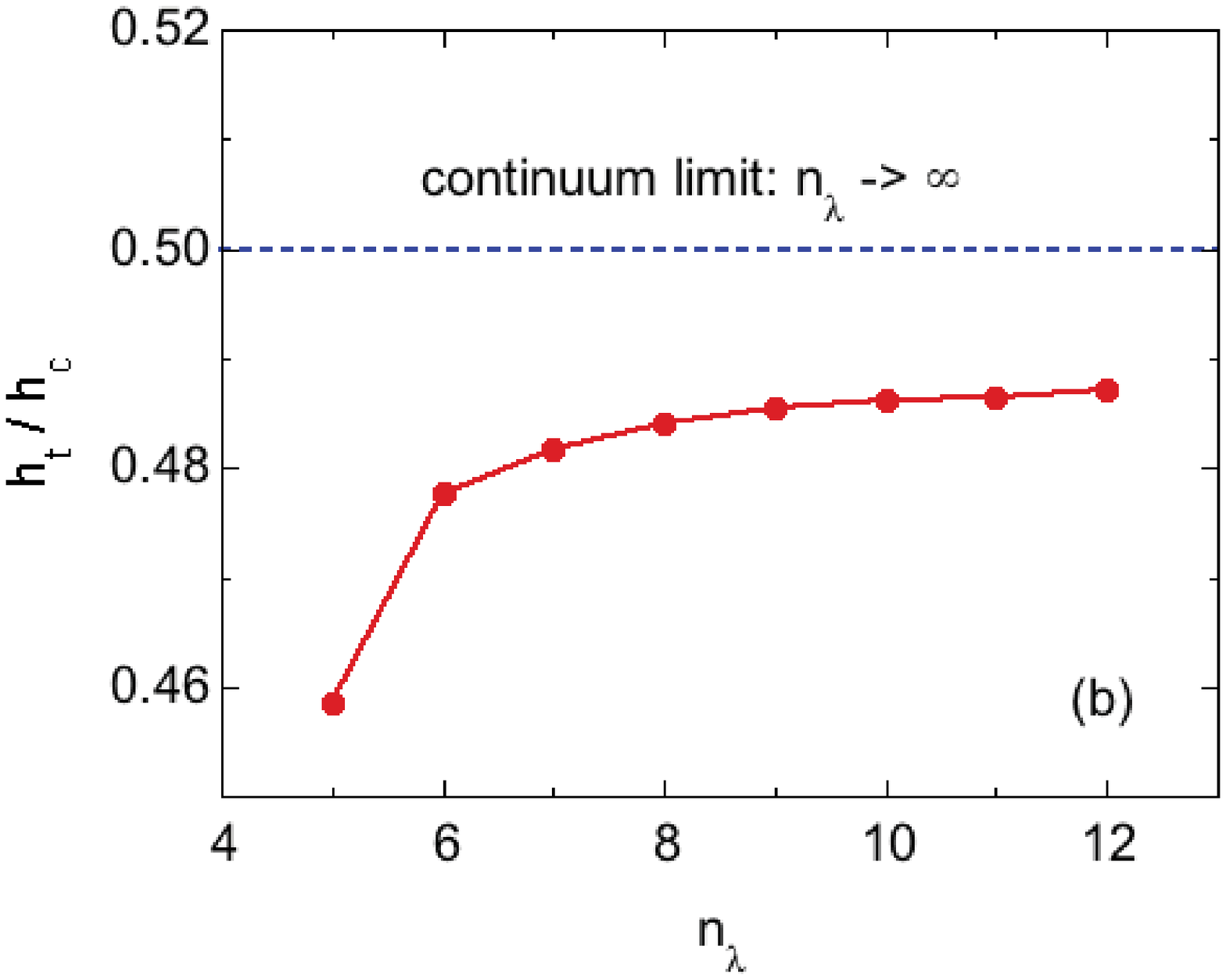}
\caption {(Color online) (a) Normalized magnetic susceptibility per spin~$\chi_x J_2/g^2\mu_{\rm B}^2$ divided by $n_\lambda^4$, where $n_\lambda$ is the number of layers per helix wavelength and the angle between the ordered moments in successive layers of the helix along the $z$~axis is $2\pi/n_\lambda$.  (b)~The ratio of the first-order helix to fan transition field $h_{\rm t}$ and the fan to paramagnet critical field $h_{\rm c}$.  The horizontal blue dashed lines in (a) and~(b) are the respective $n_\lambda\to\infty$ continuum limits in Eqs.~(\ref{Eqs:EnzLimits}) \cite{Enz1961}.}
\label{Fig:Helix_T0_Mag_Data_Chi}
\end{figure}

From the data in Table~\ref{Tab:Helix/FanData}, one sees that $\chi_x$ increases rapidly with decreasing~$kd$. For the limit $kd\to0$, Enz predicted \cite{Enz1961}
\bse
\label{Eqs:EnzLimits}
\be
\chi_x = \frac{1}{10(kd)^4}.
\ee 
For the series $kd=2\pi/n_\lambda$, one obtains
\be
\chi_x = \frac{n_\lambda^4}{10(2\pi)^4}\approx 6.416\times 10^{-5}n_\lambda^4.
\label{Eq:ChiOnLn4}
\ee
Plotted in Fig.~\ref{Fig:Helix_T0_Mag_Data_Chi}(a) are the values of $\chi_x/n_\lambda^4$ versus $n_\lambda$ from the data in Table~\ref{Tab:Helix/FanData} for the series $kd = 2\pi/n_\lambda$ with $n_\lambda = 5$ to~12.  Also shown as the horizontal dashed line is the continuum limit for $n_\lambda\to\infty$ given by Eq.~(\ref{Eq:ChiOnLn4}).  One sees that the continuum limit $n_\lambda\to\infty$ is already approached within $\sim20$\% by $n_\lambda = 12$.

Enz also made a prediction that the $kd\to0$ limit of the ratio $h_{\rm t}/h_{\rm c}$ is
\be
\frac{h_{\rm t}}{h_{\rm c}} = 1/2.
\label{Eq:hthckd0}
\ee
\ese
The ratios $h_{\rm t}/h_{\rm c}$ from Table~\ref{Tab:Helix/FanData}  are plotted in Fig.~\ref{Fig:Helix_T0_Mag_Data_Chi}(b) versus $n_\lambda$, again for the series $kd = 2\pi/n_\lambda$, and are seen to fairly closely approach the continuum limit of 1/2 (horizontal dashed line) with increasing $n_\lambda$ even by $n_\lambda=12$.

\begin{figure}[t]
\includegraphics [width=2.7in]{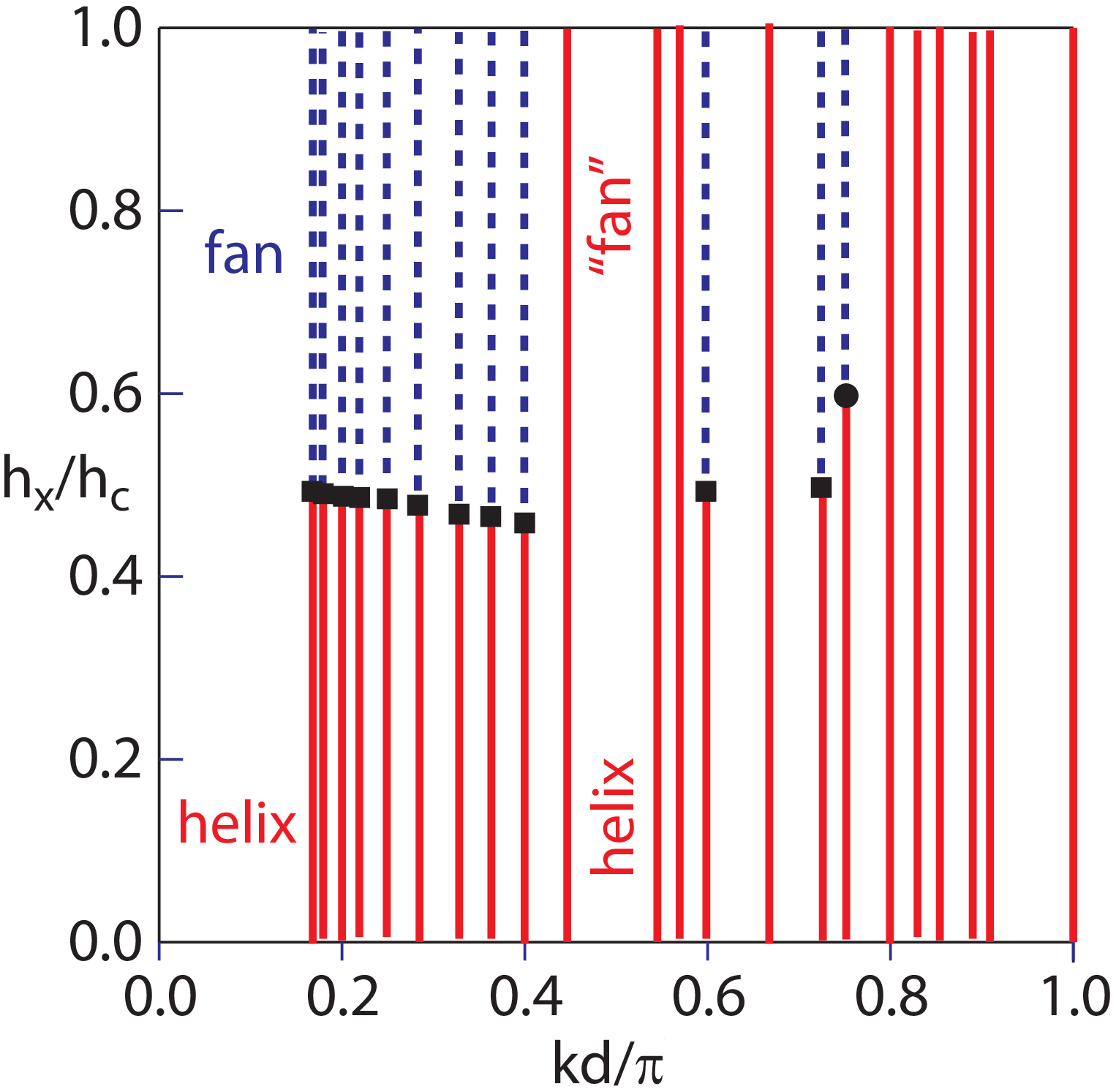}
\caption {(Color online) Reduced magnetic field $h_x/h_{\rm c}$ versus normalized turn angle~$kd/\pi$ phase diagram determined from the data in Table~\ref{Tab:Helix/FanData}.  The red solid vertical lines represent the field stability range of the (distorted) helix structure, whereas the vertical dashed blue lines represent the fan phase which can be approximated by a sinusoidal fan phase.  Field-induced first-order transitions are denoted by filled squares.  The single second-order transition is denoted by a filled circle.  The helix to ``fan'' notation denotes a continuous evolution from the helix to a fanlike structure that is quite different from a sinusoidal fan.  The structures and properties of the helix and fan are indistinguishable for each of $kd/\pi = 2/3$ and~1.}
\label{Fig:Fan_Phase_Diagram}
\end{figure}

The zero-temperature phase diagram obtained from the data in Table~\ref{Tab:Helix/FanData} is shown in Fig.~\ref{Fig:Fan_Phase_Diagram}, where the 11 first-order transitions and the single second-order transition are indicated by filled squares and a filled circle, respectively.  It appears that the region $0< kd \leq 2\pi/5$ forms a continuum of phases where the helix phase undergoes a first-order transition to a fan phase at a reduced field $h_{\rm t}/h_{\rm c}$ that smoothly increases with decreasing $kd$, reaching the continuum limit of 1/2 in Eq.~(\ref{Eq:hthckd0}) for $kd\to0$.  On the other hand, the first and second-order transitions for $3\pi/5 \leq kd \leq 3\pi/4$ are nestled between $kd$ values that show continuous crossovers between the helix and fan phases, and further surprises may be in store if additional rational values of $kd$ are explored in the range $4\pi/9 < kd < 1$.


\begin{table*}
\caption{\label{Tab:Helix/FanData} Phase transition fields for the helix to fan ($h_{\rm t}$) if present, and helix or fan to the paramagnetic phase ($h_{\rm c}$), obtained by minimizing the energy of the helix or field-induced fan.  The critical field for the fan phase with $J_{12}$ set to the value for the helix, taken from the exact data in Table~\ref{TabFanData} and Eqs.~(\ref{Eqs:hcVals}), are listed as $h_{\rm c\,Fan}$.  Also shown are the initial reduced susceptibilities of the helix phase $\chi_x \equiv \mu_{x{\rm ave}}/h_x$.  Exact solutions are given if obtained. ``N/A'' means ``not applicable''.  }
\begin{ruledtabular}
\begin{tabular}{cccccccc}
$kd$		&	$kd/\pi$	&	$n_\lambda$ 	&	 $h_{\rm t}$			& $h_{\rm c}$	& $h_{\rm c\,Fan}$	& $h_{\rm t}/h_{\rm c}$	&	$\chi_x(h_x\to0)$	\\
		&			&				&	(helix $\to$ fan)		&(helix/fan)	& (helix $J_{12}$)	&					&	(helix)				\\
\hline 
$\pi$ 	&  	1		&	2			&	none  				&  	16	  	&	16			&    N/A				&	$1/16 = 0.0625$		\\ 
$10\pi/11$  &  0.909091	&	11			&	none					&	15.359  	&	15.3585		&	N/A				&	0.0352993	\\
$8\pi/9$  &  0.888889	&	9			&	none					&	15.050  	&	15.0496		&	N/A				&	0.0374703	\\
$6\pi/7$  &  0.857143	&	7			&	none					&	14.455  	&	14.4547		&	N/A				&	0.0421042	\\
$5\pi/6$  &  0.833333	&	12			&	none					&	13.929  	&	13.9282		&	N/A				&	$(11 - 6\sqrt{3})/13 \approx0.0467458$	\\
$4\pi/5$  &  0.8		&	5			&	none					&	13.091  	&	13.0902		&	N/A				&	$(1-1/\sqrt{5})/10 \approx 0.0552786$	\\
$3\pi/4$  &  0.75		&	8			&	7.028\footnotemark[1]	&	11.657   	&	11.6569		&	0.6029 			&	$(1-1/\sqrt{2})/4 \approx 0.0732233$	\\
$8\pi/11$  &  0.727273	&	11			&	5.459\footnotemark[2]	&	10.955   	&	10.9543		&	0.4983 			&	0.0832981	\\
$2\pi/3$  &  0.666667	&	3			&	none					&	9    	&	9			&	N/A				&	$1/9 \approx 0.111111$		\\
$3\pi/5$  &  0.6		&	10			&	3.205\footnotemark[2]	&	6.860    	&	6.85410		&	0.4672			&	$(3-\sqrt{5})/6 \approx 0.127322$	\\
$4\pi/7$  &  0.571429	&	7			&	none					&	5.979    	&	5.97823		&	N/A				&	0.127887	\\
$6\pi/11$	&  0.545455	&	11			&	none					&	5.220	&	5.21953		&	N/A				&	0.126732	\\
$\pi/2$	&  0.5		&	4			&	N/A					&	N/A		&	4			&	N/A				&	N/A	\\
$4\pi/9$	&  0.444444	&	9			&	none					&	2.732	&	2.73143		&	N/A				&	0.130047	\\
$2\pi/5$	&  0.4		&	5			&	0.8758\footnotemark[2]	&	1.910	&	1.90983		&	0.4585			&	$(1+1/\sqrt{5})/10 \approx 0.144721$	\\
$4\pi/11$	&  0.363636	&	11			&	0.6475\footnotemark[2]	&	1.367	&	1.36696		&	0.4737			&	0.168098	\\
$\pi/3$	&  0.333333	&	6			&	0.4777\footnotemark[2]	&    1		&	1			&    0.4777			&	$1/5 = 0.2$	\\
$2\pi/7$	&  0.285714	&	7			&	0.2732\footnotemark[2]	&    0.5671	&	0.567040		&    0.4818			&	0.291547	\\
$\pi/4$	&  0.25		&	8			&	0.1661\footnotemark[2]	&    0.3432	&	0.343146		&    0.4840			&	$(1+1/\sqrt{2})/4 \approx 0.426777$	\\
$2\pi/9$	&  0.222222	&	9			&	0.1063\footnotemark[2]	&    0.2190	&	0.218941		&    0.4854			&	0.616267	\\
$\pi/5$	&  0.2		&	10			&	0.0710\footnotemark[2]	&    0.1460	&	0.145898		&	0.4863			&	$(3+\sqrt{5})/6 \approx 0.872678$	\\
$2\pi/11$	&  0.181818	&	11			&	0.04905\footnotemark[2]	&    0.10081	&	0.100802		&	0.4866			&	1.210426	\\
$\pi/6$	&  0.166667	&	12			&	0.03497\footnotemark[2]	&    0.07180	&	0.0717968		&	0.4870			&	$(11 + 6\sqrt{3})/13 \approx 1.645562$	\\
$0^+$	&	$0^+$	&	$\infty$ 		&	$0^+$				&	$0^+$	&	$0^+$		&	1/2 \cite{Enz1961}	&	$\infty$	\cite{Enz1961}	\\
\end{tabular}
\end{ruledtabular}
\footnotetext[1]{Second-order transition}
\footnotetext[2]{First-order transition}
\end{table*}

\clearpage

\acknowledgments

The author is grateful for collaboration and discussions about \ecp\ with N.~S.~Sangeetha.  This work was supported by the U.S. Department of Energy, Office of Basic Energy Sciences, Division of Materials Sciences and Engineering.  Ames Laboratory is operated for the U.S. Department of Energy by Iowa State University under Contract No.~DE-AC02-07CH11358.

\end{document}